\documentclass{emulateapj}
\usepackage{lscape}

\slugcomment{Accepted for publication in  Astrophysical Journal Supplement 20070912}

\shorttitle{AO IMAGING OF LBGS AT $z\sim3$}
\shortauthors{AKIYAMA ET AL.}

\begin{document}

\title{ADAPTIVE OPTICS REST-FRAME V-BAND IMAGING OF LYMAN BREAK GALAXIES AT $z\sim3$: 
HIGH-SURFACE DENSITY DISK-LIKE GALAXIES ?\altaffilmark{1}}

\author{MASAYUKI AKIYAMA\altaffilmark{2}, YOSUKE MINOWA\altaffilmark{3},
NAOTO KOBAYASHI\altaffilmark{4}, KOUJI OHTA\altaffilmark{5}, MASATAKA ANDO\altaffilmark{5},
AND IKURU IWATA\altaffilmark{6}}

\altaffiltext{1}{Based on data collected at Subaru Telescope,
which is operated by the National Astronomical Observatory of Japan.}

\altaffiltext{2}{
Subaru Telescope, National Astronomical Observatory of Japan,
Hilo, HI, 96720; akiyama@subaru.naoj.org}
\altaffiltext{3}{
Optical and Infrared Astronomy Division, NAOJ,
Mitaka, 181-8588, Japan; minoways@optik.mtk.nao.ac.jp}
\altaffiltext{4}{
Institute of Astronomy, University of Tokyo,
Mitaka, 181-0015, Japan; naoto@ioa.s.u-tokyo.ac.jp}
\altaffiltext{5}{
Department of Astronomy, Kyoto University,
Kyoto, 606-8502, Japan; ohta, andoh@kusastro.kyoto-u.ac.jp}
\altaffiltext{6}{
Okayama Astrophysical Observatory, NAOJ, 
Okayama, 719-0232, Japan; iwata@oao.nao.ac.jp}

\begin{abstract}

In order to reveal the rest-frame $V$-band morphology of galaxies
at $z\sim3$, we conducted AO-assisted $K$-band imaging
observations of $z\sim3$ LBGs whose absolute magnitues range from
$M^{*}_{V}-0.5$ mag to $M^{*}_{V}+3.0$ mag with $M^{*}_{V}$, the characteristic
absolute magnitude of $z\sim3$ LBGs, of $-24.0$ mag.
The AO observations resolve
most of the LBGs at a resolution of FWHM$\sim0.\!^{\prime\prime}2$.
The median apparent half-light radius, $r_{\rm HL}$, is $0.\!^{\prime\prime}23$, 
i.e., 1.8 kpc at $z=3$. LBGs brighter than $M^{*}_{V}$
have larger $r_{\rm HL}$ ($0.\!^{\prime\prime}40$)
than the fainter LBGs ($0.\!^{\prime\prime}23$) on average,
and there is no bright LBGs with a small $r_{\rm HL}$.
The LBGs brighter than $M^{*}_{V}$ have red rest-frame $U-V$ colors (average of $0.2$ mag) and 
most of the fainter LBGs show blue rest-frame $U-V$ colors (average of $-0.4$ mag).
The $K$-band peaks of some of the LBGs brighter than $K=22.0$ mag show 
significant shift from those in the optical images.
We fit S\'ersic profile to the images of the LBGs with $K<21.5$ mag,
taking care of the uncertainty of the final PSFs at the position
of the targets. The images of all but one of the LBGs with $K<21.5$ mag are fitted
well with S\'ersic profile with $n$ index less than 2, similar to
disk galaxies in the local universe. For the fainter LBGs, 
we examine concentration parameter instead of fitting S\'ersic profile;
the concentration parameters of the LBGs are consistent with
those of disk galaxies.
Assuming that the LBGs have a disk-shape, we
compared their size-luminosity and size-stellar mass relation with
those of $z=0$ and $z=1$ disk galaxies.
The LBGs are brighter than $z=0$ and $z=1$ disk
galaxies at the same effective radius. The rest-frame $V$-band surface brightness of the LBGs
are 2.2-2.9 mag and 1.2-1.9 mag brighter than the disk galaxies at $z=0$ and $z=1$,
respectively. The size-stellar mass relation of the LBGs 
shows that the effective radii of the LBGs do not 
depend on their stellar mass. For the LBGs brighter than $M_{V}^{*}$, the average
surface stellar mass density is 3-6 times larger than those
of the $z=0$ and $z=1$ disk galaxies. 
On the contrary, the size-stellar mass relation of the
less-luminous LBGs is similar to those of $z=0$ and $z=1$ disk galaxies.
We also examine the profiles of the serendipitously observed DRGs.
They are fitted with the S\'ersic profiles with $n<2$, and
their scatter on the $C$ vs. $r_{\rm HL}$ plane is similar to
that of the $z\sim3$ LBGs. The average surface stellar mass
density of the DRGs is even larger than that of the $z\sim3$ LBGs brighter than $M_{V}^{*}$.
The implications of the dominance of $n<2$ population among galaxies at $z\sim3$ 
and the presence of the high surface stellar mass density disk
systems are discussed.
\end{abstract}


\section{INTRODUCTION}

The statistical evolution of the morphology of galaxies is one of the most
fundamental observational clues to understand the establishing process of
the Hubble sequence of galaxies seen in the local universe.
Especially studies of the morphological evolution of galaxies in the 
rest-frame optical wavelength are important because the evolution 
reflects transformation of stellar mass distributions of galaxies
in the cosmological timescale.
High-resolution imaging observations with Hubble Space Telescope (HST),
covering rest-frame optical wavelength up to $z\sim1$, 
have shown that the Hubble sequence is already established at $z\sim1$, though the
fraction of each morphological type is different from that in the local universe 
(e.g. Conselice, Blackburne, \& Papovich 2005).
Detailed morphological parameters have been examined up to $z\sim1$, for example, 
the relation between size and luminosity of disk galaxies has been
extensively studied \citep{lil98,sim99,rav04,bar05}.
Some of the results suggest that there is an evolution in the 
size-luminosity relation of disk galaxies; disk galaxies at
$z\sim1$ are more luminous than disk galaxies at $z\sim0$ at the same
scale length \citep{lil98,bar05}, although other results indicate
that the effect of the selection bias against low-surface brightness galaxies
cannot be ignored and there is
mild or no evolution in the size-luminosity relation \citep{sim99,rav04}.
Even so, the evolution of the size-luminosity relation is thought to
be due to fading of the stellar populations in the disks in stead of
due to changes of dynamical properties of the disks, and
the size-mass relation, which reflects the surface stellar mass density of the disks, 
does not show significant evolution in the redshift range \citep{bar05}.
For elliptical galaxies, the morphological evolution has been
examined in point of view of the evolution of the fundamental plane
(e.g., \citealt{van98}).
Shifts and rotations of the fundamental plane have been observed between $z=1$ and $z=0$,
and they have been explained by a mass-dependent evolution of the mass-to-luminosity
ratios of the elliptical galaxies \citep{hol05, van05, dis05, tre05, jor06}.

Morphological studies on galaxies at higher redshifts started following
the discovery of a large number of high redshift galaxies using Lyman Break
selection and spectroscopic confirmation of their redshifts with 8-10m class telescopes
(see review by \citealt{gia02}). The spatial clustering of the Lyman Break Galaxies (LBGs)
is strong, and this implies that they are progenitors
of local elliptical galaxies (e.g., \citealt{ade05}).
Optical imaging observations
with HST, covering the rest-frame UV wavelength for high-redshift galaxies,
have revealed that the morphology of the LBGs is
completely different from that of $L^{*}$ galaxies at $z<1$.
For example $U$-dropout LBGs at $z\sim3$ have a compact core or multiple knots embedded in an extended
halo \citep{ste96, gia96, low97}. The central surface brightness in the rest-frame UV wavelength
is constant among LBGs with 
$23$ mag arcsec$^{-2}$. If we estimate the central surface brightness in the optical wavelength
assuming a flat UV-optical spectral energy distribution (SED), the central surface brightness
is much brighter than that of local galaxies \citep{ste96, gia96}.
The radial profiles of most of the LBGs can be fitted well with
either the exponential- or $r^{1/4}$-law in the same way as for the galaxies in the local universe.
The half-light radii of the LBGs become smaller at higher redshifts with $(1+z)^{-1}$
up to $z=6$ \citep{bou03,bou04,fer04,dah07}.

The peculiar UV morphology of the LBGs could only show unobscured 
star-forming regions rather than a more evolved underlying population
\citep{col96, oco97, bun00}.
In order to examine the stellar mass distributions of the LBGs, it is important to reveal
the distribution of a low-mass stellar population which reflects the
integrated star formation history of the system in the long time duration.
The long-lived low-mass stars dominate the mass of the stellar population, although
they only significantly contribute to the optical light of the stellar 
population in the wavelength range above 4000{\AA}, because of the 4000{\AA}
break in the spectra of the low-mass stars. 
With the aim to trace the distribution of the low-mass stellar population of LBGs,
$H$-band observations, which cover 4000{\AA} of LBGs at $z\sim3$, for
LBGs in the Hubble Deep Field North (HDFN) have been conducted with NICMOS 
camera on HST. The morphology of the LBGs is essentially independent
of the wavelength up to the rest-frame 4000{\AA} \citep{dic98, gia02}.
Based on the same data, 
Papovich et al. (2005) report little difference
between the rest-frame UV and optical morphologies for galaxies at $z\sim2.3$,
at slightly lower redshifts than LBGs.

Nevertheless, the NICMOS observations are not sufficient to conclude the distribution
of the low-mass stellar population in $z\sim3$ LBGs. The NICMOS observations cover 
up to the $H$-band, which is just at 4000{\AA} in the rest-frame of $z\sim3$ galaxies, and
$K$-band observations
covering rest-frame 5500{\AA} are more ideal to detect the low-mass stellar population.
In fact,
differences between optical and $K$-band morphologies of luminous galaxies at $z=2-3$ have been
reported based on ground-based $K$-band observations 
for Luminous Disk-like Galaxies (LDGs) \citep{lab03} or $z\sim2$ UV-selected galaxies \citep{for06}.
Statistical $K$-band studies of galaxies out to $z\sim3$ have been conducted based on deep
ground-based seeing-limited images \citep{tru04, tru06a}, but due
to the large seeing size, measurements of 
small distant galaxies can be systematically affected \citep{zir07}.

In addition, the NICMOS
observations are mostly limited to the $z\sim3$ LBGs in the small HDFN area and only
include objects as bright as $M_{\rm B}\sim-22$ mag \citep{pap05}, which is still
0.5 mag fainter than the characteristic absolute magnitude of the $z\sim3$ LBGs
($M^{*}_{B}\sim-22.5$ mag, if we use $B-V=0.5$ mag and $M^{*}_{\rm V}=-22.98\pm0.25$ mag 
which corresponds to $K=20.70\pm0.25$ mag for objects at $z=3$;
\citealt{sha01}; hereafter all magnitudes are in Vega-based system).
In order to fully understand the LBG population, observations of
an LBG sample covering a wide luminosity range are necessary, 
because physical properties of $z\sim3$ LBGs show dependence
on the luminosity; LBGs with weaker Ly$\alpha$ emission are more UV luminous, and
they show stronger metal interstellar absorption lines, redder
continuum emission, and larger amount of dust reddening \citep{sha01, sha03}.
Furthermore, more UV luminous LBGs show stronger clustering,
implying that they reside in more massive dark halos \citep{gia01, fou03, ade05}.

Adaptive Optics (AO) systems on 8-10m class telescopes made sufficiently high resolution
imaging observations on the $z\sim3$ galaxies in the $K$-band possible for the first time.
The diffraction limit for an 8m diameter telescope is 
$0.\!^{\prime\prime}06$ in the $K$-band,
which is smaller than that with NICMOS on HST in the $H$-band ($0.\!^{\prime\prime}25$)
and than the resolution that we require to resolve the $z\sim3$ galaxies,
$\sim0.\!^{\prime\prime}1-0.\!^{\prime\prime}2$,
corresponding to physical lengths of 0.8$-$1.5kpc at $z\sim3$ 
with $H_0$ = 70 km s$^{-1}$ Mpc$^{-1}$, 
$\Omega_{\rm M}=0.3$, and $\Omega_{\Lambda}=0.7$
(these
cosmological parameters will be maintained throughout the paper).
Additionally, the imagers of the AO systems of the ground-based telescopes usually cover
the PSF with sufficiently high spatial sampling ($0.\!^{\prime\prime}02$ pixel$^{-1}$ 
$-0.\!^{\prime\prime}06$ pixel$^{-1}$)
in comparison with the NICMOS 3 camera ($0.\!^{\prime\prime}2$ pixel$^{-1}$), which is used for
most of the HST NIR observations of the distant galaxies (e.g., \citealt{pap01}; \citealt{zir07}).
AO observations with natural 
guide stars (NGSs) have been successfully revealed high-resolution morphology of
distant field galaxies around bright guide stars, which are required to correct
a disturbed wavefront (Larkin et al. 2000; Glassman et al. 2002; Steinbring et al. 2004; 
Cresci et al. 2006; Stockton et al. 2004; Fu et al. 2005; Stockton et al. 2006;
Minowa et al. 2005; Melbourne et al. 2005; Huertas-Company et al. 2006).
Profile fittings for galaxies with magnitude down to $K\sim20$ mag are
continuously conducted with $K$-band imaging data of integration time up to 3.5h
with the Subaru Cassegrain AO system (e.g., \citealt{sto04}). In one extreme end,
Minowa et al. (2005) conducted a 26.8h very deep imaging observation of
a region around a bright star in the Subaru Deep Field with the Subaru AO system
and achieved the limiting magnitudes of $K'\sim24.7$ mag
for stellar objects and $K'\sim23.5$ mag for extended objects. Even
galaxies down to $K'\sim23$ mag are clearly resolved at the image size of
$0.\!^{\prime\prime}18$ FWHM. 

Following the success of the deep imaging observations with the Subaru AO system,
we started an imaging survey targeting the $z\sim3$ LBGs found with
the $U$-band dropout technique (Steidel et al. 2003), in order
to reveal the rest-frame 5500{\AA} (close to $V$-band) morphology of the $z\sim3$ galaxies
covering wide-luminosity range. We are concentrating on the galaxies at $z\sim3$
because $z\sim3$ is the highest redshift for which we can conduct high
resolution imaging in the rest-frame wavelength above $5000${\AA} with current facilities;
for a ground-based observation,
due to the very high background in the longer wavelength,
imaging observation in the rest-frame $>5000${\AA} for galaxies at higher redshifts is not 
feasible, on the other hand, for an observation with space facilities, still
the aperture of the telescope is much smaller than ground-based
8-10m class telescopes, thus the achieved resolution 
is limited by the diffraction limit, which is larger than $0.\!^{\prime\prime}3$ in the
longer wavelength. 
The sample of the AO observations will be described in details in Section 2 along with the
characteristic of the Subaru AO system. The AO-assisted
$K$-band observations are described in Section 3. Additionally we conduct
$J$-band and $V$-band imaging observations of a part of the targets. 
They are also shown in Section 3. Reduction processes of the data are described
in Section 4. The results and discussions are shown in Section 5 and 6, respectively.

\section{SAMPLE OF $z\sim3$ LBGS FOR AO OBSERVATIONS}

The targets of the observations are selected from the catalog of $U$-band dropout LBGs
of Steidel et al. (2003), which is the largest sample
of $z\sim3$ galaxies with spectroscopic redshifts. Based on $U-G$ and
$G-R$ colors of galaxies, 2347 $U$-band dropout
LBGs have been selected down to $R\le25.5$ mag, 55\% of them were observed
spectroscopically, and 73\% of the spectroscopic targets were confirmed to be at $z>2$ (Steidel et al. 2003). 
Most of the galaxies are distributed over $z=2.5-3.5$.
Because the $U$-band dropout galaxies are selected
not only by a red $U-G$ color but also by a blue $G-R$ color, they are thought
to be biased toward blue star-bursting galaxies in the redshift range (e.g., F\"orster Schreiber et al. 2004). However, 
as will be shown in Section 5.2, at least a part of the red
galaxy population in the redshift range is selected in the $U$-band dropout LBGs.

For the AO-assisted observation with NGSs, the
availability of a guide star brighter than a certain magnitude within 
a certain distance limits the "AO-observable" targets seriously, especially
for observations of distant galaxies in high galactic latitude regions.
We conduct the AO-assisted $K$-band imaging observations of the $z\sim3$ LBGs using the
Subaru Cassegrain AO system with a 36-elements curvature wavefront
sensor and a bimorph-type deformable mirror with the same number of
elements (AO36; Takami et al. 2004). The curvature wavefront sensor uses 
photon-counting avalanche photodiodes, in stead of CCDs,
which are usually used in Shack-Hartmann wavefront sensor AO systems, 
thus the wavefront measurement can be done without affected by a
read-out noise, which is the dominant noise source for wavefront
measurements with CCDs.
Thanks to the low-order AO system with a very sensitive wavefront
sensor, the AO36 works with faint NGSs. This greatly eases the 
difficulty in finding AO guide stars.
Significant image improvement can be achieved even with a guide star as faint as
$R=18$ mag, although the correction degrades with the magnitudes of 
the guide stars (Oya et al. 2004). In order to achieve a
$\sim 0.\!^{\prime\prime}2$ resolution at the target position,
we limit the targets with a guide star brighter than
$R=15.0$ mag.

The AO correction of blurred PSFs only work within a limited
angle from a guide star. The angle is described with the isoplanatic
angle, which is determined with the size and the height of the
dominant turbulence in the atmosphere.
With the AO36, the PSF degrades gradually with increasing the distance
from the guide star, $d$, to $d\sim40^{\prime\prime}$,
and worsens rapidly beyond $d\ge40^{\prime\prime}$ (Takami et al. 2004). 
The isoplanatic angle was estimated to be $\ge30^{\prime\prime}$ 
based on the PSF variation with the distance (Takami et al. 2004).
Although the isoplanatic angle can vary, depending
on atmospheric conditions, measurements in a number of nights
show FWHM of $0.\!^{\prime\prime}2$ can be constantly achieved up to
$d=35^{\prime\prime}$ from a guide star (Minowa et al. 2005; see Figure~\ref{PSF_param1} in this paper). 
Thus we limit 
the targets within this distance from a guide star. We also exclude
LBGs at $d\le10^{\prime\prime}$ from a guide star, because photometric
measurements in the original optical images are significantly 
affected by a bright guide star.
In summary, the targets of the observations are selected with the
criteria; $10^{\prime\prime} < d < 35^{\prime\prime}$ from a star brighter than $R=15.0$ mag, 

In the survey areas where Steidel et al. (2003) conducted the LBG survey (900 arcmin$^{2}$)
there are 84 stars brighter than $R=15.0$ mag in the U.S. Naval Observatory (USNO)
catalog made from all sky plate surveys. The summed area of $10^{\prime\prime} < d < 35^{\prime\prime}$ 
of the 84 bright stars is $76.6$
arcmin$^2$ removing overlapping regions and areas at the edge of the LBG survey. 
In the "AO-observable" area, there are 141 LBGs excluding spectroscopically
identified galactic stars. The $R$-band magnitude
distribution of the LBGs in the area is shown in Figure 1 with 
a thin solid histogram. The $R$-band magnitude distribution of the whole
LBG sample from Steidel et al. (2003) is shown with a dotted histogram.
There is no apparent difference between the ``whole'' and
the ``AO-observable'' LBG samples, thus the ``AO-observable'' sample can be 
regarded as a representative sub-sample of the whole $z\sim3$ LBGs.

In the observations, we assign a priority to a
field of view (FoV) by considering the number of LBGs with a spectroscopic redshift in the FoV,
and the brightness of the available AO guide star. We choose 12 fields, which include
36 LBGs.
The list of the fields along with the $R$-band magnitudes
of the AO guide stars is shown in Table~\ref{tab_fovs}.
The $R$-band magnitude distribution of the 36 LBGs is shown in Figure 1 with a thick
solid line. Because we select the fields for observations, considering the availability of 
the spectroscopic redshift, the "observed" LBGs are brighter than
the ``whole'' or ``AO-observable'' LBG sample on average. 
Redshifts of the 36 LBGs are listed in Table~\ref{tab_obj};
20 LBGs with a spectroscopic redshift.
The spectroscopic redshifts of the
LBGs are distributed between $z=2.4$ and $3.4$, which is consistent with
that of the whole sample of \citet{ste03}.
Because 55\% of the sample have a spectroscopic redshift, 
the observed sample is less affected by the uncertainty of redshift
which can significantly affect the discussion on galaxies selected with 
photometric redshifts (e.g., \citealt{tru06a}).
In addition, the contamination rate of foreground galaxies to the $U$-drop LBGs
without spectroscopic redshift is thought to be low.
The fraction of spectroscopically confirmed $z<2$ objects
among the spectroscopically observed $U$-dropout LBGs is 5\% (Steidel et al. 2003).

\section{OBSERVATION}

We use InfraRed Camera and Spectrograph (IRCS; Kobayashi et al. 2000)
for the AO-assisted $K$-band imaging observations. The camera side of
IRCS uses one 1024$\times$1024 InSb Alladin III detector
and has two imaging modes with different pixel scales; 
$0.\!^{\prime\prime}058$ pixel$^{-1}$ (58mas) mode and
$0.\!^{\prime\prime}023$ pixel$^{-1}$ (23mas) mode. FoVs of $59^{\prime\prime}\times59^{\prime\prime}$ 
and $23^{\prime\prime}\times23^{\prime\prime}$ are covered by 58mas and 23mas
modes, respectively. As described in the previous
section, the expected image size (FWHM) is $0.\!^{\prime\prime}20$ with faint
guide stars, the 58mas mode is sufficient to sample the PSF profile. 
It is important to estimate the shape of the PSF at the position of
the target securely, thus
in order not only to cover as many LBGs as possible, but also to
include as many PSF stars as possible, 
all but one of the fields are observed with the 58mas mode.
DSF2237a-FOV5 field is observed with the 23mas mode.
In order to remove the bad pixels, we use 9-point dithering in a
3$\times$3 grid with a width and height of $14^{\prime\prime}$.
We change the exposure
time at one position between 60s and 180s depending on the background
count rate of the observing night. Because
the linearity of the detector is 1\% at 6000 ADU, we
limit the background count below 4000 ADU.
No on-memory co-addition is done.
In order to reduce the readout noise, 16 times non-destructive read out (16-NDR)
is applied for each readout. The readout noise is estimated to be 10 $e^{-}$ rms with 16-NDR, which
is well below the Poisson noise of the thermal background emission; 150$e^{-}$ rms
with the 4000 ADU background and the gain of 5.6$e^{-}$ ADU$^{-1}$.

For broad-band imaging observations around $2.2\mu$m, there are two 
filters available in IRCS, $K^{\prime}$-band with $>50$\% transmission 
between $1.94\mu$m and $2.29\mu$m and $K$-band with $>50$\% transmission 
between $2.03\mu$m and $2.37\mu$m. Because the background level is lower
in the $K^{\prime}$-band, we primarily use $K^{\prime}$-band. For targets
at $z=2.9-3.1$, because the contamination of rest-frame [OIII]$\lambda\lambda4959/5007$ emission lines can be 
rejected by using the $K$-band, we use the $K$-band, even though the observed 
equivalent widths of the [OIII] emission lines in $z\sim3$ LBGs are at most 1/6 
of the band width of the filters \citep{pet01} and the contamination should
be negligible.
Hereafter we refer both of the $K^{\prime}$- and $K$-bands as $K$-band
unless the discrimination is crucial.

A pilot observation in two FoVs (DSF2237b-FOV1 and Q0302$-$003-FOV1)
is conducted as a normal open-use program (S03B-070) on
Oct. 14, 2003. After the successful observation, we continue
observations as an intensive program of the Subaru telescope (S04A-042)
from Apr. 4, 2004 to Oct. 25, 2004. The optical seeing size measured
with auto-guider camera, optical extinction measurement obtained though
Canada France Hawaii Telescope (CFHT), and NIR photometric condition evaluated
with the stability of the counts of bright objects in the FoV for each
observing night are summarized in Table~\ref{tab_dates}. Most of the
observing nights are photometric and the median seeing size 
is $0.\!^{\prime\prime}6$.

Most of the targets are observed at airmass smaller than 1.4
because PSF can significantly degrade at lower elevations due
to a smaller apparent isoplanatic patch.
After acquiring the target, we adjust the gain of the wavefront error 
feedback to the deformable mirror and the amplitude of the focus modulation for
the wavefront curvature measurement of the AO36 system
taking the image of the guide star with IRCS in the 23mas mode. A typical
FWHM of $0.\!^{\prime\prime}15$ is achieved for the guide star. Once the AO parameters
are fixed, we move to the 58mas mode and offset the FoV of the telescope
in order to cover the LBGs. For all of the FoVs, the AO guide stars
are covered in the same FoV, but they are saturated in the long exposure
with the 58mas mode, and they can be used neither for image alignments nor
a PSF evaluation.
The observed date, integration time, and filter used 
for each field are summarized in Table~\ref{tab_obsfield}.
For most of the fields the integration times are more than 5 hours in the $K$-band.

In some observing nights, the weather was clear but the AO correction could not 
achieve FWHM$\le0.\!^{\prime\prime}2$ during the AO parameter adjustment
process even for the AO
guide star itself due to a bad turbulence condition. In this case, we conducted
$J$-band imaging observations of the fields in order to obtain
the $J-K$ color of the targets. 
Finally 10 out of 12 FoVs are observed in the $J$-band with total exposure 
times of 0.8h - 2.7h for each FoV. The typical image size was $0.\!^{\prime\prime}4$
in the $J$-band.

In order to evaluate the optical morphologies of the $z\sim3$ LBGs with a better
resolution than the original imaging data used in Steidel et al. (2003) (typically FWHM of
$0.\!^{\prime\prime}9$), 
we obtained $V$-band imaging data for Q0302$-$003, Q2233$+$1341, DSF2237a, and DSF2237b
fields with Faint Object Camera and Spectrograph (FOCAS) attached to the Subaru telescope
from Sep. to Nov. 2004 during the open-use observations of S04B-149.
FOCAS covers $6^{\prime}$ FoV with a $0.\!^{\prime\prime}103$
pixel$^{-1}$ sampling. Each image is taken with a 300s or 600s integration, and 
a total integration time for each field is 0.5h to 1.2h. The sizes of
stellar objects are $0.\!^{\prime\prime}57$ to $0.\!^{\prime\prime}77$ in the final
reduced images. The optical observations
are summarized in Table~\ref{tab_FOCAS}.

\section{DATA REDUCTION}

\subsection{$K$-band AO Imaging Data}

We reduce the AO-assisted $K$-band imaging data basically
following the same reduction processes for deep $K$-band imaging
data of blank fields without AO; flat-fielding, sky-subtraction, image-alignment, and
image-combination. The details are as follows; at first,
we make the sky flat-field images by combining
normalized target images with median without adjusting the dithering
offsets. We normalize the raw target images by dividing with the mean
sky background value. Then, we divide the raw target images by
the sky flat-field image.
We ignore the dark counts because the count rate of the dark current is 1.8ADU per 100s, 
which is much smaller than the sky level ($\sim4000$ ADU) and
difficult to be estimated precisely. Sky background counts
are subtracted from the flat-fielded images.
The sky-subtracted images are combined to make the first pass
image of the field with the "drizzle" method using {\bf drizzle}
package in IRAF. The first pass combined image is still affected
by residuals of faint objects in the sky flat-field image which cannot be
removed with the median filter during the combination of the flat-field frames. 
Masking out objects detected
in the final image from the first pass, we make the revised
sky flat-field image again.
The original target data are divided again by the second pass
sky flat-field image. Then, sky background counts are subtracted
from the flat-fielded images. The sky-subtracted images are combined
with the "drizzle" method to make the final image.
Finally
we apply upgraded version \footnote{available from {\rm http://www.cfht.hawaii.edu/\~\ morrison/home/SExtractor.html}}
of an object detection software, SExtractor 2.3.2 (Bertin \& Arnouts 1996)
to the images. 
Photometric zero-point of each image is 
determined with observations of the UKIRT faint 
standard stars. We use FS030(J=12.021, K=11.936),
FS110(J=11.710, K=11.319), FS023(J=13.013, K=12.397),
FS134(J=11.896, K=10.715), FS130(J=12.970, K=12.260),
FS105(J=11.515, K=10.958), and FS152(J=11.632, K=11.050)
\footnote{The magnitudes are from the United Kingdom Infrared Telescope
faint standard star web page, http://www.jach.hawaii.edu/UKIRT/astronomy/calib/phot\_cal/}.

\subsection{PSF Evaluation}

Although most of the $z\sim3$ LBGs are resolved in the AO
images, the sizes of the targets are only a few times 
larger than the PSF size, thus accurate estimation of the PSF
shape is necessary in order to
reliably determine the morphological parameters of the targets.
The shape of the PSFs vary inside FoV, thus we need to evaluate
that at the position of the target itself.
In nine out of the eleven FoVs taken with the 58mas mode, there are
sufficiently bright stellar objects to estimate the shape
of the PSF within the FoV. We use these stellar objects
as the PSF reference stars. 
Their distances to the AO guide 
star are not exactly the same as those of targets, we estimate 
the PSF shape parameters by interpolating or extrapolating 
them as a function of the distance.

The AO-guide stars used in this observation are not 
sufficiently bright and the positions of the targets are 
sometimes close to the edge of the isoplanatic patch
of the AO correction, thus the effect of the atmospheric
turbulence is only partially-corrected and the PSFs of 
the images are consists of two components; a diffraction-limited
core and a seeing-limited halo. Therefore the PSF has a wing and 
is represented by a Moffat profile in stead of a simple Gaussian profile \citep{min05}.
The surface brightness, $I(r)$, of the Moffat profile \citep{mof69}
at an elliptical radial distance, $r$ ($^{\prime\prime}$),
is described by
\[ I(r) = I_{\rm total} \times 4 (2^{1/\beta}-1) \frac{\beta-1}{\pi F^2}[1+4 (2^{1/\beta}-1)(\frac{r}{F})^2]^{-\beta} \]
with the $\beta$ parameter representing the core to halo flux ratio of the PSF,
$F$ ($^{\prime\prime}$)
representing FWHM, and $I_{\rm total}$ representing the total flux \citep{tru01b}. 
The elliptical radial distance is defined by
\[ r \equiv \sqrt{ x^2 + (\frac{y}{q})^2 } \]
with the distance to the center along semi-major (semi-minor) axis, $x$ ($y$), 
and the ratio of the semi-minor over semi-major axis radius, $q$.
The Moffat profile
contains Gaussian profile as the $\beta=\infty$ case \citep{tru01b}.
The Moffat profile fit to the PSF reference stars in the final images is
done with a 2D-profile fitting software, GALFIT \citep{pen02}.
Three parameters of the Moffat profile, total magnitude,
$\beta$, and FWHM, two parameters
on ellipsoid, ellipticity and position-angle (PA), and the central position, 
are determined
by the $\chi^2$ minimization fitting algorithm. 
As an example, we show the profile
of a PSF reference star in the field of Q1422+2309-FOV12
with open squares along with that of the best fit Moffat 
profile model with solid line in Figure~\ref{Q1422FOV12K_PSF01_prof}.
The core and the wing of the PSF is well described with
the Moffat profile model down to the isophote contour level
with $10^{-3}$ of the PSF peak. It should be noted that
the plots of the LBG profiles shown in later figures only cover up to
semi-major axis (SMA) of about $1^{\prime\prime}$ and the discrepancy between
the observed PSF and the Moffat profiles seen in
Figure~\ref{Q1422FOV12K_PSF01_prof} at SMA$\ge1.\!^{\prime\prime}5$ can be neglected.

Dependences of the parameters of Moffat profile
(FWHM and $\beta$) and ellipticity to the distance from
the guide star are
shown in Figure~\ref{PSF_param1}. Although the AO guide
star magnitudes and the seeing conditions vary from
field to field, there is a clear trend that FWHMs in the final images become
larger with increasing distance from the guide stars. The 
trend is consistent with those measured with  
observations of a globular cluster, M13, with an AO-guide star of
$R=12$ mag shown with two solid lines (Minowa et al. 2005).
We fit a linear relation to the observed FWHM as a function
of distance, fixing the tilt to the mean tilt of
the solid lines. The resultant "mean" relation is shown
with a dotted line in Figure~\ref{PSF_param1}.
Although the achieved image size varies with observing
conditions, most of the measured FWHMs scatter within $0.\!^{\prime\prime}02$ 
around the "mean" relation.
There is no clear trend for the Moffat $\beta$ parameter against the distance.
The $\beta$ parameter is smaller than typical ground-based
seeing-limited observation, for example $\beta=3$ for a deep
$K$-band imaging data in Trujillo et al. (2006a). The larger $\beta$
of the seeing-limited PSF than that of the AO PSF indicate
that the former PSF is close to the Gaussian profile
with seeing limited FWHM.
The ellipticity depends on the distance; the ellipticity
becomes larger at larger distance.
The PA of the ellipticity is aligned with the PA to the
AO guide star as shown in Figure~\ref{PSF_param2}.
Between the FWHM size and the $\beta$, there is a weak
trend that the FWHM size becomes larger with increasing
$\beta$, i.e., closer to the Gaussian profile. There is
no clear dependency between $\beta$ and ellipticity.

Based on these results, we estimate the PSF shape at
the position of each target. The FWHM of the PSF is determined
with the FWHM size of the PSF reference star in the image, and
the difference of the distance from the AO guide star is corrected
with the tilt of the relation between the FWHM and the distance 
shown with the dotted line in Figure~\ref{PSF_param1}. For the FoVs without a PSF reference
star, we used the "mean" relation in the top-panel
of Figure~\ref{PSF_param1}.
The estimated FWHM for each object is shown in Table~\ref{tab_LBG}.

\subsection{$J$-band Imaging Data and $V$-band Imaging Data}

The $J$-band imaging data are reduced in the same manner 
as the reduction of the $K$-band data. The reduced images are aligned to the $K$-band
images using several bright and compact objects in the common area.
The coordinate conversions are derived with {\bf geomap} in IRAF, and the $J$-band
images are transformed with {\bf gregister} to match the $K$-band coordinates.

In order to measure $J-K$ colors of the LBGs in a fixed aperture, 
we match the shapes
of the PSFs between the $J$- and $K$-band images.
Because the $K$-band images have much better resolution than the $J$-band images,
the $K$-band images are convolved with the PSF difference 
kernel derived with {\bf psfmatch} process in IRAF.
The reliability of the PSF matching process is confirmed
that the $J-K$ colors of stellar objects derived in apertures from $1.\!^{\prime\prime}0$
to $2.\!^{\prime\prime}0$ at a $0.\!^{\prime\prime}1$ increment
agree with each other within 0.1 mag.
After the PSF matching, flux measurements for objects
in the $J$- and the PSF-matched $K$-band images are made with the double-image mode
of SExtractor by using the original $K$-band image as the object detection
image. 
We use colors in a $1.\!^{\prime\prime}0$ diameter aperture in the
discussions below. The 2$\sigma$ upper limits of $1.\!^{\prime\prime}0$
aperture magnitudes are determined by using the scatter of the background
count in the aperture size measured at various positions in the images.
They distribute between 23.90 mag and 25.72 mag for images with $0.8-2.7$
hour integration time. For the LBGs fainter than the limit,
we use the limit as the lower limit of their aperture magnitude and calculate
lower limits of their $J-K$ colors.

$V$-band imaging data are reduced as follows. At first, bias
counts are subtracted by using the overscan region of the frame.
The flat-field images are made by combining dome-flat frames.
The bias subtracted images are divided by the flat-field image.
The reduced images are aligned to the $K$-band images in 
the same way as the registration of the $J$-band images with {\bf geomap} and {\bf gregister}
in IRAF. Because we only examine the difference of peak positions of the
$z\sim3$ LBGs in the optical and NIR bands, no PSF matching is applied to 
the $V$-band images.

\section{RESULTS}

\subsection{$K$-band Magnitudes and Apparent Half-light Radii of the LBGs}
 
Postage stamp $K$-band images of the $z\sim$ 3 LBGs are shown in 
Figures~\ref{Kimage_g15_1} and ~\ref{Kimage_g15_2} in the order of the $K$-band magnitude. 
The images are centered at the cataloged positions of the $z\sim3$ LBGs. The size
of the images ($3.\!^{\prime\prime}5 \times 3.\!^{\prime\prime}5$) roughly corresponds
to the size of the aperture used in the optical color measurements by \citet{ste03} ($\sim3^{\prime\prime}$ diameter).
In the final images, 31 LBGs are detected out of the observed 36 LBGs.
The majority of the LBGs are 
isolated, but seven of the LBGs show multiple knots; Q0302$-$003-D73, Q0302$-$003-M80, Q2233$+$1341-MD46, Q0302$-$003-MD216,
DSF2237b-MD90, and DSF2237b-MD13 show two knots, and Q0302$-$003-MD373 has 3 knots. 
Considering all of the knots are within the aperture of the optical color 
measurements, we assume all of the knots belong to the LBGs.
The total magnitudes 
($K$) and the apparent half-light radii ($r_{\rm HL}$) of 
the LBGs are measured with SExtractor. We use AUTO magnitudes
from SExtractor as the total magnitudes. $r_{\rm HL}$ is defined
with the radius enclosing the half of the total flux of the object.
It should be noted that the $r_{\rm HL}$
is not corrected for the PSF convolution, thus can be affected by
PSF variation, but the variation (peak-to-peak difference of 0.05 dex for
stellar objects with $K<21$ mag)
does not significantly affect the discussion here.
For the objects with the multiple 
components, we assign a label of ``A'', ``B'', or ``C'' for each component 
and measure $K$ and $r_{\rm HL}$ individually. 
Resultant $K$ and $r_{\rm HL}$ are summarized in Table~\ref{tab_LBG}
and are plotted with large open circles in Figure~\ref{Kauto_HL} along with 
those of field (stellar) objects in the FoVs with small dots (star marks). 
The stellar objects are selected with CLASS\_STAR from SExtractor larger than 0.9.
The brightest LBGs are
$K=20.2-20.4$ mag, which are brighter than the characteristic luminosity 
of $z\sim3$ LBGs. The faintest LBG have $K=23.7$ mag. The long dashed line indicates
$M^{*}_{V}$ of the $z\sim3$ LBGs (Shapley et al. 2001). The observed sample covers a
wide range of absolute $V$-band magnitudes from $M_{V}^{*}-0.5$ mag to $M_{V}^{*}+3.0$ mag.
The $r_{\rm HL}$ of most of the LBGs are larger than stellar
objects shown with star marks, i.e., the LBGs are resolved in the 
AO-assisted $K$-band images. The median $r_{\rm HL}$ of
the detected $z\sim3$ LBGs is $0.\!^{\prime\prime}23$, which 
corresponds to 1.8kpc at $z=3$.
The average $r_{\rm HL}$ of the LBGs brighter than $M_{V}^{*}$ is $0.\!^{\prime\prime}40$
(3.1kpc at $z=3$),
which is significantly larger than the average of the fainter LBGs detected in the
$K$-band observations ($0.\!^{\prime\prime}23$, 1.8kpc at $z=3$), and there is no bright 
LBGs with a small $r_{\rm HL}$. 

Five LBGs without $K$-band detection can either be faint
or extended. The detection limit of a 5 h integration under $0.\!^{\prime\prime}2$
FWHM condition is shown with a dotted line in Figure~\ref{Kauto_HL} (in details, see
Section 5.4). Because the detection limit is determined by
the surface brightness, not only faint objects but also very extended
objects cannot be detected. However, all but one of the five 
LBGs are fainter than 25.0 mag in the $R$-band, and belong to the
faintest end of the sample of the observed LBGs, thus most likely
they are not detected because they are faint.

\subsection{$J-K$ Colors and Optical-NIR Difference of the LBGs}

The observed $1.\!^{\prime\prime}0$ aperture $J-K$ colors of the LBGs show clear dependence
on the $K$-band magnitude.
The $J-K$ colors are summarized
in Table~\ref{tab_LBG} and $J-K$ colors against $K$-band magnitudes are plotted in 
Figure~\ref{Fig_JKcolmag} with open circles. The LBGs brighter than $M_{V}^{*}$
show redder color than the fainter LBGs among the detected LBGs. The average $J-K$ color of
the LBGs brighter (fainter) than $M_{V}^{*}$ is 2.2 (1.4) mag (if we exclude LBGs that are
not detected in the $J$-band). There is no LBGs bluer than 
$J-K=1.8$ mag and brighter than $M_{V}^{*}$. The distribution is consistent
with the results obtained with $J$- and $K$-band observations of a
larger sample of $z\sim3$ LBGs (Shapley et al. 2001) shown with small open squares.
Because the $J$- and $K$-bands corresponds to rest-frame 
$U$- and $V$-bands for $z\sim3$ galaxies, the observed
$J-K$ color corresponds to the rest-frame $U-V$ color for
the $z\sim3$ LBGs, and the color reflects the size of the 4000{\AA}
break and/or the amount of dust extinction of the galaxies.
We derive the relationship between the observed $J-K$ colors and
the rest-frame $U-V$ colors using
a library of model SEDs
made by PEGASE spectral synthesis code (Fioc \& Rocca-Volmerange 1997).
We use star formation histories with constant starformation and 
exponentially decreasing starformation models with ages from 10Myr to
2Gyr. Metallicity range from 0.1 $Z_{\odot}$ to 1 $Z_{\odot}$ is
considered. All SED models at $z=2.4-3.4$ except for exponentially decreasing models with
decaying time scale shorter than 100Myr follow
\[ (U-V)_{\rm Vega, rest} = 0.74(J-K)_{\rm Vega, obs} -1.39 \]
within $\Delta (U-V)=\pm0.20$ mag in the observed $J-K$ color range.
The relation calculated with the response curve of the $K^{\prime}$-band filter
is consistent that calculated with the response curve of the $K$-band filter
within the scatter.
The effect of the internal extinction changes the rest-frame $U-V$
color and the observed $J-K$ color in parallel to the relation.
Though the exponentially decreasing models with short time scale
do not follow the relation, considering that the constant starformation
models describe the SEDs of $z\sim3$ LBGs sufficiently well (e.g., \citealt{sha01}),
we use the relation.
The average $J-K$ color of the LBGs brighter (fainter) than $M_{V}^{*}$  can
be converted to 0.2 ($-0.4$) mag in the rest-frame $U-V$ color.
The colors of the bright and faint LBGs are close to those of red massive 
and blue less-massive galaxies at $1.9<z<2.7$, respectively, selected 
with the photometric redshifts
(0.6 and $-0.3$ mag typically; Kajisawa \& Yamada 2006). Kajisawa \& Yamada (2006) 
show the transition of 
the red massive and blue less massive galaxies happens
at a stellar mass of $\sim 2\times10^{10} M_{\odot}$ in the redshift range, 
which corresponds to $K=21.3$ mag if we use the mean relationship
between the observed $V$-band absolute magnitude and the stellar mass
for $z\sim3$ LBGs (\citealt{sha01}, their Figure 13). The observed transition 
magnitude agrees roughly with that expected from Kajisawa \& Yamada (2006).
The agreements of the colors and the transition magnitude imply that
the $z\sim3$ LBG sample can cover not only blue less-massive galaxies but also 
red massive galaxies in the redshift range selected with photometric 
redshift method, although the reddest galaxies ($U-V>0.5$ mag) can be missed.

The $K$-band images of the LBGs are compared with the optical $V$-, $G$-, or 
$R$-band images in Figures~\ref{VKimage_g15_1} and \ref{VKimage_g15_2}.
We do not match the PSF size of the $K$-band images to that of the optical images, thus, 
only the positions of the peaks should be compared. Some of the bright
and red LBGs show significant shifts between the peak positions in the 
$K$-band and optical images. The shifts are shown as a function of
$K$-band magnitude in Figure~\ref{Opt_K_diff}. In order to examine
the uncertainties of the shift measurements, we also plot the differences of
$K$-band and optical positions of optically-bright field objects with crosses.
We exclude extended large galaxies from the field objects. Considering
all but one of the compact field objects show differences smaller
than $0.\!^{\prime\prime}20$, we regard the shift larger than $0.\!^{\prime\prime}25$
and $0.\!^{\prime\prime}20$ as significant and marginal shifts, respectively.
Among the $z\sim3$ LBGs with $K\le22$ mag,
DSF2237b-MD81, Q1422$+$2309-C36,
and DSF2237b-D5 show significant shifts of $0.\!^{\prime\prime}44$ (3.4kpc), $0.\!^{\prime\prime}29$ (2.2kpc),
and $0.\!^{\prime\prime}26$ (2.0kpc), respectively and DSF2237b-C10($0.\!^{\prime\prime}22$; 1.7kpc),
DSF2237b-D27($0.\!^{\prime\prime}21$; 1.6kpc), Q0302-003-D73A($0.\!^{\prime\prime}20$; 1.5kpc),
and DSF2237b-MD19($0.\!^{\prime\prime}20$; 1.5kpc) also show marginal shifts.
For the LBGs with $K\ge22$ mag, no significant shift is observed.
The shifts and the red colors of the bright LBGs can be explained with an existence of an
old and/or reddened stellar population which dominates the galaxy light in the $K$-band but
not significant in the optical band, and the UV-bright star-forming regions lie within
the region of optical light but they are centered at different positions. 
It is worth noting that DSF2237b-MD81 with the largest shift has a strange
SED and cannot be fitted well with one component SED model due to an excess in 
the NIR bands (Shapley et al. 2001). The NIR excess may be explained with dust reddened
component with young stellar age. On the contrary, for blue less-massive galaxies, the morphologies
of the UV-bright star-forming regions follow the galaxy main body. 

\subsection{S\'ersic Profile Fitting for the LBGs with $K<21.5$ mag}

As can be seen in Figures~\ref{Kimage_g15_1} and ~\ref{Kimage_g15_2}, the
$K$-band morphologies of the $z\sim3$ LBGs are smooth and not irregular.
In order to examine whether
their light profiles are similar to those of local galaxies or not,
we apply one component S\'erisc profile fitting to the AO-assisted $K$-band 
images of the $z\sim3$ LBGs with $K<21.5$ mag.
Considering that the spatial resolution and the signal-to-noise (SN) ratio
of the images are not sufficient to deconvolve multiple components, for example
disk and bulge, of the targets, we only consider one component in the fitting.
One component S\'ersic profile fitting describes the light profiles of
the $z\sim3$ LBGs well at the current resolution and the SN ratio, as shown below.

The S\'ersic profile is given by
\[ I(r)=I_{e}\exp(-\kappa_n((r/r_e)^{1/n}-1)) \]
with the surface brightness $I(r)$
at an elliptical radial distance $r$ ($^{\prime\prime}$), 
and the surface brightness $I_{e}$ at $r_{e}$ ($^{\prime\prime}$),
the effective radius or the half-light radius.
The index $n$ determines the shape of the profiles;
a profile with a larger $n$ shows a more concentrated core and a more extended wing,
while a profile with a smaller $n$ shows a flatter core with a sharper cutoff.
The exponential-profile of the disk component and the $r^{1/4}$-profile
of the spheroidal component of local galaxies are equivalent to S\'ersic
profiles with $n=1$ and $n=4$, respectively.
$\kappa_n$ is a function of $n$ determined so as to half of the total light is
enclosed within $r_{e}$.
Hereafter we refer effective radius in arcsec and in kpc units
with $r_{e}$ and $R_{e}$, respectively.

One-component S\'ersic-profile fitting is
extensively used to fit the profiles of nearby galaxies (e.g., \citealt{she03})
and of galaxies from intermediate to high redshifts (e.g., \citealt{tru06a}),
and successfully discriminates between galaxies dominated by disk-like profiles (small $n$) and
bulge-like profiles (large $n$). It needs to be noted that kinematic examination
has not established the correspondence between the profile (exponential
or $r^{1/4}$) and the kinematics (rotation-supported or velocity-dispersion-supported)
for high-redshift galaxies.

For the 2D-profile fitting process, we use GALFIT software.
Three parameters of the S\'ersic profile, total magnitude,
$n$, and $r_{e}$, two parameters on ellipsoid, ellipticity
and position-angle (PA), and the central position, are determined
by the $\chi^2$ minimization.
The S\'ersic profile fitting is only applied to the LBGs with $K<21.5$ mag,
because for the fainter LBGs, the fitting results are not conclusive.
Among the LBGs with $K<21.5$ mag, Q0302$-$003-D73 has two knots.
Both of them are brighter than $K=21.5$ mag, and we fit
each knot separately.
We correct the best fit $\log R_{e}$ by $+0.13$ ($+0.25$) dex considering the systematic
shift expected from the simulations of the fitting for galaxies with best-fit $n<2$ ($n>2$). The details are 
described in the next section.

The rms noise of each pixel is determined with the rms value
of the sky region around each object because the data are background
limited. However, in order to put higher weight to the brighter
region of objects, we put two times higher weight for the peak.

Since
the shape of the PSF at the position of the target has the uncertainty
as described in Section 4.2, we make various model PSFs using GALFIT with Moffat parameters
within the uncertainty of the true PSF, and use the model PSFs
in the fitting process. We use the PSF shape parameters as follows:
we cover FWHM of 
$\pm0.\!^{\prime\prime}02$ of the estimated FWHM (see Table~\ref{tab_LBG}) for each object,
Moffat $\beta$ index between 1.3-1.7, and ellipticities of 0.00, 0.10, and 0.15
for all objects. Most of the Moffat $\beta$s and ellipticities of
the PSF reference stars are enclosed in the parameter range as shown in Figure~\ref{PSF_param1}.
Considering the result shown in Figure~\ref{PSF_param2},
we fix the PA of 
the ellipticity to the PA to the AO guide star.
The scatters of the best fit parameters due to the variation of the model
PSF are added to the final uncertainties of the best fit parameters.

The effects of changing the model PSF for the fitting process for two objects,
the brightest object DSF2237b-MD81 ($K=20.21$ mag) and 
one the faintest objects Q0302$-$003-D73A ($K=21.38$ mag),
are presented in Figures~\ref{DSF2237bFOV1K_MD81_check} and \ref{Q0302FOV3K_D73_check}, 
respectively.
The horizontal
axes are the used shape parameters of the model PSF and
the vertical axes are the obtained best fit S\'ersic parameters with the model PSF.
The left-hand (right-hand)
panels show the effect of changing PSF FWHM ($\beta$) parameter.  
It should be noted that
in these figures, the FWHMs outside of the estimated FWHM uncertainty are
also plotted.
The vertical dashed and dotted lines indicates the best-estimate FWHM and the 
uncertainties for each target.
The uncertainty of the PSF does not seriously affect the best fit
parameters of DSF2237b-MD81. On the contrary, for Q0302$-$003-D73A,  
the best-fit S\'ersic $n$ index varies with the FWHM of the model PSF used.
However, the index is still well constrained ($\pm1.0$), if we consider the range
of the FWHM uncertainties. The best-fit $R_{e}$ does not vary with the PSF shape
change.

The resulting fitting parameters for LBGs with $K<21.5$ mag are shown
in Figure~\ref{LBG_FIT_ser}. Each point with error bar represents
a set of best-fit parameters with a model PSF. The error bar indicates
the uncertainty of each of the fitting. The variations of the best-fit parameter set
with changing PSF
are represented with the scatter of the points. In Figure~\ref{LBG_FIT_prof},
we show the observed profiles of the LBGs along with the best-fit S\'ersic models
obtained with the model PSF with the best estimate parameters. The red-solid,
blue-dashed, and green-dotted lines show the best fit S\'ersic (with free $n$),
$r^{1/4}$-law (S\'ersic with $n=4$) and exponential (S\'ersic profile with $n=1$)
profiles, respectively. The profiles are derived with the elliptical isophote fitting
package {\bf ellipse} in IRAF. We determine the shape of the ellipse at each 
contour with the best-fit model of the S\'ersic profile, and this shape of the ellipse is
used to derive the profiles of the observed image, the models with
$r^{1/4}$-law profile and with exponential-law profile. The profiles are shown along
the semi-major axis (SMA). The error bar of the observed data is 
determined by the rms scatter of intensity data along the fitted ellipse.
Outside of the galaxies, the fluctuation of the derived isophotal intensity,
which reflects the uncertainty of the background determination, 
consistent with the derived rms scatter.

The resultant effective radius and the $n$ index are summarized in
Table~\ref{tab_LBG} and shown in Figure~\ref{LBG_DRG_Re_N} with filled
squares.
The error bars include not only the uncertainty of each fitting
but also uncertainty due to the PSF model. All but one of the LBGs are fitted
with low n value. They locate similar part of the $R_{e}$ vs. $n$ plane
to the disk galaxies.

\subsection{Simulating $z=3$ Galaxy Images from HST ACS Images of Galaxies at Intermediate Redshifts}

In order to examine the reliabilities and uncertainties of the S\'ersic profile fitting, we
make simulated AO $K$-band images of $z\sim3$ galaxies from
high-resolution images of galaxies at intermediate redshifts,
and compare the results obtained by fitting to the original images and
the simulated images. 
The simulated images are made from
HST ACS F775W- and F850LP-bands images of galaxies at intermediate redshifts
in the GOODS-North and -South regions.
The wavelength coverages of the F775W and F850LP filters are
7480-7910{\AA} and 8780-9320{\AA}, respectively, thus cover the rest-frame 5500{\AA},
which corresponds to $K$-band for $z\sim3$ galaxies,
of galaxies at $z=0.36-0.69$. We limit the sample for the simulation
to the galaxies with spectroscopic redshift in the redshift range.
For the sample selection,
we use spectroscopic redshifts 
from Cowie et al. (2004) and Le F\`evre et al. (2004)
for galaxies in the GOODS-North and GOODS-South regions, respectively.
The ACS images cover 701 (GOODS-North) and 211 (GOODS-South) galaxies
in the redshift range. 

We make the simulated images,
basically following the method used in "cloning" high-redshift galaxies
by Bouwens, Broadhurst, \& Illingworh (2003).
At first, we construct the rest-frame 5500{\AA} images of the
$z=0.36-0.69$ galaxies by linearly interpolating the HST ACS F775W-
and F850LP-band images ("original" images). Secondly, we normalize and scale
the images considering the luminosity distance and the angular-size 
distance. We make
pure-luminosity evolution (PLE) models with 1 mag and 2 mag by
simply multiplying 2.5 and 6.3 to the model images, respectively, 
as well as no evolution model.
We do not consider the evolution of internal structures of the galaxies.
After that, the scaled images are resampled to match the
sampling of the AO $K$-band images of $0.\!^{\prime\prime}029$ pixel$^{-1}$.
The original pixel scale of the archived ACS images is $0.\!^{\prime\prime}03$ pixel$^{-1}$
and the angular physical scale difference is 1.26 between $z=0.5$ and $z=3.0$,
i.e., the AO $K$-band images for the $z=3$ galaxies have 1.26 times
larger spatial sampling in physical scale (kpc per pixel) than that
of the ACS images for $z=0.5$ galaxies.
The kernel describing the difference between the PSF in the rescaled 
ACS image and the PSF in a typical AO image is applied to the resampled 
images ("model" images). In order to model the PSF of the AO images, we use the image of the brightest 
stellar object among objects not saturated (the stellar object 
shown in Figure~\ref{Q1422FOV12K_PSF01_prof}).
The FWHM of the PSF of the ACS image is $0.\!^{\prime\prime}1$, which
is 2 times better than the typical FWHM of the AO images.
Finally, noise is added to the image considering typical
background noise of the AO $K$-band images ("simulated" images). We assume the 5 h integration
for the simulation.

We apply the SExtractor with the same object detection criteria to the simulated images
as to the AO-assisted images.
The simulated $K$ and $r_{\rm HL}$ distribution
of the simulated galaxies are shown in Figure~\ref{Kauto_HL_all} along
with the observed $z\sim3$ LBGs with open circles. The left panel shows the
$K$-band magnitudes and $r_{\rm HL}$s measured in the simulated
images without the additional noise ("model" image). Most of the model
galaxies are fainter than the observed $z\sim3$ LBGs at the same size without
any evolution.
The $K$-band magnitudes
and $r_{\rm HL}$s measured from the "simulated" images are shown with filled squares in 
the right panel for 2 mag PLE case.
In the panel, we also plot the objects
that are not detected in the "simulated" images with crosses, 
using $K$-band magnitudes and $r_{\rm HL}$s
of the "model" images. 
Based on the distribution of the detected and non-detected objects in the diagram,
we estimate the detection limit
of the $K$-band imaging observation as shown with the thick dotted
lines in the panel. The detection limit roughly corresponds to
the surface brightness of $22.0$ mag arcsec$^{-2}$ in the $K$-band.
The estimated detection limit well reproduces the envelope of the observed distribution of
the $z\sim3$ LBGs.

Then, we apply the same profile fitting procedure to both of the "original"
and the "simulated" images, and compare the obtained best fit parameters
from the two images for each object. We only consider the simulated
galaxies with $K<21.5$ mag in the no, 1 mag, and 2 mag PLE models,
same as for the $z\sim3$ LBGs.
Figure~\ref{org_sim_comp_n} shows the distribution of $n$ measured in
the "original" and the "simulated" images. The left and right panels shows
galaxies with $\log R_{e} ({\rm kpc}) <1$ and $\log R_{e} ({\rm kpc})>1$ in the "original"
images, respectively.
The $n$ index derived from the "simulated" image shows
a larger scatter among original galaxy with a larger $n$,
while the scatter for objects with $n<2$ is small. The scatter is
larger for galaxies with $\log R_{e} ({\rm kpc}) >1$ than those with $\log R_{e} ({\rm kpc}) <1$. 
This is because
in order to accurately determine the $n$ index for larger $n$ and/or larger $R_{e}$ galaxies,
we need to detect more extended component with higher SN \citep{hau07}.
In addition, the $n$ index from the simulated image is systematically
smaller than that from the original image; for example, 
the galaxies with $n=3-5$ are fitted well with $n=2-4$ in the simulated images.
However, it is worth noting that most of the galaxies fitted well with $n>2$
in the original images are fitted well with $n>2$ in the simulated images.
Therefore the SN ratio and the spatial resolution of the AO-assisted
$K$-band images of the $z\sim3$ LBGs are sufficient to discriminate
between $n<2$ (disk-like profile) and $n>2$ (spheroid-like profile) galaxies.

For $R_{e}$, the results are shown separately for galaxies with $n<2$ (left) and $n>2$ (right) in 
the "original" images in Figure~\ref{org_sim_comp_Re}.
Again the scatter is larger for $n>2$ galaxies due to the same reason
mentioned above.
In addition, the values derived with the simulated images are systematically smaller than those measured
in the original images for both cases; the offset is $-0.13$ ($-0.25$) dex in $\log R_e$ 
for $n<2$ ($n>2$) galaxies. The smaller $\log R_e$ obtained from the profile
fitting to the simulated images can be owing to the fact that the
outer part of the original image is missed in the simulated image
due to the $(1+z)^{-4}$ surface brightness dimming and the high 
background in the $K$-band observation. Such systematic effect is
discussed by \citet{tru01a} and the larger offset obtained for 
galaxies with larger $n$ is consistent with the trend obtained
in their calculations. Because the systematic offset can affect the
comparison between the distributions of $z\sim3$ galaxies and 
galaxies at low to intermediate redshifts from literature, 
we correct the best fit $R_{e}$ of the $z\sim3$ LBGs
obtained from the S\'ersic profile fitting by $+0.13$ ($+0.25$)
dex for $n<2$ ($n>2$), adapting
these offset values as already mentioned.


Because the $R_{e}$ and $n$ of the simulated objects are derived
assuming the same observational condition and the same fitting method 
as used for the $z\sim3$ LBGs, the distributions of the $R_{e}$ and
$n$ of the simulated objects can be directly compared with those of the $z\sim3$ LBGs.
The scatter of the corrected $R_{e}$ and $n$ of the simulated objects with
2 mag PLE are shown in Figure~\ref{LBG_DRG_Re_N} with blue small crosses
There are two groups in the simulated objects;
one around $n\sim1$ with $\log R_{e}\sim0.7$
and the other around $n\sim3$ with $\log R_{e}\sim0.5$. 
The original images of the galaxies indicate that the
former (latter) group represents disk (elliptical) galaxies as expected.
The different morphologies in the original images successfully recovered
on the diagram as the two groups, i.e., the two groups can be distinguished
with the spatial resolution, the SN
ratio, and the fitting method of the AO $K$-band observations 
The distribution of the latter group is 
consistent with the $n$ - $R_{e}$ distribution of the elliptical galaxies
in the local universe \citep{cao93, don94, gut04}.
The observed scatter of the $z\sim3$ LBGs 
is close to that of the disk galaxies, and all but one LBGs belong to the
group of disk galaxies. 

\subsection{Concentration Parameter and Exponential-law Fitting for the Fainter LBGs}

For $z\sim3$ LBGs fainter than $K=21.5$ mag,
no reliable profile fitting can be done with
the S\'ersic profile with free $n$. Therefore, we examine the profile of the LBGs with more 
simple parameters, concentration $C$, and $r_{\rm HL}$. The concentration parameter
is closely related to the $n$ of the S\'ersic profile (Abraham et al. 1994; Abraham et al. 1996a;
Abraham et al. 1996b); the profile with a larger $n$ has larger $C$. 
We follow the definition by Barshady (2000), i.e., $C$ parameter defined with the ratio between
the radii enclosing 20\% and 80\% of the total flux, i.e, ratio between $r(20\%)$ and $r(80\%)$, 
\[ C_{80/20}\equiv\log[r(80\%)/r(20\%)]. \]

We plot the distribution of the simulated galaxies with $K<22.5$ mag on
$C_{80/20}$ vs. $r_{\rm HL}$ plane in Figure~\ref{LBG_DRG_C_HL}. The symbols are
coded with the $n$ of the original image; red crosses for $n<2$, 
and blue open circles for $n>2$. The division is not perfect, however, the
lower right (upper left) region is dominated by $n<2$ ($n>2$) objects.
The $z\sim3$ LBGs brighter than $K=22.5$ mag
are plotted on the same diagram with filled squares.
The distribution of the LBGs is similar to $n<2$ population, and all
but one of the LBGs have small concentration parameters. 
Q0302$-$003-D73A
has the largest $C_{80/20}$ among the LBGs and it is consistent with the fact that
the object is fitted well with S\'ersic profile with $n=3.0$.
The distribution 
indicates that the LBGs on average have profile similar to $n<2$ population.
The distribution is consistent with the profile fitting results for the LBGs with $K<21.5$ mag.
Based on the results, we fit the profiles of LBGs with $21.5<K<22.5$ mag
with exponential law, i.e., S\'ersic profile with $n=1$ and determined $R_e$ for each
object. 
We consider the same uncertainty of
the PSF parameters described in Section 5.3. The fitting results are summarized
in Table~\ref{tab_LBG}. We correct the $R_{e}$s obtained by the $n=1$ fitting
by $+0.13$ dex as described in Section 5.4.

\subsection{Surface Brightnesses and Surface Stellar Mass Densities of the LBGs}

Because the $z\sim3$ LBGs show the low concentration profiles similar to
the disk galaxies in the local universe, we examine
the properties of the $z\sim3$ LBGs comparing with those of the disk galaxies at $z=0$ and $z=1$
from Barden et al. (2005). We assume that the $z\sim3$ LBGs have a disk shape.
In order to distinguish a spheroidal shape with $n=1$ profile 
from the disk shape, we need to examine the statistical 
distribution of the ellipticities of the 
$z\sim3$ LBGs, but the current sample is limited in number, especially for
bright LBGs, we leave the discussion for later studies.
Barden et al. (2005) have selected
the $z\sim0$ and $z\sim1$ disk galaxies from 
the second data release of the Sloan Digital Sky Survey and 
HST/ACS images of Galaxy Evolution from Morphologies and SEDs survey, respectively.
They have applied 1 component elliptical S\'ersic profile fitting to the images of the galaxies \citep{hau07}, and
select disk galaxies with the best-fit S\'ersic $n$ index smaller than 2.5, with which
they can discriminate between spheroid-dominated (E/S0/Sa) and disk-dominated (Sb/Sc/Irr) galaxies.

The size-luminosity relation of the $z\sim3$ LBGs is
shown in Figure~\ref{LBG_DRG_Mv_Re}. In the right panel of 
the Figure, we compare the scatter of the $z\sim3$ LBGs with those of the $z\sim0$ (blue dashed contour)
and $z\sim1$ (red solid contour) disk galaxies.
The $M_{V}$s of the $z\sim3$ LBGs are brighter than the average of the $z=0$ and $z=1$
disk galaxies at the same $R_{e}$.
For $z\sim0$ and $z\sim1$ galaxies, the relations between the
$R_{e}$ and $M_{V}$ are examined as the surface brightness distributions.
The absolute mean surface brightness of the disk galaxies within
the effective radius in the rest-frame $V$-band, 
$\mu_{V}$ (mag arcsec$^{-2}$), is related to
$M_{V}$ and $R_{e}$ by,
\[ \mu_{V} {\rm (mag\ arcsec^{-2})} = M_{V} ({\rm mag}) + 5 \log R_{e} ({\rm kpc}) +2.5 \log q + 38.568 \]
\citep{bar05}.
We use two values of the average of the $q$s in order to calculate the average surface brightness; 
$q=0.5$ following Barden et al. (2005) and
$q=1.0$ representing an extreme case with the faintest surface brightness.
The average $\mu_{V}$ of the 4 LBGs brighter than $M_{V}^{*}$($z=3$) is $\mu_{V}=17.64$ (18.38) mag arcsec$^{-2}$
with $q=0.5$ ($q=1.0$). The standard deviation is 0.71 mag arcsec$^{-2}$. If we use the LBGs with $K<22.5$ mag,
$\mu_{V}$ is $17.92$ (18.67) mag arcsec$^{-2}$ with
$q=0.5$ ($q=1.0$) and the standard deviation is 1.11 mag arcsec$^{-2}$.
The constant surface brightnesses of 18.38 and 18.67 mag arcsec$^{-2}$ are shown with dotted
and dashed lines, respectively, in Figure~\ref{LBG_DRG_Mv_Re}. The surface brightnesses are significantly 
brighter than the surface brightness limit observed in the local galaxies 
($\mu_{V}=22.4$ mag arcsec$^{-1}$)
derived with the Freeman law, $\mu_{0,B}=21.65\pm0.30$ mag arcsec$^{-1}$ \citep{fre70}, using
$B-V=0.3-0.5$ for disk galaxies \citep{rob94},
and the relation between the mean surface brightness and the central surface brightness for S\'ersic profile with $n=1$ ($\mu=\mu_{0}+1.1$).
The mean surface brightnesses derived from the distributions of $z=0$ and $z=1$ galaxies
shown in Figure~\ref{LBG_DRG_Mv_Re} are 20.84$\pm$0.03 and 19.84$\pm$0.07 mag arcsec$^{-1}$ with
$q=0.5$, respectively \citep{bar05}. If we apply the $q=0.5$ ($q=1.0$), the surface brightnesses of the $z\sim3$ LBGs
are 2.9 (2.2) mag and 1.9 (1.2) mag brighter than the disk galaxies at $z=0$ and $z=1$, respectively.

The evolution of the surface brightness from $z=1$ to $z=0$ is thought to be due
to fading of the stellar population in the disks, and the size-mass
relation, which reflects the surface stellar mass density of the disks, 
does not show significant evolution in the redshift range (Barden et al. 2005).
In order to examine the size-stellar mass relation of the $z\sim3$ LBGs,
we estimate their stellar mass with the $V$-band luminosity.
Following Rudnick et al. (2003), we use
the dependence of the stellar mass ($M_{*}$) to $V$-band luminosity ($L_{V}$) ratio, 
$\log (M_{*}/L_{V})$ ($M_{\odot}/L_{\odot}$), on the rest-frame $U-V$ color.
Using the same SED library used to derive the relation between
the observed $J-K$ colors and the rest-frame $U-V$ colors, 
we derive the relation,
\[ \log (M_{*}/L_{V})=0.85\times(U-V)_{\rm rest}-0.43 , \]
for constant and exponentially-decaying starformation histories by assuming the Kroupa initial
mass function (IMF; Kroupa et al. 1993). 
The relation depends on the assumed IMF; if we use Salpeter IMF (Salpeter 1955),
the relation is
\[ \log (M_{*}/L_{V})=0.95\times(U-V)_{\rm rest}-0.28 . \]
The tilt is steeper than the Kroupa IMF case.
The stellar masses of 39 $z\sim3$ LBGs derived with the Salpeter
IMF relation match with those estimated with
optical-to-NIR multi-color photometry including
$J$- and $K$-bands by Shapley et al. (2001) with Salpeter IMF,
within a standard deviation of $0.25$ dex. 
The stellar masses of the 3 red $z>2$ galaxies derived with the relation
are also consistent with those derived with a SED fitting \citep{kri06a}, 
within the standard deviation.
We refer the standard
deviation as the uncertainty of the stellar mass estimation with the relation. For objects
without spectroscopic redshifts, the uncertainties become $0.35$ dex
due to the uncertainty of their redshift.
The effect of the internal extinction changes the $\log (M_{*}/L_{V})$ and
$(U-V)$ in parallel to the relation (Rudnick et al. 2003), and
the effect is smaller than the scatter of the relation itself within 
the observed range of the $U-V$ color.
We use the relation with the Kroupa IMF, because the
IMF is used in Barden et al. (2005).

The derived stellar masses of the $z\sim3$ LBGs are summarized in Table~\ref{tab_LBG}
and the size-stellar mass relation is shown in Figure~\ref{LBG_DRG_Ms_Re}.
The size-stellar mass relation of the $z\sim3$ LBGs shows that their $R_{e}$
does not depend on their $M_{*}$ in the observed mass range.
In the right panel of the figure, we overplot the distributions
of $z=0$ (blue dotted contour) and $z=1$ (red solid contour) disk galaxies.
Bright LBGs with $\sim M_{V}^{*}$, i.e., $\log M_{*} (M_{\odot}) \ge 11$,
have smaller $\log R_{e}$ than the $z=0$ and $z=1$ disk galaxies with
similar mass. On the contrary, fainter LBGs with smaller stellar masses
follow the distributions of $z=0$ and $z=1$ disk galaxies.
We follow the definition of the surface stellar mass density, 
$\Sigma_{M}$, which represents the average surface stellar mass
density within $R_{e}$, from Barden et al. (2005);
\[ \log \Sigma_{M} (M_{\odot}\ {\rm kpc}^{-2}) = \log M_{*} (M_{\odot}) - 2 \log R_{e} ({\rm kpc}) - \log (2 \pi q). \]
The mean $\log \Sigma_{M} (M_{\odot}\ {\rm kpc}^{-2})$ is 9.3 (9.0) for the four $z=3$ LBGs
brighter than $M_{V}^{*}$ with $q=0.5$ ($q=1.0$) with a standard deviation of 0.3. For the LBGs with $K<22.5$ mag, 
the mean $\log \Sigma_{M} (M_{\odot} {\rm kpc}^{-2}) $ is 8.8 (8.5) with $q=0.5$ ($q=1.0$)
with standard deviation of 0.7.
For $z=0$ and $z=1$ disk galaxies, the $\log \Sigma_{M}$ is
constant with $8.50\pm0.03$ assuming $q=0.5$ \citep{bar05}.
As seen in the distributions, the average surface stellar
mass density of the LBGs brighter than $M_{V}^{*}$ is 3 ($q=1.0$) - 6 ($q=0.5$) times
larger than the averages of the $z=0$ and $z=1$ disk galaxies, and the
fainter LBGs have similar $\Sigma_{M}$ with the $z=0$ and $z=1$ disk 
galaxies.

\section{DISCUSSION}

\subsection{$n<2$ Galaxies at High-redshifts}

All but one of the LBGs with $K<21.5$ mag are fitted well with S\'ersic
profile with low $n$ values, and the scatter of the LBGs with $K<22.5$ mag
on the $C$-$r_{\rm HL}$ plane 
implies that the profiles of the LBGs are
more close to $n<2$ profiles than $n>2$ profiles as a whole.
These results indicate the light profiles of the $z\sim3$ LBGs are similar
to the exponential-law of disk galaxies, and less concentrated 
than those of elliptical galaxies. 

Motivated by the fact that the LBG that has the
largest best fit $n$ parameter has a red $J-K$ color (Q0302$-$003-D73A; $J-K=2.46$ mag),
we examine the profiles of red galaxies at similar redshifts
to the $z\sim3$ LBGs. The criterion $J-K>2.3$ mag for
Distant Red Galaxies (DRGs; \citealt{fra03}) can select the
red galaxy population at similar or slightly lower redshifts to the $z\sim3$ LBGs;
spectroscopic follow-ups of DRGs found among deep surveys have shown that they are
galaxies at $z=2-3$ \citep{van03,red05}, though there can
be contamination of $1<z<2$ galaxies
among $K$-band bright DRGs found in shallow wide area surveys (64\% at $K_s<18.7$ mag; \citealt{con06}).
Using the $1.\!^{\prime\prime}0$ diameter aperture $J-K$ colors of galaxies
in the 10 AO observed fields with $J$-band data, we select DRGs within
$10^{\prime\prime} < d < 35^{\prime\prime}$ 
from the AO guide stars. There are 7 DRGs excluding 3 LBGs that meet the 
DRG criterion in the sample down to $K\sim22$ mag (Figure~\ref{Fig_JKcolmag}). 
All but one of them are fainter than $K=20$ mag.
The number density of the DRGs with $K\sim22$ mag is
consistent with previous studies (e.g., Kajisawa et al. 2006).
The postage stamp images of the DRGs are shown in Figure~\ref{Kimage_DRG_g15}, and
the $K$-band magnitudes, $J-K$ colors, and $r_{\rm HL}$s of the DRGs are
listed in Table~\ref{tab_DRG}. The $r_{\rm HL}$ of the DRGs are
similar to those of the LBGs in the same magnitude range (shown with open squares in
Figure~\ref{Kauto_HL}).
We apply the same S\'ersic profile fitting method by changing
the model PSF to the DRGs with $K<21.5$ mag.
The details of the fitting results are 
summarized in Figures~\ref{DRG_FIT_ser} and ~\ref{DRG_FIT_prof}.
All of the DRGs with $K<21.5$ mag are fitted well with S\'ersic profile with $n<2$.
The best-fit S\'ersic parameters of the DRGs are 
plotted in Figure~\ref{LBG_DRG_Re_N} with open squares. The distribution
of the DRGs is not significantly different from that of the $z\sim3$ LBGs.
The $C$ parameters of the DRGs including DRGs with $21.5<K<22.0$ mag are plotted
in Figure~\ref{LBG_DRG_C_HL} with open squares. The distribution
of the DRGs on the $C$ vs. $r_{\rm HL}$ plane is similar to that of the $z\sim3$ LBGs, and 
favors the $n<2$ profile as a whole.

In addition to the LBGs and the DRGs, we observed one radio galaxy 4C28.58 at
$z=2.891$ during the observing program. We used $K$-band filter in order
to avoid being affected by redshifted strong H$\alpha$ and [OIII] emission lines of 
the radio galaxy. The AO-assisted $K$-band
image of 4C28.58 is shown in the left panel of Figure~\ref{RG_FIT_ser_mod}. 
The image consists of one bright and one faint peaks, and
SExtractor detected two components. The $K$-band magnitude, $J-K$
color, and $r_{\rm HL}$ of each component are measured separately. They
are summarized in Table~\ref{tab_DRG} and plotted in Figure~\ref{Kauto_HL} with 
open triangles.
Because the two peaks are enclosed
in one extended halo, we fit one component S\'ersic profile centered at
the bright peak to the image. The results are shown in the middle and right panels of
Figure~\ref{RG_FIT_ser_mod}
and summarized in Table~\ref{tab_DRG}.
Although the central part of the galaxy has a concentrated core, 
the overall profile is also fitted well with the S\'ersic profile with small $n$ index of $0.3\pm0.2$.

Not only the LBGs,
but also the DRGs and a radio galaxy are dominated with S\'ersic profile with $n<2$.
The result implies that there are less $n>2$ galaxies at $z\sim3$
even among the red and massive galaxies (it should be noted that even $K$-band bright 
LBGs have red $U-V$ colors),
which are dominated by $r^{1/4}$ profiles
in the local universe, 
though the sample of the $z\sim3$ galaxies with $K<21.5$ mag for which S\'ersic profile
fitting with free $n$ is reliable is still quite small in order to draw a rigid conclusion.
The presence of disk-like galaxies at $z\sim3$ is
already reported by \citet{lab03}.
Moreover, the disk-like profiles
of old-red massive galaxies are reported for individual red galaxies at
$z\sim1.5-2.5$ (Iye et al. 2003; Stockton et al. 2004;
Daddi et al. 2005; Toft et al. 2005). Their photometric properties are consistent 
with those of progenitors of elliptical galaxies in the local universe;
their SEDs are well described
with a passive stellar population with a few Gyr age, 
and estimated stellar masses are $\sim10^{11}M_{\odot}$. However
their rest-frame optical profiles are elongated and fitted well with S\'ersic profiles
with $n<2$. In addition to these individual results, it is reported
that the fraction of high-concentration spheroidal galaxies decreases at higher
redshifts using the rest-frame optical data; from $z=0$ to $z=1$ with HST/ACS data \citep{bar05},
from $z=1$ to $z=2$ with HST/NICMOS data covering rest-frame $B$-band \citep{pap05}, and
from $z=0.5$ to $z=3$ with HST/NICMOS data \citep{con05}. The cosmic stellar
mass density in morphologically "early-type" galaxies show a significant decrease
from $z=0.0$ to $z=1.7$ \citep{abr07}.
The observed dominance of disk-like profiles among the $z\sim3$ LBGs and the DRGs  
follows the decreasing trend of the spheroidal galaxies at $z>1$.

\subsection{Dense Disks of High Redshift Galaxies ?}

If we assume that the $z\sim3$ LBGs have disk morphology,
as shown in Figure~\ref{LBG_DRG_Mv_Re}, the $M_{V}$ of the $z\sim3$
LBGs are brighter than $z=0$ and $z=1$ disk galaxies at
the same $R_{e}$. The average surface brightness of the $z\sim3$ LBGs
is 2.2-2.9 mag and 1.2-1.9 mag brighter than those of $z=0$ and
$z=1$ disk galaxies, respectively. The DRGs are plotted in the figure
with open squares. The $R_{e}$s of the DRGs are similar to those
of the $z\sim3$ LBGs at the same $M_{V}$. The average $V$-band surface
brightness of the 7 DRGs is 17.53 (18.28) mag arcsec$^{-2}$ with $q=0.5$ ($q=1.0$), 
which is close to that of the $z\sim3$ LBGs brighter than $M_{V}^{*}$.
We also plot 
LDGs at $2.5<z<3.0$ from \citet{lab03} and a red disk galaxy at $z\sim2.5$
from \citet{sto04} with large crosses and a large asterisk, respectively.
All of them have similar $R_{e}$ to the $z\sim3$ LBGs at the same $M_{V}$.

If we examine $R_{e}$
as a function of $M_{*}$, the $R_{e}$ does not depend on $M_{*}$ as shown in Figure~\ref{LBG_DRG_Ms_Re},
and for the $z\sim3$ LBGs brighter than $M_{V}^{*}$, the average surface stellar mass density estimated
from $M_{*}$ and $R_{e}$ is 3-6 times larger than that 
of $z=0-1$ disk galaxies. The
large surface stellar mass density of the LBGs brighter than $M_{V}^{*}$ implies they have different
stellar mass distribution from those of disk galaxies at $z=0-1$.
On the contrary, for less-massive
$z\sim3$ LBGs, the $\Sigma_{M}$ distribution is similar to
those of $z=0$ and $z=1$ galaxies.
The DRGs are plotted with open squares in the figure.
We estimate their stellar mass using the same relation between the
stellar mass to $V$-band luminosity ratio and the rest-frame $U-V$ color
as shown in Section 5.6. The $R_{e}$s of the DRGs are similar to the $z\sim3$ LBGs
brighter than $M_{V}^{*}$, while their estimated stellar masses are
larger than those of the LBGs on average. The average surface
stellar mass density of the DRGs is calculated to be $\log \Sigma_{M}(M_{\odot}$ {\rm kpc}$^{-2}$)
$=9.8 (9.5)$ with $q=0.5$ ($q=1.0$). The average is even larger than that
of the $z\sim3$ LBGs brighter than $M_{V}^{*}$.

The presence of disk galaxies with high surface stellar mass density is already reported among objects at $z\sim1$
\citep{sim99, rav04}, but their fraction among the whole sample
is not high and no evolution is detected in the
average $\Sigma_{M}$ of disk galaxies below $z<1$ (\citealt{bar05}; Figure~\ref{LBG_DRG_Ms_Re}).
At higher redshifts, a significant fraction of red galaxies show
large surface stellar mass density \citep{tru06b, zir07}; for example 
"quiescent" DRGs have $R_{e}<1$ kpc and stellar mass of
$4\times10^{9}-2\times10^{10} M_{\odot}$ with the Kroupa IMF.
Their $R_{e}$s are smaller than the $z\sim3$ LBGs in the same
mass range, and the estimated surface stellar mass density is
comparable to those of the $z\sim3$ LBGs brighter than $M_{V}^{*}$.

Can we miss low surface stellar mass disks among massive $z\sim3$ galaxies ?
For the LBGs brighter than $M_{V}^{*}$, as shown in 
Figures~\ref{Kauto_HL} and \ref{Kauto_HL_all}, the detection limit of our observation 
is deep enough to detect galaxies two times larger than 
the detected LBGs, therefore if there are many $M_{V}^{*}$ LBGs 
around $\log \Sigma_{M} (M_{\odot} {\rm kpc}^{-2}) = 8.5$ like $z=0$ and $z=1$ galaxies, 
then we expect to detect more LBGs with $\log \Sigma_{M} (M_{\odot} {\rm kpc}^{-2}) \sim 8.5$ among $M_{V}^{*}$
LBGs than observed, even though the current sample size is quite limited.
Because we examine 
$z\sim3$ galaxies selected not only with $U$-dropout but also with 
DRG criterion, if there are many disk 
galaxies with $\log \Sigma_{M} (M_{\odot} {\rm kpc}^{-2}) = 8.5$ at $z\sim3$, we are missing a large number of
such galaxies by both of the LBG and the DRG selections.

The higher stellar mass surface density of the disk systems at 
higher redshift is consistent with simple evolution models of 
rotationally-supported disks in virialized dark-matter 
halos \citep{fal80, mo98, bou02, som06}. Because a halo virialized 
at a higher redshift ($z_{\rm vir}$) has a higher density ($\rho_{\rm vir}\propto H(z_{\rm vir})^{2}$ 
where $H(z_{\rm vir})$ is the Hubble constant at the time of the virialization) 
and a smaller size ($r_{\rm vir}\propto H(z_{\rm vir})^{-2/3}$) for a fixed 
dark-halo mass, if a rotationally-supported disk forms with a fixed
fraction of mass and angular momentum of the dark matter halo, the disk
is expected to be more compact and more dense.
In a simple model, for disks with a fixed mass,
the surface mass density is predicted to follow $\Sigma \propto \rho r \propto H(z)^{4/3}$ \citep{mo98}.
If we assume that the stellar disk follows the relation, we expect that the
surface stellar mass density of disks form at $z=3$ is 7 and 3 times higher
than the disks form at $z=0$ and $z=1$, respectively \citep{mo98}.
Recently a model with a milder evolution of the surface stellar mass 
density is proposed
by Somerville et al. (2006) based on results of N-body simulations.
Because the dark halo build up from the inside out, the mass density 
in the central region of the dark halo in the scale of galaxies evolves slower than that expected
from the simple model shown above in the redshift range between $z=0-3$. 
The model with the mild evolution predicts the size of the disks of $z=3$ galaxies with 
stellar masses greater than $3\times10^{10} M_{\odot}$ is 2 times smaller
than those of $z=0$ galaxies with the same stellar mass.
The observed high surface stellar mass density of the $z\sim3$ LBGs 
brighter than $M_{V}^{*}$ is slightly smaller than that expected from the former
simple model and closer to the latter model with the milder evolution. 

Because in the local universe, such high surface stellar mass density
disk as the massive $z\sim3$ LBGs is rare, we expect
that the disks of the massive $z\sim3$ LBGs are destroyed between $z\sim3$ to $z\sim1$.
The strong spatial clustering of the LBGs implies that they reside in
massive dark halos and that they evolve into local elliptical
galaxies \citep{ste98, mo99, gia02, fou03, ade05}. Therefore, the high
surface stellar mass density disks would evolve into local elliptical galaxies
through merging events.

\subsection{Placing the $z\sim3$ LBGs in the Growth History of Galaxies}

Considering the difference between
the mass of the dark matter halos the $z\sim3$ LBGs
reside ($2-6\times10^{11} M_{\odot}$; \citealt{ade05})
and the Jeans mass at the time of reionization ($10^{10} M_{\odot}$),
we naively expect that the $z\sim3$ LBGs form from building blocks through
several major merges.
Although initially a disk structure can be formed through 
continuous accretion of inter galactic matter \citep{ste02},
a merging event is thought to
transform the disk structure to spheroidal structure through violent
mixture, which has been simulated by many authors with N-body codes 
(see review by \citealt{bar92rev}). 
How does the flat profiles of the LBGs and other high-redshift massive galaxies
similar to local disk-galaxies
survive through the merging events ? Why is the fraction of galaxies
with highly-concentrated profiles similar to local spheroidal galaxies
small 
at high redshifts ? 
The high fraction of disk-like
galaxies in the high-redshift 
universe would be explained with a hypothesis that they have been going through only
gas-rich "wet" merges at that time. 
Recent smoothed particle hydrodynamics simulations of mergers of gas-rich (gas mass fraction
larger than 0.5) galaxies show that disk-like ($n=1$) distribution
of stellar component can be reproduced again after the "wet" merging
\citep{ste02, spr05, rob06, cox06}.
The sizes ($2-6$ kpc) of the remnant disks in the simulations \citep{rob06} are broadly
consistent with those observed in the $z\sim3$ LBGs with similar stellar masses 
($\sim$ a few $\times 10^{10} M_{\odot}$).
We expect the gas mass fraction of the $z\sim3$ LBGs can be high,
based on the fact that the fraction of the $z\sim2$ UV-selected star-forming galaxies, 
which are similar to the LBGs, is really estimated to be 
0.5 on average \citep{erb06}.

How do the $z\sim3$ LBGs evolve into spheroids of local galaxies ?
The disk-like property of $z\sim3$ galaxies conflicts with the
scenario at the first glance.
However, once the gas fraction of galaxies decreases at lower redshifts, 
the disk-like galaxies would evolve into
spheroidal galaxies through gas-poor "dry" merging (e.g., \citealt{ste02}).
The $z\sim3$ galaxies with stellar mass of $10^{10} M_{\odot}$ are
expected to experience still several major merges from $z=3$ to $z=0$ \citep{nag02, con05b}.
The "dry" merging events do not induce intense starformation (e.g., \citealt{van05b}),
and the scenario is consistent with the observational evidences suggesting
the large part of the stars in elliptical galaxies in the nearby universe
is formed at $z>1.5$ (see review by \citealt{ren06}). 

\section{SUMMARY}

In order to examine the rest-frame $V$-band morphology of
the $z\sim3$ LBGs covering a wide luminosity range ($M_{V}^{*}-0.5$ mag $-$
$M_{V}^{*}+3.0$ mag), we conduct AO-assisted imaging observations
of 36 of them in the $K$-band, which corresponds to the rest-frame $V$-band
in $z\sim3$ galaxies. Thirty one of the LBGs are detected in the deep
imaging observations. The AO observations clearly resolve most of
the $z\sim3$ LBGs at the resolution of FWHM$\sim0.\!^{\prime\prime}2$.

1. The median $r_{\rm HL}$ is $0.\!^{\prime\prime}23$, 
which corresponds to 1.8 kpc at $z=3$. LBGs brighter than $M_{V}^{*}$
have larger $r_{\rm HL}$ ($0.\!^{\prime\prime}40$) than the fainter LBGs 
($0.\!^{\prime\prime}23$)
on average, and there is no bright LBGs with small $r_{\rm HL}$.

2. We also obtain $J$-band data of thirty of the LBGs in order to
examine their $J-K$ colors, which correspond to the rest-frame $U-V$ colors.
The bright LBGs show red rest-frame $U-V$ colors 
(average of $0.2$ mag), while
most of the fainter LBGs show blue rest-frame $U-V$ color (average of $-0.4$ mag).
The red and blue colors are close to the colors of red massive and blue less massive
galaxies at $2<z<3$ selected with photometric redshifts, respectively.

3. The AO-assisted $K$-band images of the LBGs are compared with the
optical images with the seeing-limited resolution. 
The peaks in the $K$-band images of 7 of the LBGs with $K<22$ mag
show significant or marginal shifts from those in the optical images.
The presence of the shifts implies that the UV-bright star-forming regions
are not necessarily centered at their main body observed in the $K$-band.
For the LBGs with $K>22$ mag, no shift between the peaks in 
the $K$-band and the optical images is observed.

4. We fit S\'ersic profiles to the images of the LBGs with $K<21.5$ mag,
taking care of the uncertainty of the final PSF at the position
of the targets. The images of all but one of the LBGs with $K<21.5$ mag are fitted
well with S\'ersic profile with $n$ index less than 2, similar to
disk galaxies in the local universe. The uncertainties of the
obtained parameters are examined with the simulated images
of $z\sim3$ galaxies made from HST/ACS images of galaxies at $z=0.36-0.69$.
The SN and the spatial resolution of the AO-assisted $K$-band 
images of the $z\sim3$ LBGs are sufficient to discriminate between
$n<2$ and $n>2$ galaxies.

5. For LBGs with $21.5<K<22.5$ mag, 
we examined concentration parameter instead of fitting S\'ersic profile.
The scatter of the LBGs on the $C$ vs. $r_{\rm HL}$ plane is consistent with
those of the disk galaxies.

6. Assuming that the $z\sim3$ LBGs have a disk shape, we
compare their size-luminosity and size-stellar mass relation with
those of $z=0$ and $z=1$ disk galaxies from Barden et al. (2005).
The $z\sim3$ LBGs are brighter than $z=0$ and $z=1$ disk
galaxies at the same $R_{e}$. The surface brightness of the LBGs,
which are estimated from $M_{V}$ and $R_{e}$, are 2.2-2.9 mag
and 1.2-1.9 mag brighter than those of the disk galaxies at $z=0$ and $z=1$,
respectively. 

7. The size-stellar mass relation of the $z\sim3$ LBGs 
shows that the $R_{e}$ of the $z\sim3$ LBGs does not 
depend on their $M_{*}$. The $z\sim3$ LBGs brighter than $M_{V}^{*}$ have
the average
surface stellar mass density 3-6 times larger than those 
of the $z=0$ and $z=1$ disk galaxies. 
On the contrary, for less-luminous $z\sim3$ LBGs, 
their size-stellar mass relation is consistent with
those of $z=0$ and $z=1$ disk galaxies.

8. We examine the profiles of the serendipitously observed DRGs.
They are also fitted with the S\'ersic profiles with $n<2$, and
their scatter on the $C$ vs. $r_{\rm HL}$ plane is similar to
that of the $z\sim3$ LBGs. Their $R_e$s are similar to those of
the $z\sim3$ LBGs at the same $M_{V}$. Because their estimated
stellar masses are larger than those of the $z\sim3$ LBGs on
average, their average surface stellar mass density is even
larger than that of the $z\sim3$ LBGs brighter than $M_{V}^{*}$.

Using the deep AO imaging observations in the $K$-band, the light
profiles of the $z\sim3$ galaxies in the rest-frame optical
band can be examined with the high spatial resolution for the
first time, though
the number of the observed bright $z\sim3$ galaxies
is still quite limited by the availability of the bright guide star
required for the AO observation with a NGS.
AO systems with laser guide stars (LGSs) are establishing their
importance in the field of
the high resolution NIR imaging observations of galaxies in the 
high galactic latitude regions. We start an LGS AO observing program to
extend the sample of bright $z\sim3$ galaxies, in order to
establish the morphological evolution between $z=3$ to $z=0$.
The high-resolution $K$-band imaging observations of the galaxies at $z\ge3$
should be one of unique fields which are explored only with LGS AO on
ground-based 8-10m class telescopes until the launch of the James Webb Space Telescope.

\acknowledgments

We would like to thank through reviewing and valuable comments by 
the refree.
We thank staff members of the Subaru Telescope 
for their support during observations, especially
support scientists, 
Drs. Hiroshi Terada, Shin Oya, Michihiro Takami, and
Miki Ishii. We would like to thank Dr. Masahiro Nagashima
for a valuable discussion. M.A. was supported by a Research
Fellowship of the Japan Society for the 
Promotion of Science (JSPS) for Young Scientists
during some part of this program. Part of the
program is also supported by a Grant-in-Aid for Young Scientists (B)
from JSPS (18740118).

\clearpage

\begin{deluxetable}{lccccc}
\tabletypesize{\scriptsize}
\tablecaption{Observed Field of Views \label{tab_fovs}}
\tablewidth{0pt}
\tablehead{
\multicolumn{1}{c}{Field ID} &
\multicolumn{2}{c}{Center} &
\multicolumn{2}{c}{Guide Star} &
\multicolumn{1}{c}{$R$ \tablenotemark{a}} \\
\multicolumn{1}{c}{} &
\multicolumn{1}{c}{RA} &
\multicolumn{1}{c}{DEC} &
\multicolumn{1}{c}{RA} &
\multicolumn{1}{c}{DEC} &
\multicolumn{1}{c}{(mag)} \\
\multicolumn{1}{c}{} &
\multicolumn{1}{c}{(2000)} &
\multicolumn{1}{c}{(2000)} &
\multicolumn{1}{c}{(2000)} &
\multicolumn{1}{c}{(2000)} &
\multicolumn{1}{c}{} 
}
\startdata
Q0302$-$003-FOV1    & 03:04:28.9 & $-$00:07:55.0 & 03:04:29.71 & $-$00:07:39.8 & 14.7 \\
Q0302$-$003-FOV3    & 03:04:45.3 & $-$00:13:26.1 & 03:04:44.71 & $-$00:13:38.5 & 14.5 \\
Q1422$+$2309-FOV12  & 14:24:34.1 & $+$22:50:48.0 & 14:24:34.43 & $+$22:50:53.8 & 14.5 \\
Q1422$+$2309-FOV4   & 14:24:49.2 & $+$22:51:30.0 & 14:24:48.98 & $+$22:51:40.0 & 11.6 \\
Q1422$+$2309-FOV5   & 14:24:28.3 & $+$22:56:38.0 & 14:24:29.76 & $+$22:56:41.0 & 13.7 \\
Q1422$+$2309-FOV7   & 14:24:29.2 & $+$22:53:05.0 & 14:24:29.03 & $+$22:52:59.6 & 13.0 \\
Q2233$+$1341-FOV1   & 22:36:24.6 & $+$13:58:54.2 & 22:36:24.04 & $+$13:58:51.1 & 13.7 \\
DSF2237b-FOV1       & 22:39:20.5 & $+$11:55:27.0 & 22:39:20.69 & $+$11:55:35.6 & 13.1 \\
DSF2237b-FOV2       & 22:39:22.3 & $+$11:48:42.0 & 22:39:22.13 & $+$11:48:40.4 & 14.1 \\
DSF2237b-FOV3       & 22:39:23.7 & $+$11:47:50.0 & 22:39:23.76 & $+$11:47:50.0 & 13.0 \\
DSF2237a-FOV4       & 22:39:58.6 & $+$11:50:49.6 & 22:39:58.55 & $+$11:50:50.0 & 13.3 \\
DSF2237a-FOV5       & 22:39:49.3 & $+$11:52:59.4 & 22:39:49.61 & $+$11:52:59.7 & 14.3 \\
4C28.58\tablenotemark{b}    & 23:52:00.8 & $+$29:10:29.0 & 23:52:02.03 & $+$29:10:27.9 & 12.1 \\
\enddata
\tablenotetext{a}{$R$-band magnitude of the AO-guide star from USNO catalog.}
\tablenotetext{b}{The radio galaxy is observed in order to compare with the LBGs.
See Section 6.1.}
\end{deluxetable}

\begin{deluxetable}{llcccrrl}
\tabletypesize{\scriptsize}
\tablecaption{Observed Objects \label{tab_obj}}
\tablewidth{0pt}
\tablehead{
\multicolumn{1}{c}{Field ID} &
\multicolumn{1}{c}{Name} &
\multicolumn{1}{c}{RA\tablenotemark{a}} &
\multicolumn{1}{c}{DEC\tablenotemark{a}} &
\multicolumn{1}{c}{$R$\tablenotemark{a}} &
\multicolumn{1}{c}{$z_{\rm em}$\tablenotemark{a}} &
\multicolumn{1}{c}{$z_{\rm abs}$\tablenotemark{a}} &
\multicolumn{1}{c}{Note} \\
\multicolumn{1}{c}{} &
\multicolumn{1}{c}{} &
\multicolumn{1}{c}{(2000)} &
\multicolumn{1}{c}{(2000)} &
\multicolumn{1}{c}{(mag)} &
\multicolumn{1}{c}{} &
\multicolumn{1}{c}{} &
\multicolumn{1}{c}{} 
}
\startdata
Q0302$-$003-FOV1   & Q0302$-$003-MD373   & 03:04:28.47 & $-$00:07:45.9 & 24.36 &  2.787 &  9.999 &       \\
                   & Q0302$-$003-D113    & 03:04:30.33 & $-$00:08:11.4 & 24.64 &  2.920 &  9.999 & QSO   \\
                   & Q0302$-$003-MD346   & 03:04:27.84 & $-$00:07:41.1 & 24.14 &  9.999 &  9.999 &       \\ \hline
Q0302$-$003-FOV3   & Q0302$-$003-M80     & 03:04:45.70 & $-$00:13:40.6 & 24.12 &  3.416 &  9.999 &       \\
                   & Q0302$-$003-MD216   & 03:04:44.78 & $-$00:13:01.8 & 24.43 &  9.999 &  2.401 &       \\
                   & Q0302$-$003-M86     & 03:04:45.18 & $-$00:13:17.4 & 25.43 &  9.999 &  9.999 &       \\
                   & Q0302$-$003-D73     & 03:04:44.91 & $-$00:13:21.0 & 24.46 &  9.999 &  9.999 &       \\
                   & Q0302$-$003-MD204   & 03:04:44.81 & $-$00:13:24.8 & 24.42 &  9.999 &  9.999 &       \\ 
                   & Q0302$-$003-MD192   & 03:04:45.35 & $-$00:13:51.1 & 24.37 &  9.999 &  9.999 &       \\
\hline
Q1422$+$2309-FOV12  & Q1422$+$2309-C36     & 14:24:33.11 & $+$22:51:15.3 & 24.08 &  3.261 &  3.251 &       \\ 
                    & Q1422$+$2309-MD68    & 14:24:33.24 & $+$22:50:52.2 & 25.17 &  9.999 &  2.801 &       \\
                    & Q1422$+$2309-MD61    & 14:24:34.74 & $+$22:50:25.7 & 25.68 &  2.580 &  9.999 &       \\
                    & Q1422$+$2309-C26     & 14:24:34.98 & $+$22:50:23.6 & 24.95 &  3.088 &  3.083 &       \\
                    & Q1422$+$2309-MD59    & 14:24:33.86 & $+$22:50:22.4 & 24.70 &  9.999 &  9.999 &       \\
                    & Q1422$+$2309-MD60    & 14:24:33.62 & $+$22:50:25.1 & 25.36 &  9.999 &  9.999 &       \\
                    & Q1422$+$2309-D25     & 14:24:33.04 & $+$22:50:35.7 & 25.09 &  9.999 &  9.999 &       \\ \hline
Q1422$+$2309-FOV4   & Q1422$+$2309-C35     & 14:24:50.20 & $+$22:51:14.4 & 24.68 &  3.423 &  3.412 &       \\ \hline
Q1422$+$2309-FOV5   & Q1422$+$2309-MD133   & 14:24:29.12 & $+$22:56:24.6 & 23.24 &  2.748 &  2.745 &       \\ \hline
Q1422$+$2309-FOV7   & Q1422$+$2309-C52     & 14:24:30.29 & $+$22:52:49.9 & 24.61 &  9.999 &  3.072 &       \\
                    & Q1422$+$2309-MD90    & 14:24:26.93 & $+$22:53:21.6 & 24.37 &  2.743 &  9.999 &       \\ 
                    & Q1422$+$2309-C57     & 14:24:27.22 & $+$22:53:10.3 & 24.98 &  9.999 &  9.999 &       \\
                    & Q1422$+$2309-MD88    & 14:24:26.96 & $+$22:53:16.2 & 25.19 &  9.999 &  9.999 &       \\ \hline
Q2233$+$1341-FOV1   & Q2233$+$1341-MD46    & 22:36:24.26 & $+$13:59:12.5 & 23.81 &  2.716 &  2.711 &       \\ \hline
DSF2237b-FOV1 & DSF2237b-MD81 & 22:39:21.72 & $+$11:55:10.4 & 24.16 &  9.999 &  2.823 &       \\ 
              & DSF2237b-D28  & 22:39:20.25 & $+$11:55:11.3 & 24.46 &  2.938 &  2.926 &       \\
              & DSF2237b-MD90 & 22:39:19.73 & $+$11:55:40.8 & 23.60 &  9.999 &  9.999 &       \\
              & DSF2237b-D27  & 22:39:19.79 & $+$11:55:08.0 & 24.92 &  9.999 &  9.999 &       \\
              & DSF2237b-MD80 & 22:39:22.22 & $+$11:55:08.3 & 25.37 &  9.999 &  9.999 &       \\ \hline  
DSF2237b-FOV2 & DSF2237b-MD22 & 22:39:23.07 & $+$11:48:57.5 & 24.36 &  9.999 &  2.925 &       \\
              & DSF2237b-MD19 & 22:39:21.08 & $+$11:48:27.7 & 24.48 &  2.616 &  2.614 &       \\
              & DSF2237b-C10  & 22:39:24.33 & $+$11:48:28.5 & 25.10 &  9.999 &  9.999 &       \\ \hline
DSF2237b-FOV3 & DSF2237b-D5   & 22:39:24.97 & $+$11:47:57.5 & 23.51 &  9.999 &  2.618 &       \\ 
              & DSF2237b-MD13 & 22:39:23.14 & $+$11:47:26.0 & 24.97 &  9.999 &  9.999 &       \\
              & DSF2237b-MD14 & 22:39:24.77 & $+$11:47:42.5 & 25.04 &  9.999 &  9.999 &       \\ \hline
DSF2237a-FOV4 & DSF2237a-C11  & 22:39:57.69 & $+$11:50:32.7 & 24.69 &  3.152 &  9.999 &       \\ \hline 
DSF2237a-FOV5 & DSF2237a-C15  & 22:39:50.06 & $+$11:52:45.7 & 23.04 &  3.151 &  3.138 &       \\ \hline
4C28.58       & 4C28.58       & 23:51:59.20 & $+$29:10:29.0 &       &  2.891\tablenotemark{b} &  ---   &       \\ \hline
\enddata
\tablenotetext{a}{Taken from Steidel et al. (2003). 9.999 in $z_{\rm em}$ and $z_{\rm abs}$ means not available.}
\tablenotetext{b}{Taken from R\"ottgering et al. (1996).}
\end{deluxetable}

\begin{deluxetable}{lrcccccccl}
\tabletypesize{\scriptsize}
\tablecaption{Journal of Observations \label{tab_dates}}
\tablewidth{0pt}
\tablehead{
\multicolumn{2}{c}{Date\tablenotemark{a}} &
\multicolumn{5}{c}{Seeing size\tablenotemark{b}} &
\multicolumn{1}{c}{Atmospheric extinction\tablenotemark{c}} &
\multicolumn{1}{c}{Photometric condition\tablenotemark{d}} &
\multicolumn{1}{c}{Note} \\
\multicolumn{2}{c}{(HST)} &
\multicolumn{1}{c}{($^{\prime\prime}$)} &
\multicolumn{1}{c}{} &
\multicolumn{1}{c}{($^{\prime\prime}$)} &
\multicolumn{1}{c}{} &
\multicolumn{1}{c}{($^{\prime\prime}$)} &
\multicolumn{1}{c}{(mag)} &
\multicolumn{1}{c}{} &
\multicolumn{1}{c}{} 
}
\startdata
2003/10/14 & 11.0h & 0.49     & / & 0.49     & / & 0.53      & 0.05(0.00-0.10) & whole night & thin cirrus \\
2004/04/04 &  7.0h & 1.06     & / & 0.74     & / & 0.51      & 0.10(0.05-0.15) & whole night & clear \\
2004/04/05 &  7.0h & 0.72     & / & 0.59     & / & $\cdots$  & 0.08(0.05-0.10) & whole night & clear \\
2004/05/28 &  5.5h & 0.48     & / & 0.48     & / & $\cdots$  & 0.08(0.05-0.10) & whole night & thin cirrus \\
2004/06/13 &  7.0h & $\cdots$ & / & 0.86     & / & 0.65      & 0.10(0.00-0.20) & last half   & thick cirrus \\
2004/06/14 & 10.0h & 0.59     & / & 0.69     & / & $\cdots$  & 0.30(0.00-1.00) & last half   & thin cirrus \\
2004/06/30 &  5.0h & 0.68     & / & 0.71     & / & $\cdots$  & 0.50(0.00-1.00) & no          & thick cirrus, cloudy \\
2004/07/02 & 10.0h & 0.52     & / & 0.52     & / & 0.69      & 0.05(0.00-0.07) & whole night & clear \\
2004/07/03 & 10.0h & 0.65     & / & 0.45     & / & 0.38      & 0.03(0.00-0.05) & whole night & clear \\ 
2004/07/04 & 10.0h & 0.75     & / & 0.80     & / & $\cdots$  & 0.03(0.00-0.10) & whole night & clear \\
2004/07/05 &  5.0h & 0.67     & / & 0.60     & / & $\cdots$  & 0.02(0.00-0.06) & whole night & clear \\
2004/07/26 &  5.0h & 0.74     & / & 0.85     & / & $\cdots$  & 0.04(0.03-0.05) & whole night & clear \\
2004/07/27 &  5.0h & 0.87     & / & 0.61     & / & 0.53      & 0.05(0.00-0.07) & whole night & clear \\
2004/07/28 &  5.0h & 0.46     & / & 0.38     & / & 0.49      & 0.02(0.00-0.05) & whole night & clear \\
2004/07/29 &  5.0h & 0.66     & / & 0.38     & / & 0.52      & 0.10(0.00-0.20) & whole night & cirrus,clear \\
2004/09/06 &  7.0h & 0.41     & / & 1.11     & / & $\cdots$  & 0.06(0.04-0.08) & whole night & cirrus,clear \\
2004/09/23 &  6.0h & 1.18     & / & 0.87     & / & 0.44      & 0.06(0.04-0.08) & whole night & clear \\
2004/09/24 &  8.0h & 0.58     & / & 0.71     & / & 0.96      & 0.06(0.00-0.10) & whole night & clear \\
2004/09/25 &  7.0h & 0.57     & / & 0.59     & / & 0.47      & 0.08(0.00-0.10) & whole night & clear \\
2004/09/26 &  9.0h & 0.55     & / & 1.32     & / & 1.52      & 0.06(0.02-0.08) & whole night & fog, clear \\
2004/09/29 &  7.0h & 0.72     & / & 0.79     & / & $\cdots$  & 0.05(0.00-0.10) & no           & thick cirrus \\
2004/10/25 &  2.0h & $\cdots$ & / & $\cdots$ & / & $\cdots$  & $\cdots$ & $\cdots$   & cloudy \\
\enddata
\tablenotetext{a}{Date and allocated observing hours. Dates are in Hawaiian Standard Time (HST).}
\tablenotetext{b}{Seeing size (FWHM) measured with the optical auto-guiding (AG) camera at 
evening twighlight, midnight, and dawn are separated by $/$. The sensitivity of the AG camera peaks in the $R$-band.}
\tablenotetext{c}{Atmospheric extinction in the optical band measured at CFHT from 
http://www.cfht.hawaii.edu/Instruments/Skyprobe/. A typical value
and the range of the value shown in the parentheses in mag. Large atmospheric extinction means
non-photometric condition.}
\tablenotetext{d}{Photometric period during the night. 
We examined the photometric condition from the time variation of the 
count rates of bright objects in the observed fields. }
\end{deluxetable}

\begin{deluxetable}{lllrll}
\tabletypesize{\scriptsize}
\tablecaption{Observations for Each Field\label{tab_obsfield}}
\tablewidth{0pt}
\tablehead{
\multicolumn{1}{c}{Field ID} &
\multicolumn{1}{c}{Date} &
\multicolumn{1}{c}{Band} &
\multicolumn{1}{c}{Integ. Time\tablenotemark{a}} &
\multicolumn{1}{c}{Note, FWHM\tablenotemark{b}} &
\multicolumn{1}{c}{} \\
\multicolumn{1}{c}{} &
\multicolumn{1}{c}{(HST)} &
\multicolumn{1}{c}{} &
\multicolumn{1}{c}{} &
\multicolumn{1}{c}{} &
\multicolumn{1}{c}{} 
}
\startdata
Q0302$-$003-FOV1    & 2003/10/14 & $K$          &  90$^s$$\times$126 & 0.18 (30,13) \\ 
                    & 2004/09/25 & $K$          &  90$^s$$\times$40  & No PSF reference.  \\
                    & 2004/09/26 & $K$          &  90$^s$$\times$75  & No PSF reference. \\ 
                    &            &              &    total 6.03$^h$  & \\ \cline{2-5}
                    & 2004/09/06 & $J$          & 120$^s$$\times$40  & No PSF reference. \\
                    & 2004/09/26 & $J$          & 120$^s$$\times$27  & No PSF reference. \\ 
                    &            &              &    total 2.23$^h$  & \\ \hline
Q0302$-$003-FOV3    & 2004/09/23 & $K^{\prime}$ & 100$^s$$\times$74  & No PSF reference. \\ 
                    & 2004/09/24 & $K^{\prime}$ &  90$^s$$\times$27  & No PSF reference. \\ 
                    & 2004/09/25 & $K^{\prime}$ & 120$^s$$\times$54  & No PSF reference. \\ 
                    &            &              &    total 4.53$^h$  & \\ \cline{2-5}
                    & 2004/09/24 & $J$          & 120$^s$$\times$45  & No PSF reference. \\ 
                    &            &              &    total 1.50$^h$  & \\ \hline
Q1422$+$2309-FOV12  & 2004/04/05 & $K$          &  90$^s$$\times$63  & 0.17(12,6), 0.23(14,13) \\ 
                    & 2004/04/04 & $K^{\prime}$ & 120$^s$$\times$42  & 0.16(12,6), 0.20(14,13) \\ 
                    & 2004/05/28 & $K^{\prime}$ & 120$^s$$\times$97  & 0.16(12,6), 0.18(14,13) \\ 
                    &            &              &   total 6.21$^h$   & 0.17(12,6), 0.19(14,13) \\ \cline{2-5}
                    & 2004/06/13 & $J$          & 120$^s$$\times$36  & 0.47(12,6), 0.50(14,13)\\
                    & 2004/06/14 & $J$          & 120$^s$$\times$45  & 0.41(12,6), 0.48(14,13)\\ 
                    & 2004/07/03 & $J$          &  60$^s$$\times$9   & For photometric calibration.\\
                    &            &              &   total 2.70$^h$   & 0.44(12,6),0.49(14,13) \\ \hline
Q1422$+$2309-FOV4   & 2004/07/02 & $K^{\prime}$ & 100$^s$$\times$45  & 0.19(14,14),0.25(19,10) \\
                    & 2004/07/04 & $K^{\prime}$ & 100$^s$$\times$89  & 0.21(14,14),0.22(19,10) \\ 
                    &            &              &   total 3.72$^h$   & 0.21(14,14),0.22(19,10) \\ \cline{2-5}
                    & 2004/07/02 & $J$          & 120$^s$$\times$18  &  \\
                    & 2004/07/03 & $J$          &  60$^s$$\times$9   & For photometric calibration.\\
                    & 2004/07/04 & $J$          & 100$^s$$\times$36  & 0.42(19,10) \\ 
                    &            &              &   total 1.60$^h$   & \\ \hline
Q1422$+$2309-FOV5   & 2004/04/05 & $K^{\prime}$ &  90$^s$$\times$44  & No PSF reference. \\
                    & 2004/06/14 & $K^{\prime}$ & 120$^s$$\times$80  & No PSF reference. \\
                    & 2004/07/05 & $K^{\prime}$ & 100$^s$$\times$89  & No PSF reference. \\ 
                    &            &              &   total 6.24$^h$   & \\ \cline{2-5}
                    & 2004/06/13 & $J$          & 120$^s$$\times$36  & No PSF reference. \\ 
                    & 2004/07/03 & $J$          &  60$^s$$\times$9   & For photometric calibration. \\
                    &            &              &   total 1.20$^h$   & \\ \hline
Q1422$+$2309-FOV7   & 2004/04/04 & $K$          & 120$^s$$\times$79  & 0.20(1,20) \\  
                    & 2004/07/03 & $K$          &  90$^s$$\times$98  & 0.19(1,20) \\ 
                    &            &              &   total 5.08$^h$   & 0.20(1,20) \\ \cline{2-5}
                    & 2004/07/03 & $J$          &  60$^s$$\times$9   & For photometric calibration.\\
                    & 2004/07/05 & $J$          & 100$^s$$\times$27  & \\  
                    &            &              &   total 0.75$^h$   & \\ \hline
Q2233$+$1341-FOV1   & 2004/07/04 & $K^{\prime}$ &  60$^s$$\times$170 & 0.18(4,7),0.20(11,18),0.21(6,35)\\
                    & 2004/09/06 & $K^{\prime}$ & 100$^s$$\times$89  & 0.18(4,7),0.27(11,18),0.28(6,35)\\
                    & 2004/09/23 & $K^{\prime}$ & 100$^s$$\times$70  & 0.23(4,7),0.25(11,18),0.26(6,35)\\
                    & 2004/09/24 & $K^{\prime}$ &  90$^s$$\times$25  & 0.23(5,26),0.22(6,35)\\ 
                    &            &              &   total 7.87$^h$   & 0.19(4,7),0.22(5,26),0.23(5,26) \\ \cline{2-5}
                    & 2004/09/24 & $J$          & 120$^s$$\times$54  & 0.39(4,7),0.38(5,26),0.35(6,35)\\ 
                    &            &              &   total 1.80$^h$   &  \\ \hline
DSF2237b-FOV1       & 2003/10/14 & $K$          &  90$^s$$\times$149 & 0.20(33,4) \\
                    & 2004/07/26 & $K$          &  90$^s$$\times$79  & 0.22(33,4) \\ 
                    &            &              &   total 5.71$^h$   & 0.22(33,4) \\ \cline{2-5}
                    & 2004/09/29 & $J$          & 120$^s$$\times$81  & 0.45(33,4)\\  
                    &            &              &   total 2.70$^h$   & \\ \hline
DSF2237b-FOV2       & 2004/07/03 & $K$          &  60$^s$$\times$162 & 0.28(11,9) \\
                    & 2004/07/27 & $K$          &  90$^s$$\times$80  & 0.17(11,9) \\ 
                    &            &              &   total 2.00$^h$   & 0.25(11,9) \\ \cline{2-5}
                    & 2004/06/13 & $J$          & 120$^s$$\times$72  & 0.40(11,9),0.40(4,19),0.45(34,24)\\ 
                    &            &              &   total 2.40$^h$   & \\ \hline
DSF2237b-FOV3       & 2004/05/28 & $K^{\prime}$ & 120$^s$$\times$16  & 0.14(4,8),0.17(25,5),0.19(31,20) \\ 
                    & 2004/07/26 & $K^{\prime}$ & 120$^s$$\times$44  & 0.13(4,8),0.18(25,5),0.20(31,20) \\ 
                    & 2004/07/30 & $K^{\prime}$ & 120$^s$$\times$61  & 0.13(4,8),0.21(25,5),0.23(31,20) \\
                    & 2004/09/25 & $K^{\prime}$ &  90$^s$$\times$61  & 0.13(4,8),0.19(25,5),0.24(31,20) \\
                    &            &              &  total 5.56$^h$    & 0.13(4,8),0.19(25,5),0.23(31,20) \\  \cline{2-5}
                    & 2004/06/14 & $J$          & 120$^s$$\times$63  & 0.30(4,8),0.37(25,5),0.46(31,20) \\ 
                    &            &              &  total 2.10$^h$    &  \\ \hline
DSF2237a-FOV4       & 2004/07/02 & $K^{\prime}$ &  90$^s$$\times$75  & 0.17(27,7)\\ 
                    &            &              &  total 1.88$^h$    &  \\ \hline
DSF2237a-FOV5       & 2004/07/28 & $K^{\prime}$ & 180$^s$$\times$81  & 0.15(0,0) \\ 
                    &            &              &  total 4.05$^h$    &  \\ \hline
4C28.58             & 2004/07/30 & $K$          &  90$^s$$\times$45  & 0.14(10,8), 0.17(28,5)\\  
                    &            &              &  total 1.13$^h$    &  \\ \hline
\enddata
\tablenotetext{a}{
Integration time for each frame in second and numbers of the frames
for each field. Total integration time in hours for each field
is shown in the total line.
}
\tablenotetext{b}{
FWHM of stellar objects in arcsec with the distance
from the guide star in X and Y directions of the final image in parenthesis. FWHM and
distances are in arcsec unit.
}
\end{deluxetable}

\clearpage

\begin{deluxetable}{lccc}
\tabletypesize{\scriptsize}
\tablecaption{Optical $V$-band Imaging Observation with FOCAS \label{tab_FOCAS}}
\tablewidth{0pt}
\tablehead{
\multicolumn{1}{c}{Field ID} &
\multicolumn{1}{c}{Date} &
\multicolumn{1}{c}{Integ. Time\tablenotemark{a}} &
\multicolumn{1}{c}{FWHM\tablenotemark{b}} \\
\multicolumn{1}{c}{} &
\multicolumn{1}{c}{(HST)} &
\multicolumn{1}{c}{(s)} &
\multicolumn{1}{c}{($^{\prime\prime}$)} 
}
\startdata
Q0302$-$003-FOV1      & 2004/10/16 & $600^s\times3$  &        \\ 
                      &            &   total 0.5$^h$ & $0.77$ \\ \hline
Q0302$-$003-FOV3      & 2004/11/10 & $600^s\times2$  &        \\ 
                      & 2004/11/11 & $600^s\times3$  &        \\ 
                      &            &   total 0.8$^h$ & $0.70$ \\ \hline
Q2233$+$1341-FOV1     & 2004/10/15 & $600^s\times7$  &        \\ 
                      &            &   total 1.2$^h$ & $0.69$ \\ \hline
DSF2237b-FOV1         & 2004/10/15 & $600^s\times3$  & \\ 
                      & 2004/10/16 & $600^s\times3$  & \\ 
                      &            &   total 1.0$^h$ & $0.57$ \\ \hline
DSF2237b-FOV2,FOV3    & 2004/10/15 & $600^s\times4$  & \\ 
                      & 2004/10/16 & $600^s\times1$  & \\ 
                      &            &   total 0.9$^h$ & $0.57$ \\ \hline
DSF2237a-FOV4,FOV5    & 2004/09/20 & $300^s\times6$  & \\ 
                      &            &   total 0.5$^h$ & $0.62$ \\ \hline
\enddata
\tablenotetext{a}{
Integration time for each frame in second and numbers of the frames
for each field. Total integration time in hours for each field
is shown in the total line.
}
\tablenotetext{b}{FWHM of stellar objects in arcsec measured in the final images.}
\end{deluxetable}

\begin{deluxetable}{lcccccccccr}
\tabletypesize{\scriptsize}
\tablecaption{Results for $z\sim3$ LBGs Observations \label{tab_LBG}}
\tablewidth{0pt}
\tablehead{
\multicolumn{1}{c}{Name} &
\multicolumn{1}{c}{Distance\tablenotemark{a}} &
\multicolumn{1}{c}{FWHM\tablenotemark{b}} &
\multicolumn{1}{c}{$K$\tablenotemark{c}} &
\multicolumn{1}{c}{$J-K$\tablenotemark{d}} &
\multicolumn{1}{c}{$r_{\rm HL}$} &
\multicolumn{1}{c}{$C_{80/20}$} &
\multicolumn{1}{c}{$n$} &
\multicolumn{1}{c}{$\log R_{e}$\tablenotemark{e}} &
\multicolumn{1}{c}{$M_V$\tablenotemark{f}} &
\multicolumn{1}{c}{$\log M_{*}$\tablenotemark{g}} \\
\multicolumn{1}{c}{} &
\multicolumn{1}{c}{($^{\prime\prime}$,$^{\prime\prime}$)} &
\multicolumn{1}{c}{($^{\prime\prime}$)} &
\multicolumn{1}{c}{(mag)} &
\multicolumn{1}{c}{(mag)} &
\multicolumn{1}{c}{($^{\prime\prime}$)} &
\multicolumn{1}{c}{} &
\multicolumn{1}{c}{} &
\multicolumn{1}{c}{(kpc)} &
\multicolumn{1}{c}{(mag)} &
\multicolumn{1}{c}{($M_{\odot}$)}
}
\startdata
Q0302$-$003-MD373A  & $(-18,  -6)$ & 0.14 & 23.12 & $1.03\pm 0.08$ &  0.20 & 0.40 & $\cdots$            & $\cdots$       & $-20.43$     & 9.1 \\
Q0302$-$003-MD373B  & $(-18,  -6)$ & 0.14 & $\cdots$ &    $\cdots$ &  $\cdots$ & $\cdots$ & $\cdots$    & $\cdots$       & $\cdots$     & $\cdots$ \\
Q0302$-$003-MD373C  & $(-18,  -6)$ & 0.14 & 23.63 &        $>2.34$ &  0.14 & 0.46 & $\cdots$            & $\cdots$       & $-19.92$     & 9.8   \\
Q0302$-$003-D113    & $(  9, -31)$ & 0.18 & 22.12 & $1.50\pm 0.07$ &  0.18 & 0.47 & 1.0FIX & $0.14^{+0.06}_{-0.07}$ & $-21.51$ & 9.9   \\
Q0302$-$003-MD346   & $(-27,  -1)$ & 0.16 & 22.88 & $0.43\pm 0.04$ &  0.14 & 0.45 & $\cdots$ & $\cdots$ & $-20.80$ & 8.9   \\ 
\hline
Q0302$-$003-M80A    & $( 14,  -2)$ & 0.20\tablenotemark{m} & 21.96 & $1.54\pm 0.09$ & 0.41\tablenotemark{h} & 0.53\tablenotemark{h} & 1.0FIX & $0.48^{+0.07}_{-0.09}$  & $-21.95$ & 10.1    \\
Q0302$-$003-M80B    & $( 14,  -2)$ & 0.20\tablenotemark{m} & 22.96 & $1.43\pm 0.12$ & 0.22\tablenotemark{h} & 0.47\tablenotemark{h} & $\cdots$            & $\cdots$   & $-20.95$ &  9.6    \\
Q0302$-$003-MD216A  & $(  1,  36)$ & 0.23\tablenotemark{m} & 22.12 & $1.15\pm 0.05$ & 0.25 & 0.47 & 1.0FIX   & $0.24^{+0.06}_{-0.08}$ & $-21.15$ & 9.5 \\
Q0302$-$003-MD216B  & $(  1,  36)$ & 0.23\tablenotemark{m} & 22.64 &        $>2.22$ & 0.26 & 0.54 & $\cdots$ & $\cdots$               & $-20.63$ & 10.0 \\
Q0302$-$003-M86     & $(  7,  21)$ & 0.19\tablenotemark{m} & 23.59 & $1.29\pm 0.16$ & 0.17 & 0.49 & $\cdots$ & $\cdots$               & $-20.09$ & 9.2 \\
Q0302$-$003-D73A    & $(  3,  17)$ & 0.18\tablenotemark{m} & 21.38 & $2.46\pm 0.11$ & 0.32 & 0.60 & $3.0^{+1.1}_{-0.7}$ & $0.58^{+0.12}_{-0.15}$ & $-22.30$ & 10.8 \\
Q0302$-$003-D73B    & $(  3,  17)$ & 0.18\tablenotemark{m} & 21.40 & $1.36\pm 0.04$ & 0.30 & 0.51 & $0.5^{+0.1}_{-0.1}$ & $0.48^{+0.03}_{-0.03}$ & $-22.30$ & 10.1 \\
Q0302$-$003-MD204   & $(  1,  13)$ & 0.17\tablenotemark{m} & 22.21 & $1.14\pm 0.06$ & 0.27 & 0.55 & 1.0FIX & $0.11^{+0.08}_{-0.10}$ & $-21.47$ & 9.6 \\
Q0302$-$003-MD192   & $(  9, -12)$ & 0.17\tablenotemark{m} & 22.77 & $1.11\pm 0.09$ & 0.23 & 0.38 & $\cdots$ & $\cdots$ & $-20.91$ & 9.4 \\
\hline
Q1422+2309-C36   & $(-17,  21)$ & 0.21 & 21.22 & $2.75\pm 0.13$ & 0.21 & 0.49  & $0.6^{+0.3}_{-0.5}$ & $0.17^{+0.03}_{-0.03}$ & $-22.60$ & 11.1 \\
Q1422+2309-MD68  & $(-16,  -1)$ & 0.17 & 22.61 & $>1.41$ & 0.20 & 0.52  & $\cdots$ & $\cdots$     & $-20.95$ & 9.6 \\
Q1422+2309-MD61  & $(  4, -28)$ & 0.21 & $\cdots$ &    $\cdots$ & $\cdots$ & $\cdots$ & $\cdots$ & $\cdots$    & $\cdots$            & $\cdots$ \\
Q1422+2309-C26   & $(  8, -30)$ & 0.21 & 22.63 & $1.25\pm 0.12$ & 0.25 & 0.47 & $\cdots$   & $\cdots$       & $-21.10$ & 9.5 \\
Q1422+2309-MD59  & $( -7, -31)$ & 0.22 & 21.94 & $>1.83$ & 0.31 & 0.43  & 1.0FIX        & $0.53^{+0.06}_{-0.08}$  & $-21.74$  & 10.2 \\
Q1422+2309-MD60  & $(-10, -28)$ & 0.21 & $\cdots$ &    $\cdots$ & $\cdots$ & $\cdots$ & $\cdots$ & $\cdots$    & $\cdots$            & $\cdots$ \\
Q1422+2309-D25   & $(-18, -18)$ & 0.20 & $\cdots$ &    $\cdots$ & $\cdots$ & $\cdots$ & $\cdots$ & $\cdots$    & $\cdots$            & $\cdots$ \\
\hline
Q1422+2309-C35    & $( 16, -25)$ & 0.24 & 21.81 & $>1.85$ & 0.26 & 0.54  & 1.0FIX    & $0.31^{+0.06}_{-0.01}$ & $-22.10$ & 10.3 \\
\hline
Q1422+2309-MD133  & $( -9, -16)$ & 0.18\tablenotemark{m} & 20.41 & $1.97\pm 0.03$ & 0.32 & 0.52  & $1.0^{+0.1}_{-0.1}$ & $0.38^{+0.06}_{-0.08}$ & $-23.11$ & 10.8 \\
\hline
Q1422+2309-C52    & $( 17,  -9)$ & 0.19 & 22.84 & $>0.75$ &  0.18 & 0.49 & $\cdots$  & $\cdots$   & $-20.88$ & 9.1 \\
Q1422+2309-MD90   & $(-28,  21)$ & 0.24 & 23.11 & $>0.68$ &  0.20 & 0.41 & $\cdots$  & $\cdots$   & $-20.41$ & 8.9 \\
Q1422+2309-C57    & $(-24,  10)$ & 0.21 & 22.26 & $>1.16$ &  0.33 & 0.47 & 1.0FIX    & $0.85^{+0.05}_{-0.06}$   & $-21.42$ & 9.6 \\
Q1422+2309-MD88   & $(-28,  16)$ & 0.23 & 23.16 & $>0.37$ &  0.16 & 0.40 & $\cdots$  & $\cdots$   & $-20.52$ & 8.8 \\
\hline
Q2233+1341-MD46A & $(  3,  21)$ & 0.20 & 22.84 & $1.28\pm 0.12$ & 0.17\tablenotemark{h} & 0.60\tablenotemark{h} & $\cdots$ & $\cdots$ & $-20.66$ & 9.9  \\
Q2233+1341-MD46B & $(  3,  21)$ & 0.20 & 22.03 & $>2.15$ & 0.21\tablenotemark{h} & 0.51\tablenotemark{h} & 1.0FIX  & $0.10^{+0.10}_{-0.26}$  & $-21.47$ & 9.7  \\
\hline
DSF2237b-MD81  & $( 14, -25)$ & 0.17 & 20.21 & $2.19\pm 0.04$ & 0.40 & 0.48  & $0.9^{+0.1}_{-0.0}$ & $0.58^{+0.04}_{-0.05}$ & $-23.36$ & 11.0 \\
DSF2237b-D28   & $( -6, -24)$ & 0.21 & 22.53 & $1.11\pm 0.09$ & 0.21 & 0.55 & $\cdots$   & $\cdots$   & $-21.11$ & 9.5 \\
DSF2237b-MD90A & $(-14,   5)$ & 0.22 & 22.74 & $0.56\pm 0.07$ & 0.22 & 0.54 & $\cdots$   & $\cdots$   & $-20.94$ & 9.0 \\
DSF2237b-MD90B & $(-14,   5)$ & 0.22 & $\cdots$ &    $\cdots$ & $\cdots$ & $\cdots$ & $\cdots$ & $\cdots$    & $\cdots$            & $\cdots$ \\
DSF2237b-D27   & $(-13, -27)$ & 0.19 & 21.92 & $1.21\pm 0.06$ & 0.30 & 0.47 & 1.0FIX     & $0.35^{+0.06}_{-0.06}$ & $-21.76$  & 9.8 \\
DSF2237b-MD80  & $( 22, -27)$ & 0.18 & 23.70 & $1.86\pm 0.22$ & 0.15 & 0.39 & $\cdots$   & $\cdots$   & $-19.98$ & 9.5 \\
\hline
DSF2237b-MD22  & $( 14,  17)$ & 0.20 & 22.01 & $1.40\pm 0.05$ & 0.28 & 0.46 & 1.0FIX & $0.74^{+0.04}_{-0.05}$ & $-22.01$ & 10.0 \\
DSF2237b-MD19  & $(-14, -12)$ & 0.19 & 20.36 & $2.58\pm 0.04$ & 0.39 & 0.49 & $1.8^{+0.3}_{-0.2}$ & $0.69^{+0.07}_{-0.09}$ & $-23.07$ & 11.2 \\
DSF2237b-C10   & $( 32, -11)$ & 0.23 & 21.91 & $2.23\pm 0.06$ & 0.21 & 0.39 & 1.0FIX & $0.27^{+0.06}_{-0.09}$ & $-22.06$ &  10.5 \\
\hline
DSF2237b-D5    & $( 17,   7)$ & 0.16 & 20.35 & $2.02\pm 0.03$ & 0.47 & 0.50 & $1.1^{+0.1}_{-0.1}$ & $0.73^{+0.04}_{-0.05}$ & $-23.08$ &  10.8 \\
DSF2237b-MD13A & $( -9, -24)$ & 0.17 & 23.14 & $1.35\pm 0.11$ & 0.20\tablenotemark{h} & 0.45\tablenotemark{h} & $\cdots$ & $\cdots$  & $-20.54$ & 9.4 \\
DSF2237b-MD13B & $( -9, -24)$ & 0.17 & 23.01 & $1.21\pm 0.10$ & 0.16\tablenotemark{h} & 0.58\tablenotemark{h} & $\cdots$ & $\cdots$  & $-20.67$ & 9.3 \\
DSF2237b-MD14  & $( 14,  -7)$ & 0.16 & $\cdots$ &    $\cdots$ & $\cdots$ & $\cdots$ & $\cdots$ & $\cdots$    & $\cdots$            & $\cdots$ \\
\hline
DSF2237a-C11   & $(-13, -17)$ & 0.18 & $\cdots$ &    $\cdots$ & $\cdots$ & $\cdots$ & $\cdots$ & $\cdots$    & $\cdots$            & $\cdots$ \\
\hline
DSF2237a-C15   & $(  6, -14)$ & 0.20\tablenotemark{m} & 22.29 & $\cdots$ & 0.28 & 0.44 & 1.0FIX   & $0.65^{+0.05}_{-0.06}$  & $-21.49$ & 8.9 \\
\hline
\enddata
\tablenotetext{a}{RA and DEC distances from the guide star in arcsec.}
\tablenotetext{b}{Esimated Moffat FWHM of the PSF at the position of the target. See detailes in the text.
"m" indates the FWHM value estimated from the mean relation shown in Figure 3.}
\tablenotetext{c}{AUTO magnitudes from SExtractor.}
\tablenotetext{d}{$1.\!^{\prime\prime}0$ diameter aperture $J-K$ colors.}
\tablenotetext{e}{The correction of $+0.13$ ($+0.25$) dex is applied to the best fit $R_{e}$ of an $n<2$ ($n>2$) galaxy, see details in Section 5.4.}
\tablenotetext{f}{$V$-band absolute magnitude. For objects without spectroscopic redshifts, $z=3$ is assumed.}
\tablenotetext{g}{Stellar mass of the galaxies. Kroupa IMF is assumed. The uncertainties are estimated to be 0.25 and 0.35 dex
for objects with and without spectroscopic redshifts, respectively. See details in Section 5.6.}
\tablenotetext{h}{The value measured with masking the neighboring object.}
\end{deluxetable}

\clearpage

\clearpage
\begin{landscape}
     \begin{deluxetable}{lcccccccccrrr}
\tabletypesize{\scriptsize}
\tablecaption{Results for DRGs and a radio galaxy in our sample \label{tab_DRG}}
\tablewidth{0pt}
\tablehead{
\multicolumn{1}{c}{Name} &
\multicolumn{1}{c}{RA} &
\multicolumn{1}{c}{DEC} &
\multicolumn{1}{c}{Distance\tablenotemark{a}} &
\multicolumn{1}{c}{FWHM\tablenotemark{b}} &
\multicolumn{1}{c}{$K_{\rm auto}$} &
\multicolumn{1}{c}{$J-K$} &
\multicolumn{1}{c}{$r_{\rm HL}$} &
\multicolumn{1}{c}{$C_{80/20}$} &
\multicolumn{1}{c}{$n$} & 
\multicolumn{1}{c}{$\log R_{e}$\tablenotemark{c}} &
\multicolumn{1}{c}{$M_{V}$} &
\multicolumn{1}{c}{$\log M_{*}$\tablenotemark{d}} \\
\multicolumn{1}{c}{} &
\multicolumn{1}{c}{(2000)} &
\multicolumn{1}{c}{(2000)} &
\multicolumn{1}{c}{($^{\prime\prime}$,$^{\prime\prime}$)} &
\multicolumn{1}{c}{($^{\prime\prime}$)} &
\multicolumn{1}{c}{(mag)} &
\multicolumn{1}{c}{(mag)} &
\multicolumn{1}{c}{($^{\prime\prime}$)} &
\multicolumn{1}{c}{} &
\multicolumn{1}{c}{} & 
\multicolumn{1}{c}{(kpc)} &
\multicolumn{1}{c}{(mag)} &
\multicolumn{1}{c}{($M_{\odot}$)} \\
}
\startdata
Q0302-003-1-AO023   & 03:04:30.50 & $-$00:07:58.6 & $( 11, -18)$ & 0.15 & 21.54 &       $>3.14$ & 0.22 & 0.46  & 1.0FIX              & $0.34^{+0.07}_{-0.15}$ & $-22.14$ & 11.1 \\
Q0302-003-1-AO062   & 03:04:28.30 & $-$00:07:34.1 & $(-20,   5)$ & 0.15 & 21.55 & $2.86\pm0.20$ & 0.32 & 0.66  & 1.0FIX              & $0.36^{+0.07}_{-0.23}$ & $-22.13$ & 11.0 \\
Q0302-003-1-AO071   & 03:04:28.26 & $-$00:07:36.7 & $(-21,   3)$ & 0.15 & 21.57 &       $>4.97$ & 0.42 & 0.58  & 1.0FIX              & $0.78^{+0.08}_{-0.65}$ & $-22.11$ & 11.0 \\
Q1422+2309-5-AO022  & 14:24:29.24 & $+$22:56:08.6 & $( -7, -32)$ & 0.22 & 21.25 & $2.39\pm0.09$ & 0.27 & 0.44  & $0.5^{+0.3}_{-0.1}$ & $0.36^{+0.04}_{-0.07}$ & $-22.43$ & 10.8 \\
DSF2237b-2-AO015    & 22:39:23.38 & $+$11:48:51.7 & $( 18,  11)$ & 0.20 & 20.95 & $3.20\pm0.07$ & 0.19 & 0.54  & $1.3^{+0.9}_{-1.2}$ & $-0.02^{+0.07}_{-0.07}$ & $-22.73$ & 11.3 \\
DSF2237b-3-AO015    & 22:39:24.40 & $+$11:48:03.3 & $(  8,  13)$ & 0.15 & 20.85 & $3.19\pm0.11$ & 0.32 & 0.47  & $0.4^{+0.1}_{-0.1}$ & $0.48^{+0.03}_{-0.03}$ & $-22.83$ & 11.3 \\
DSF2237b-3-AO022    & 22:39:21.90 & $+$11:47:47.1 & $(-27,  -2)$ & 0.18 & 19.98 & $3.35\pm0.08$ & 0.48 & 0.59  & $1.9^{+0.3}_{-0.2}$ & $0.92^{+0.04}_{-0.04}$ & $-23.70$ & 11.7 \\
\hline
4C28.58A            & 23:51:50.20 & $+$29:10:29.0 & $(-35,  -0)$ & 0.22\tablenotemark{m} & 19.63 & $\cdots$ &  0.48 & 0.52 & $0.3^{+0.2}_{-0.2}$ & $0.83^{+0.02}_{-0.01}$ & $-24.00$ & $\cdots$  \\
4C28.58B            & 23:51:50.20 & $+$29:10:29.0 & $(-35,  -0)$ & 0.22\tablenotemark{m} & 20.37 & $\cdots$ &  0.38 & 0.57 & $\cdots$            & $\cdots$  & $-23.26$ & $\cdots$ \\
\hline
\enddata
\tablenotetext{a}{RA and DEC distances from the guide star in arcsec.}
\tablenotetext{b}{Esimated Moffat FWHM of the PSF at the position of the target. See detailes in the text.
"m" indates the FWHM value estimated from the mean relation shown in Figure 3.}
\tablenotetext{c}{The correction of 0.13 dex is applied to the best fit $R_{e}$, see details in Section 5.4.}
\tablenotetext{d}{Stellar mass of the galaxies. Kroupa IMF is assumed. The uncertainties are estimated to be 0.35dex.}
\end{deluxetable}

\clearpage
\end{landscape}

\clearpage

\begin{figure}
\begin{center}
\begin{minipage}{0.5\linewidth}
\plotone{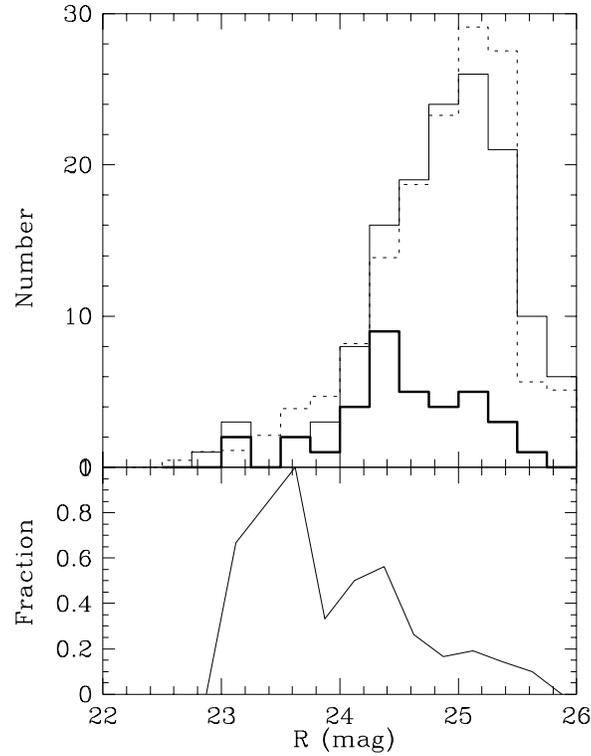}
\end{minipage}
\end{center}
\figcaption{
Upper)
$R$-band magnitude distribution of the "observed" LBGs (36 objects), 
"AO-observable" LBGs (141 objects at $10^{\prime\prime}<r<35^{\prime\prime}$ from $R<15.0$ mag star), 
and the "whole" $U$-drop LBG sample (2462 objects including objects with $R>25.5$ mag from Steidel et al. 2003)
shown with thick solid, thin solid, dotted lines,
respectively. The histogram of the "whole" 
sample is divided by 17 in order to match the sample
size of the "AO-observable" LBGs. 
Lower) 
Fraction of the "observed" LBGs among
the "AO-observable" LBGs shown as a function 
of $R-$band magnitude.
\label{sample_mag}}
\end{figure}

\begin{figure}
\begin{center}
\begin{minipage}{0.5\linewidth}
\plotone{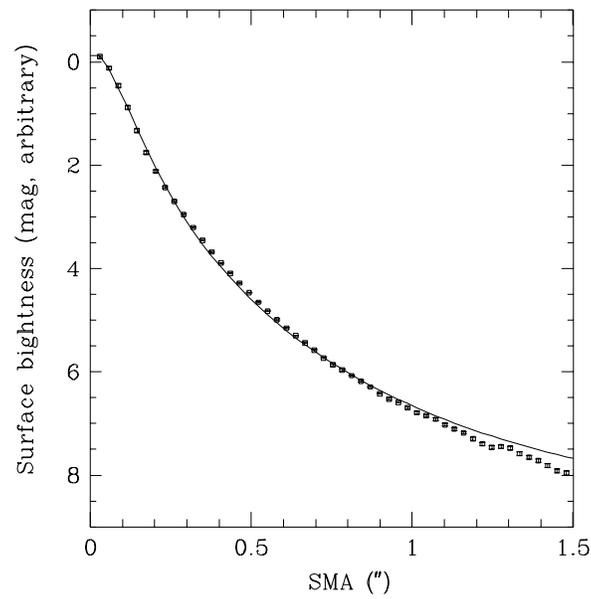}
\end{minipage}
\end{center}
\figcaption{
Radial profile of a stellar object in 
the Q1422+2309 FOV12 image shown with 
open square. Solid line represents the radial
profile of the best-fit Moffat model.
\label{Q1422FOV12K_PSF01_prof}}
\end{figure}

\begin{figure}
\begin{center}
\begin{minipage}{0.35\linewidth}
\plotone{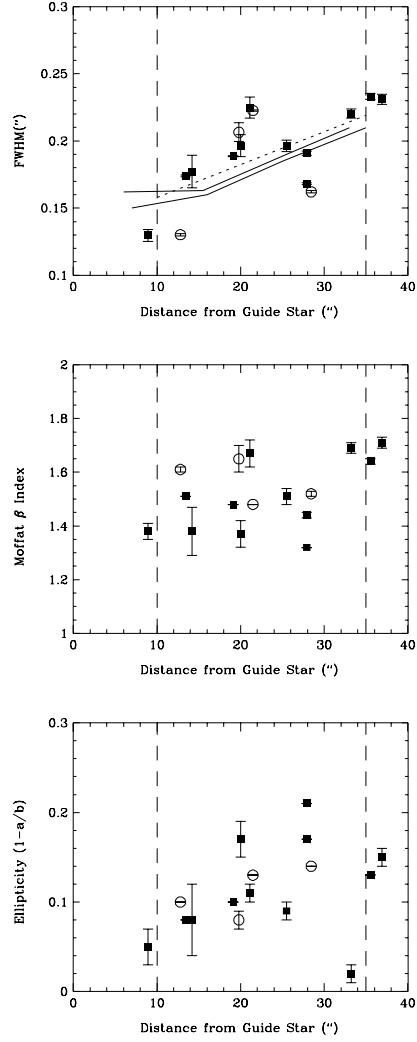}
\end{minipage}
\end{center}
\figcaption{
Measured shape parameters of the PSF reference stars in the final
images as a function of distance from the guide star.
Top) FWHM vs. distance, middle) Moffat $\beta$ parameter
vs. distance, and bottom) ellipticity vs. distance.
The PSF reference stars in the FoV with AO-guide star brighter
(fainter) than $13$ mag are plotted with open circles (filled square).
The solid lines indicate the
relation between the FWHM and the distance from Figure 18 of
Minowa et al. (2005). The relations are obtained from
M13 observations with $R=12.7$ mag guide star.
The dotted line shows the mean relation between 
FWHM and distance described in Section 4.2. The relation is used to estimate
the FWHM for FoVs without PSF reference stars.
The vertical dashed lines indicate the range of the
distances of the LBGs.
\label{PSF_param1}}
\end{figure}

\begin{figure}
\begin{minipage}{0.5\linewidth}
\plotone{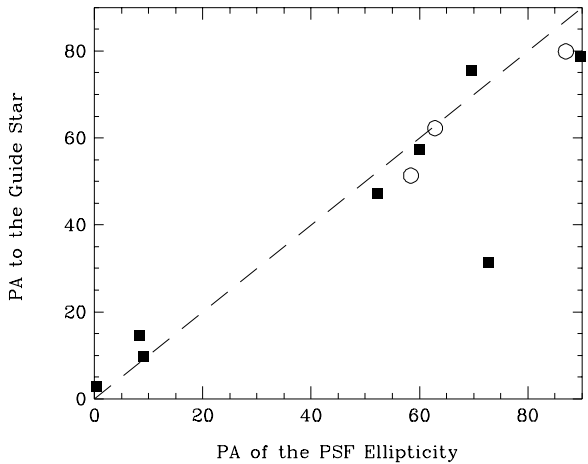}
\end{minipage}
\figcaption{
PAs of the ellipticity of the PSF reference stars vs.
PAs to the guide star. Symbols are the same as those in Figure 3. 
Only the PSF reference stars with
$e>0.08$ are plotted.
The dashed line shows the relation that the PA of the ellipticity
is equal to the PA to the guide star.
\label{PSF_param2}}
\end{figure}


\begin{figure}
\begin{center}
\begin{minipage}{0.9\linewidth}
\plotone{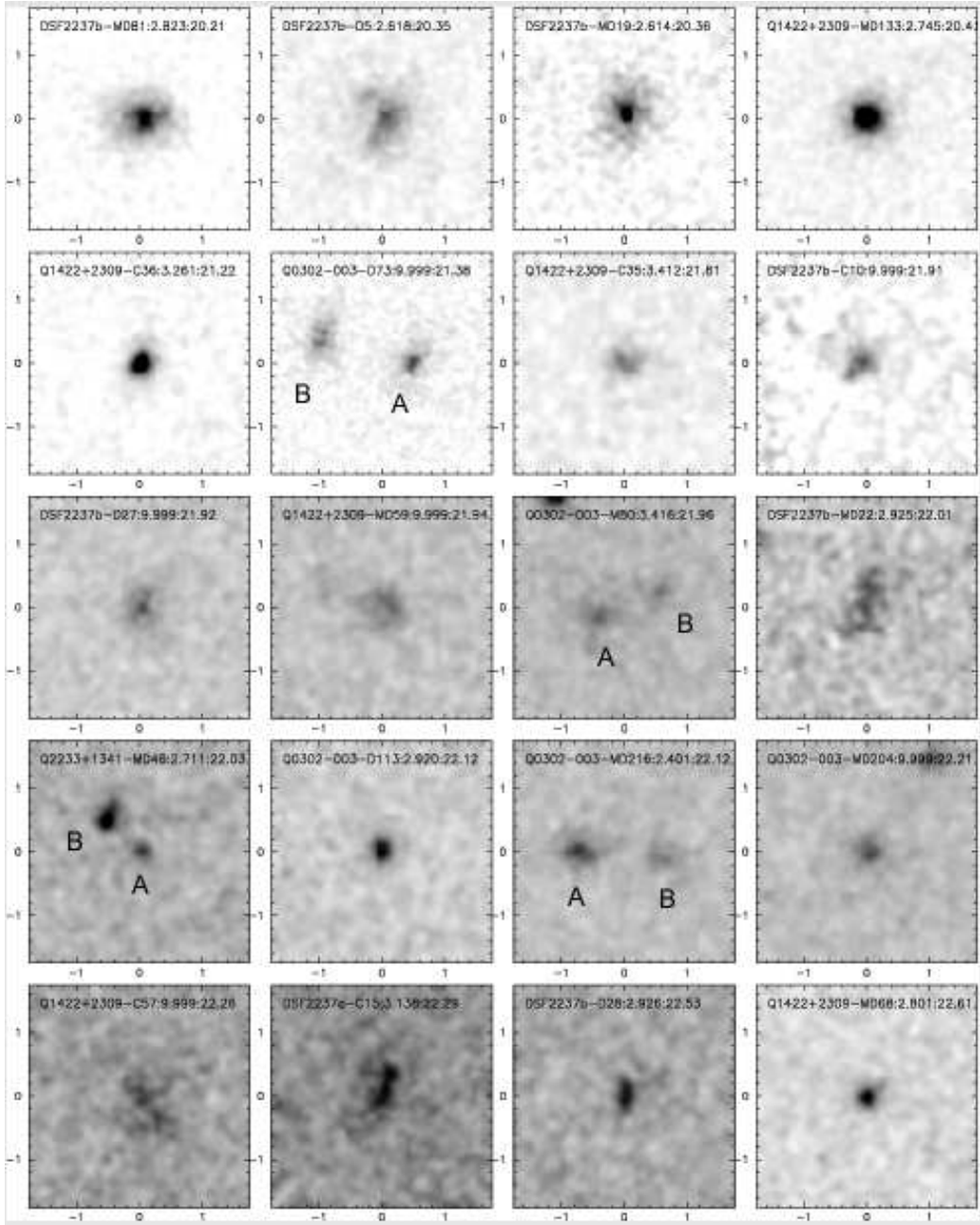}
\end{minipage}
\end{center}
\figcaption{
$K$-band images of the $z\sim3$ LBGs. 
A $3.\!^{\prime\prime}5\times3.\!^{\prime\prime}5$ FoV
is shown in the order of $K$-band magnitude. North is to the top and
east is to the left. FoVs are centered at 
the cataloged positions of the $z\sim3$ LBGs.
Gaussian convolution with $\sigma$ of 1 (3) pixel is applied for
the images of the detected (non-detection) LBGs. 
Name, spectroscopic redshift, and $K$-band magnitude are
shown at the top of each panel. Redshift of 9.999 refers to no
spectroscopic redshift available. For LBGs with multiple knots,
each knot is labeled with "A", "B", and "C".
\label{Kimage_g15_1}}
\end{figure}

\begin{figure}
\begin{center}
\begin{minipage}{0.9\linewidth}
\plotone{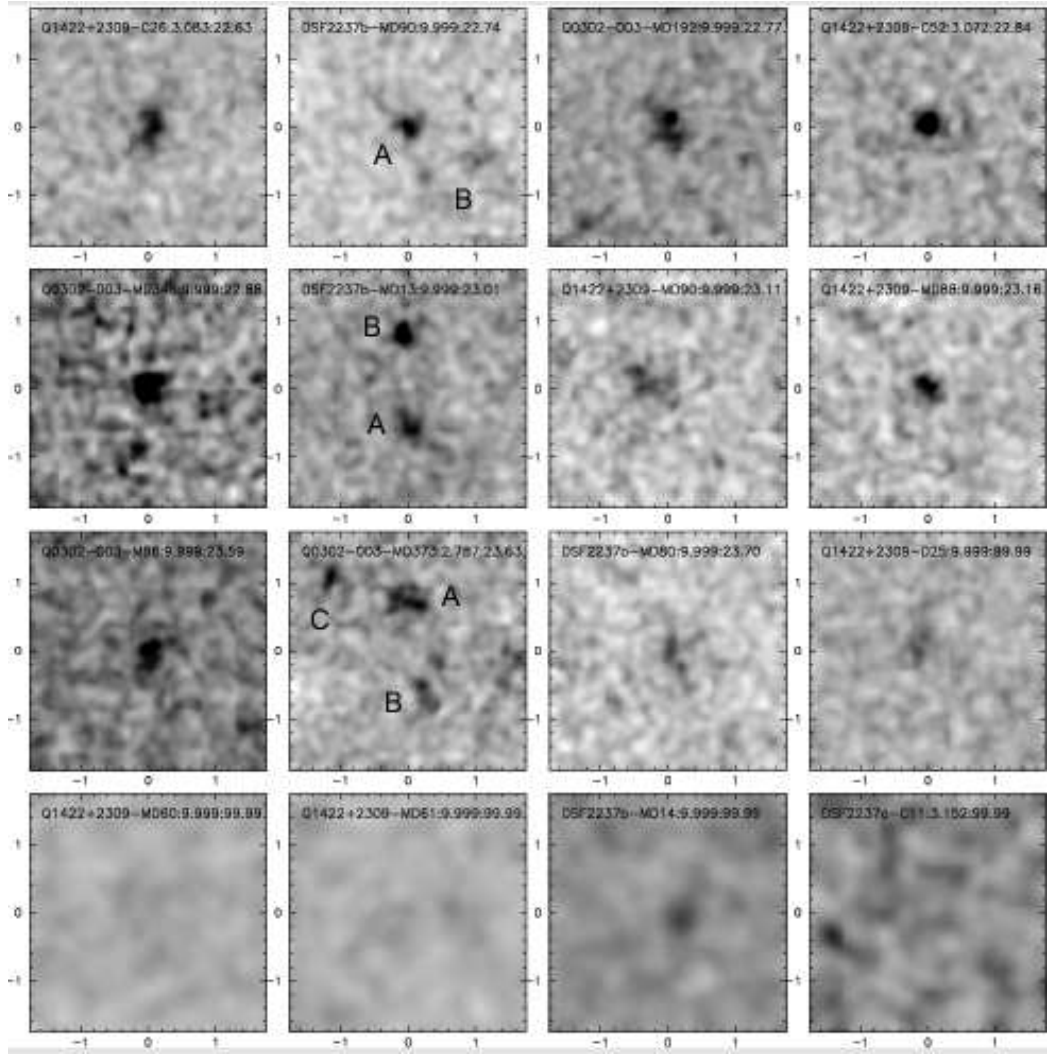}
\end{minipage}
\end{center}
\figcaption{
Continued. The last five objects are not detected in the observation, 
and "99.99" are used for their $K$-band magnitudes.
\label{Kimage_g15_2}}
\end{figure}

\begin{figure}
\begin{minipage}{0.5\linewidth}
\plotone{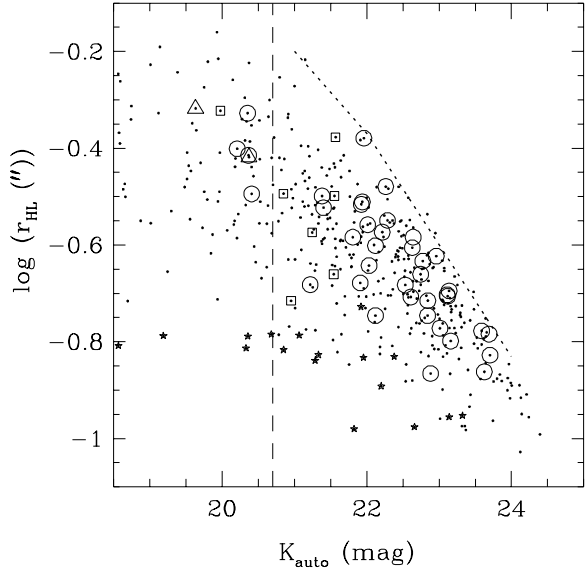}
\end{minipage}
\figcaption{
$K$-band magnitude vs. apparent $r_{\rm HL}$
of the $z\sim3$ LBGs (open circles). 
Objects detected at $10^{\prime\prime} < d < 35^{\prime\prime}$
from the AO guide stars are plotted with small dots.
Star marks indicate stellar objects in the field.
Detection limit for extended objects with 5h integration
is shown with a dotted line, for details see Section 5.3.
The long-dashed line shows
$m_{K}^{*}$ of the $z\sim3$ LBGs (Shapley et al. 2001).
Objects which meet DRG criterion and two knots of 4C28.58 are marked with open squares and open
triangles, respectively, see Section 6.1 for details.
\label{Kauto_HL}}
\end{figure}

\begin{figure}
\begin{minipage}{0.5\linewidth}
\plotone{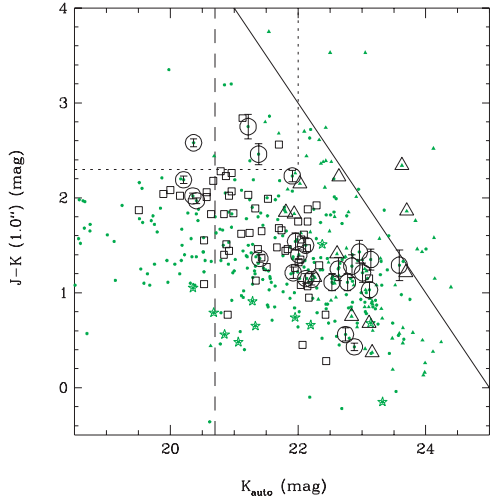}
\end{minipage}
\figcaption{
$K$-band magnitude vs. $J-K$ color 
of the $z\sim3$ LBGs (open circles, triangles for lower limits).
The long dashed line indicates $m_{K}^{*}$ of the $z\sim3$ LBGs.
Typical detection limit in the $J$-band is shown with the solid line.
Small open squares show the color-magnitude distribution of the
$z\sim3$ LBGs from Shapley et al. (2001). Their observation limit
is $K\sim22.5$ mag.
$K$-band detected objects $10^{\prime\prime} < d < 35^{\prime\prime}$
from the AO guide stars are plotted with green small dots and green small filled triangles
for lower limits. Stellar objects are marked with green star marks. 
The dotted lines indicate the selection criteria of DRGs ($J-K>2.3$ mag and $K<22$ mag).
\label{Fig_JKcolmag}}
\end{figure}

\begin{figure}
\begin{center}
\begin{minipage}{0.9\linewidth}
\plotone{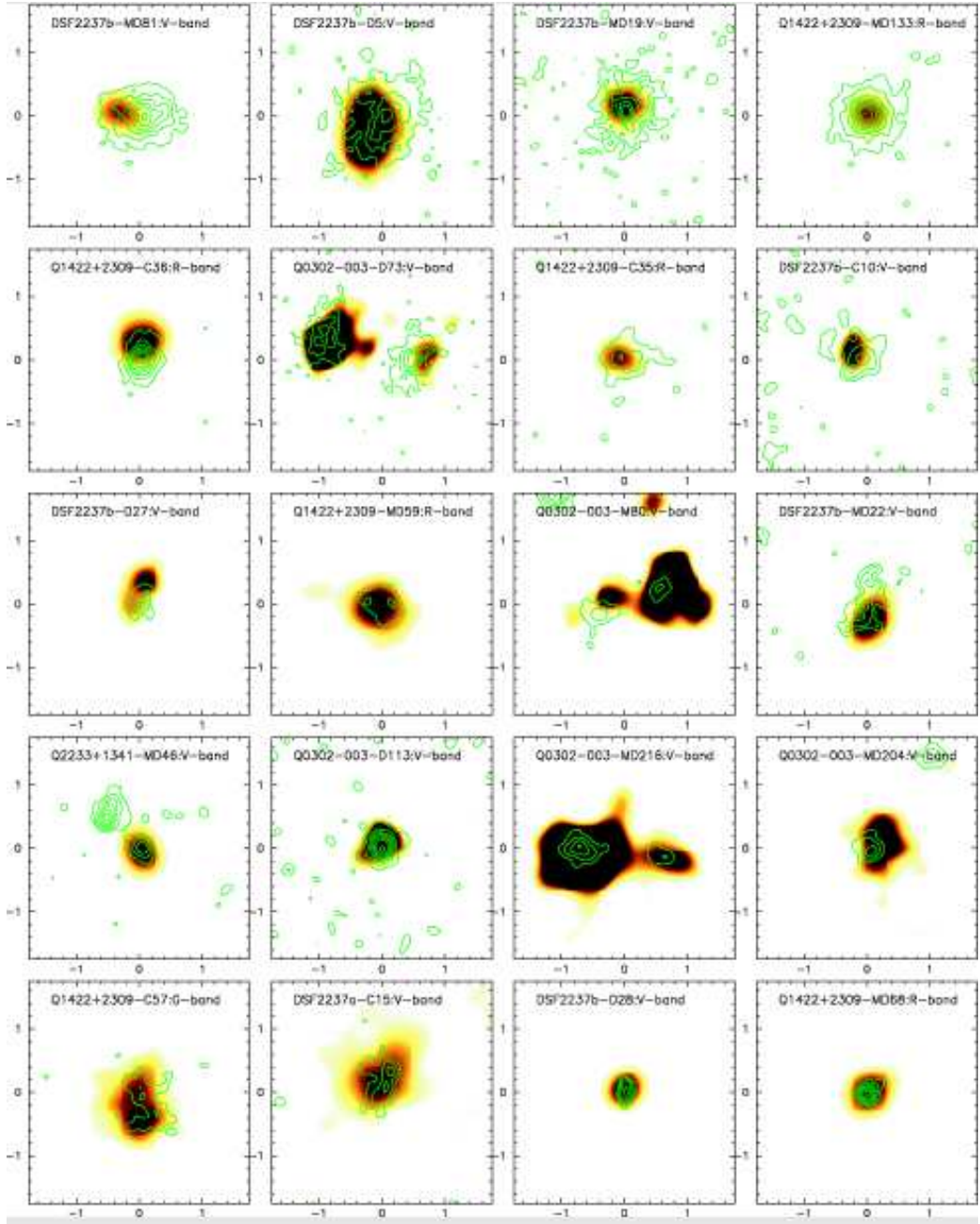}
\end{minipage}
\end{center}
\figcaption{
Optical images ($V$-, $G$-, or $R$-band) of the $z\sim3$ LBGs with
$K$-band contours superposed.
A FoV of $3.\!^{\prime\prime}5\times3.\!^{\prime\prime}5$ is shown.
North is to the top and east is to the left, same as in Figure 5.
The positions are roughly centered at 
the cataloged positions of the $z\sim3$ LBGs.
\label{VKimage_g15_1}}
\end{figure}

\begin{figure}
\begin{center}
\begin{minipage}{0.9\linewidth}
\plotone{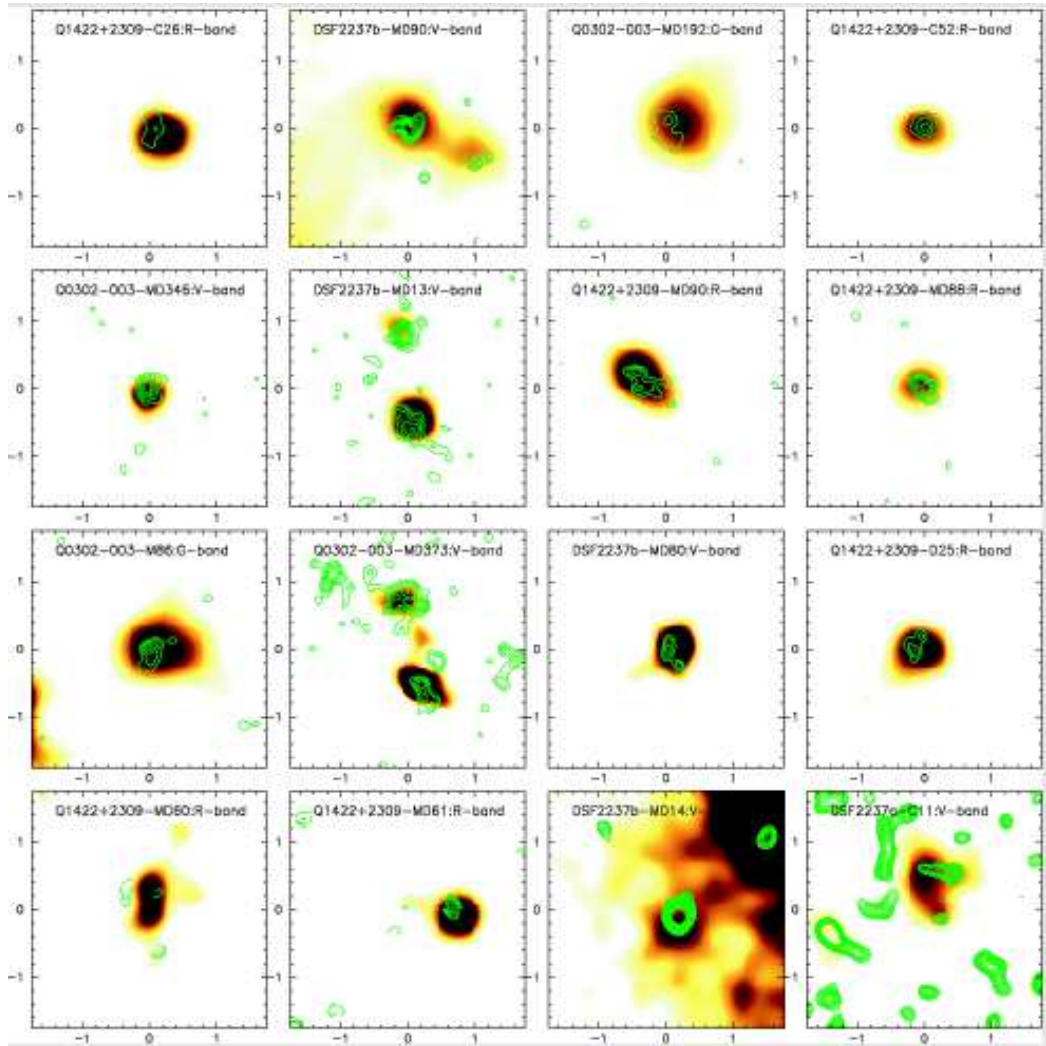}
\end{minipage}
\end{center}
\figcaption{
Continued. The last five objects are not detected in the $K$-band images.
\label{VKimage_g15_2}}
\end{figure}
\clearpage

\begin{figure}
\begin{center}
\begin{minipage}{0.5\linewidth}
\plotone{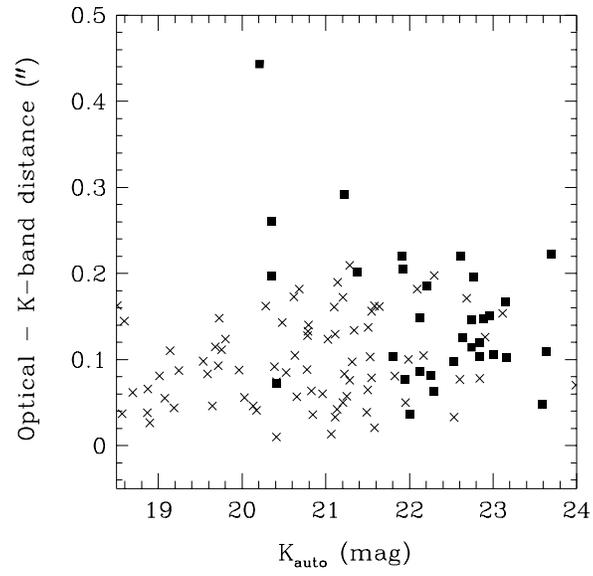}
\end{minipage}
\end{center}
\figcaption{
Distance between $K$-band and optical peaks
as a function of $K$ magnitudes for the
$z\sim3$ LBGs (filled squares) and field objects (crosses).
\label{Opt_K_diff}}
\end{figure}
\clearpage

\begin{figure}
\begin{minipage}{0.6\linewidth}
\plotone{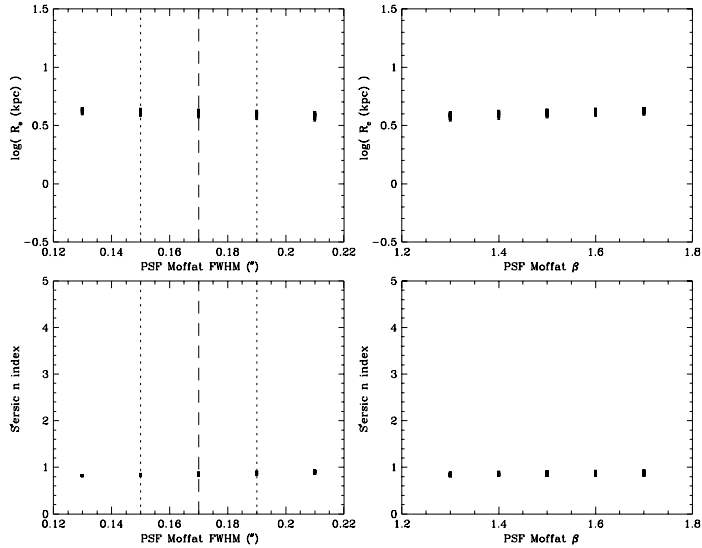}
\end{minipage}
\figcaption{
Dependence of the best fit S\'ersic parameters on the model
PSF used in the S\'ersic fit for DSF2237b-MD81.
Panels in the left hand side show the dependence of the best
fit $R_{e}$ (upper) and S\'ersic $n$ index (lower) on the 
Moffat FWHM of the model PSF. Points with error bars at one
certain Moffat FWHM show the results with changing the Moffat
$\beta$ parameter and the ellipticity of the model PSF.
The vertical dashed and dotted lines show the estimated
FWHM size at the target position and the uncertainty as explained in Section 4.2.
Panels in the right hand side show the effect of changing the
Moffat $\beta$ parameter of the model PSF. See details
Section 5.3.
\label{DSF2237bFOV1K_MD81_check}}
\end{figure}

\begin{figure}
\begin{minipage}{0.6\linewidth}
\plotone{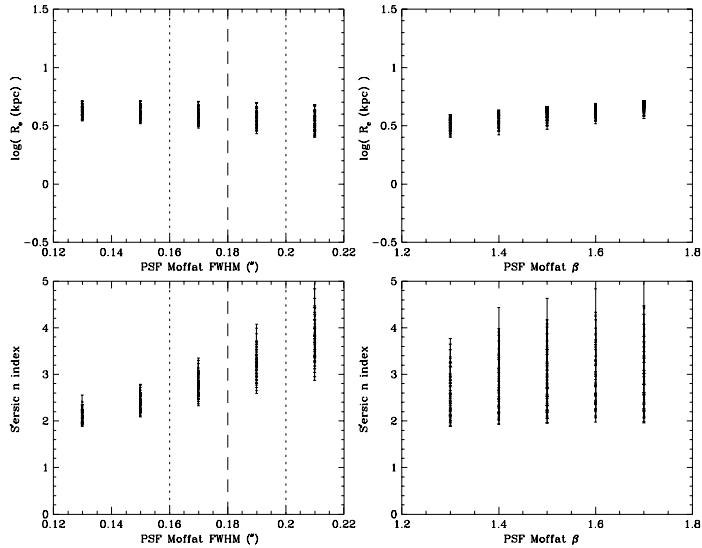}
\end{minipage}
\figcaption{
Same as Figure 13, but for Q0302$-$003-D73A.
\label{Q0302FOV3K_D73_check}}
\end{figure}

\begin{figure}
\begin{minipage}{0.9\linewidth}
\plotone{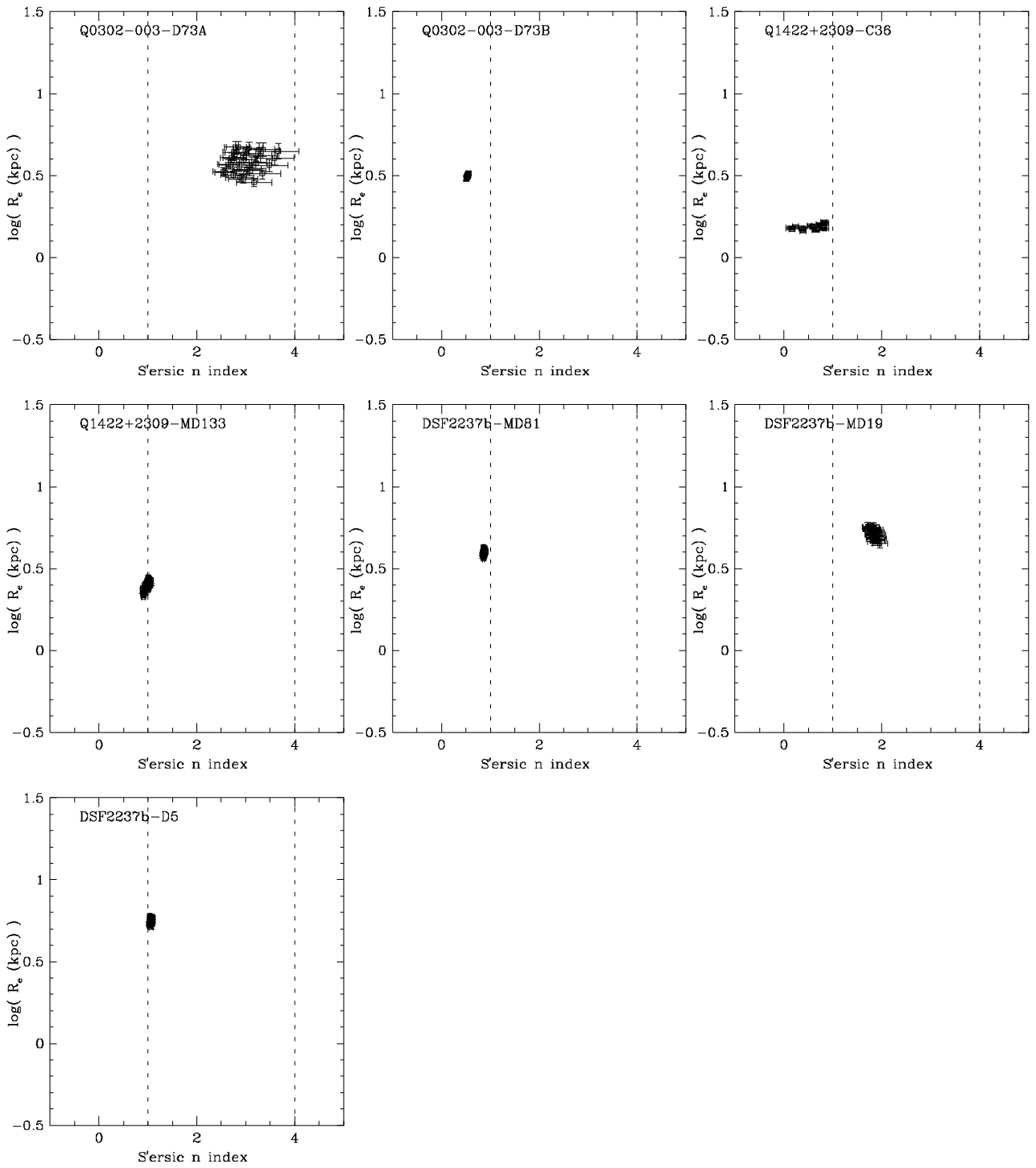}
\end{minipage}
\figcaption{
Distribution of the
best fit S\'ersic $n$ index and $R_{e}$ for
the LBGs with $K<21.5$ mag. 
Fitting for each object is done with changing
the shape of the PSF.
Each point with errorbars
represents a fitting result with one PSF model.
The ranges of the shape parameters of the PSFs
covered in the fittings are described in Section 5.3.
Dashed lines indicate $n=1$ (exponential-law) 
and $n=4$ ($r^{1/4}$-law).
\label{LBG_FIT_ser}}
\end{figure}

\begin{figure}
\begin{minipage}{0.9\linewidth}
\plotone{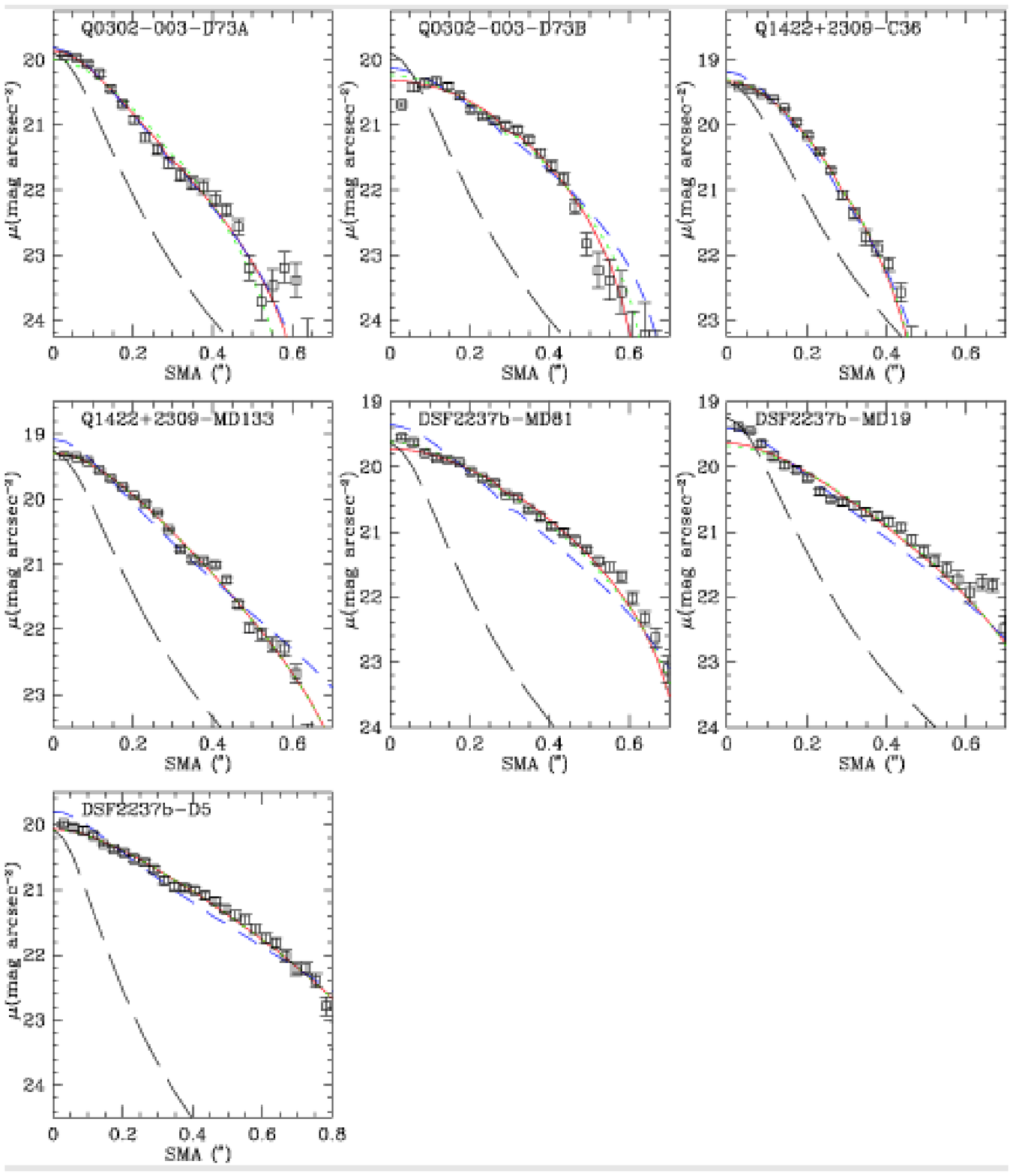}
\end{minipage}
\figcaption{
$K$-band profiles of the LBGs with $K<21.5$ mag along
the semi-major axis. The red-solid, blue-dashed, and
green-dotted lines show the best fit S\'ersic (with free n),
$r^{1/4}$ (S\'ersic with $n=4$), and exponential (S\'ersic with $n=1$)
profiles, respectively. Long dashed lines show the 
estimated profiles of the PSFs at the object positions.
\label{LBG_FIT_prof}}
\end{figure}

\begin{figure}
\begin{minipage}{0.7\linewidth}
\plotone{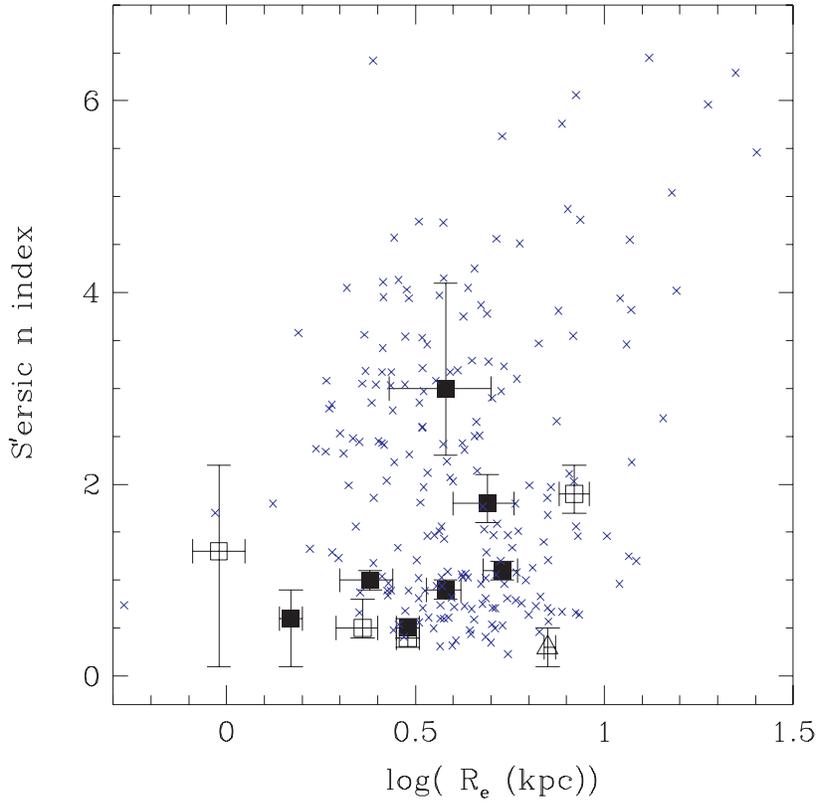}
\end{minipage}
\figcaption{
$R_e$ vs. $n$ for the $z\sim3$ LBGs with $K<21.5$ mag (filled squares).
DRGs and a radio galaxy 4C28.58 are shown with open squares and
an open triangle, respectively.
Simulated $z=3$ galaxies that are brighter than $K=21.5$
mag in the 2 mag PLE model are plotted with blue small crosses.
\label{LBG_DRG_Re_N}}
\end{figure}

\begin{figure}
\begin{minipage}{0.9\linewidth}
\plotone{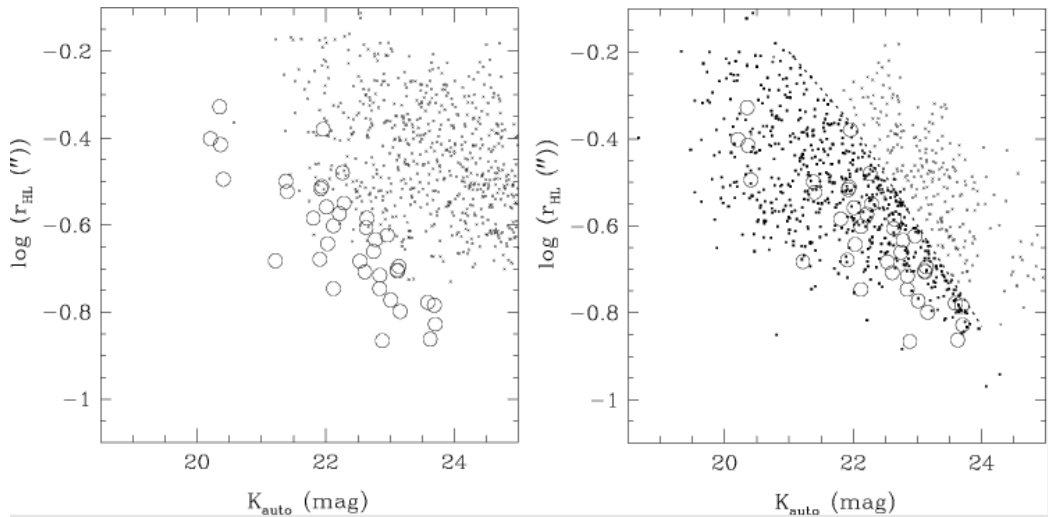}
\end{minipage}
\figcaption{
Left) $K$-band magnitude vs. $r_{\rm HL}$ of the simulated
galaxies without noise and evolution, i.e. "model" images.
Measured $K$-band magnitudes and $r_{\rm HL}$ of the
"model" images are shown with crosses.
The observed $K$-band magnitudes and $r_{\rm HL}$s of
the $z\sim3$ LBGs are shown with open circles.
Most of the model galaxies are fainter than the
observed $z\sim3$ LBGs at the same size without any evolution.
Right) Same figure for "simulated" images with noise
and PLE of 2 mag. Measured $K$-band magnitudes
are $r_{\rm HL}$ of the "simulated" images are shown
with filled squares. The model galaxies which are not
detected in the simulation are plotted with crosses.
The detection limit of the observation is determined
with the envelop of the distribution of the detected
objects as shown with thick dotted line.
\label{Kauto_HL_all}}
\end{figure}

\begin{figure}
\begin{minipage}{0.95\linewidth}
\plottwo{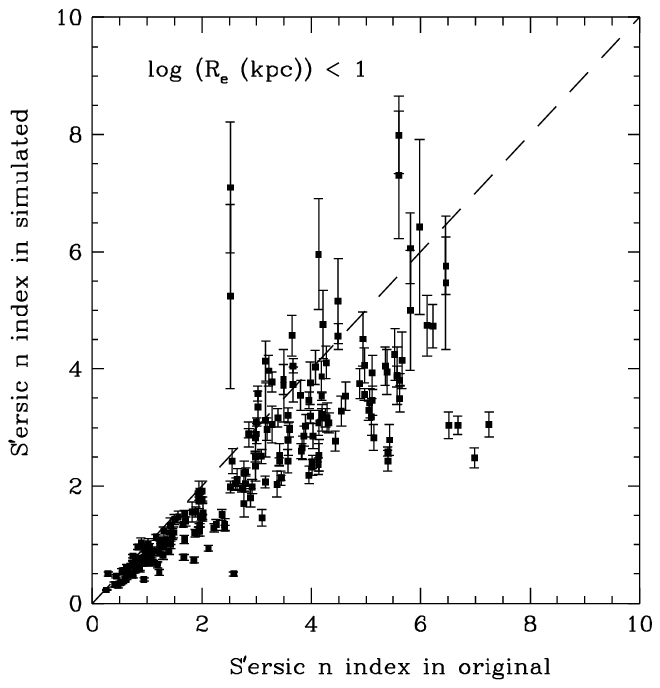}{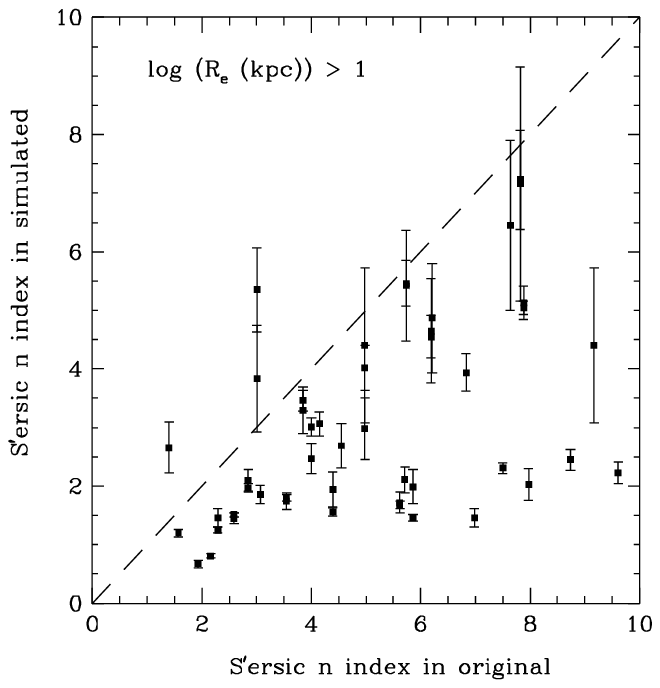}
\end{minipage}
\figcaption{
S\'ersic $n$ index derived with the original images
of the $z=0.36-0.69$ galaxies vs. that estimated
with the $z=3$ simulated images. 
Galaxies with $\log R_{e} ({\rm kpc}) <1$ ($\log R_{e} ({\rm kpc}) >1$) in the original images
are shown in the
left (right) panel.
The long dashed line indicate the relation that
$n$ in the original images equal to those in the simulated images.
\label{org_sim_comp_n}}
\end{figure}

\begin{figure}
\begin{minipage}{0.95\linewidth}
\plottwo{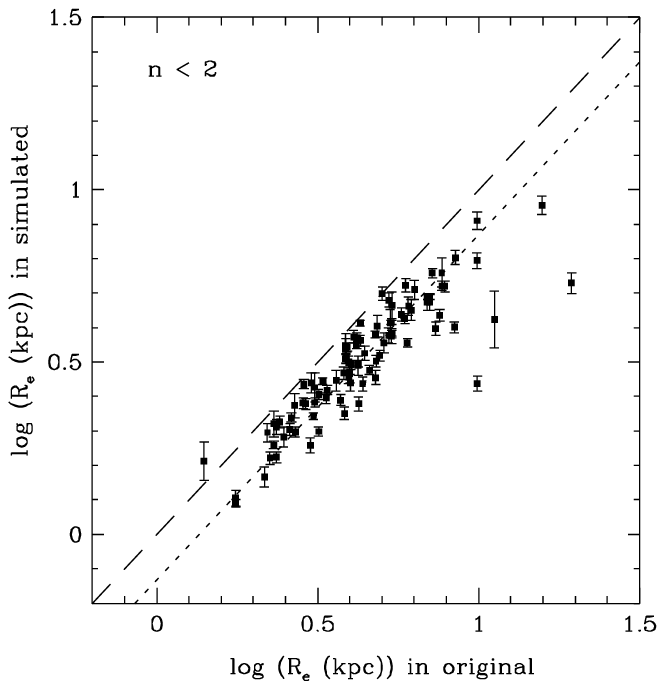}{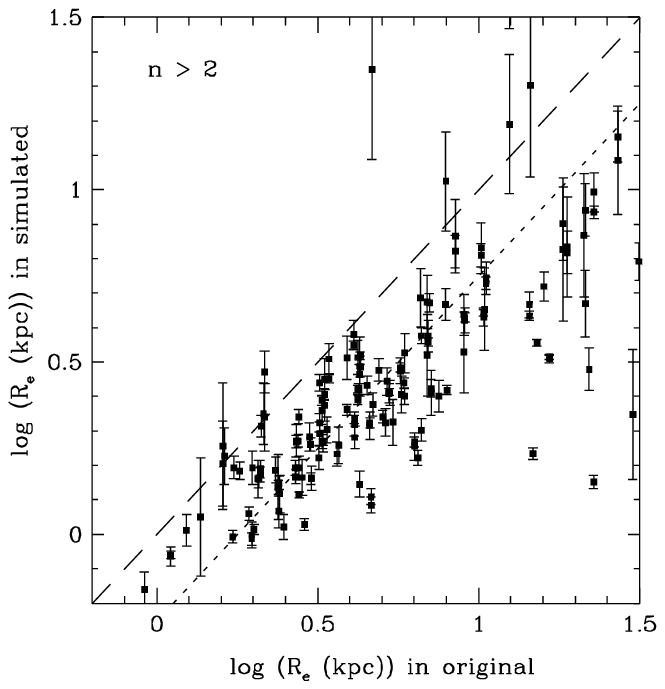}
\end{minipage}
\figcaption{
$R_{e}$ derived with the original images
of the $z=0.36-0.69$ galaxies vs. that estimated
with the $z=3$ simulated images.
Galaxies with $n<2$ ($n>2$) in the original images are shown in the left (right) panel.
The long dashed line indicate the relation that 
$R_{e}$ in the original images equal to those in the simulated images.
The dotted lines indicate the systematic offset of 0.13 (0.25) dex for
the $\log R_{e} ({\rm kpc}) $ of $n<2$ ($n>2$) galaxies.
\label{org_sim_comp_Re}}
\end{figure}

\clearpage

\begin{figure}
\begin{minipage}{0.48\linewidth}
\plotone{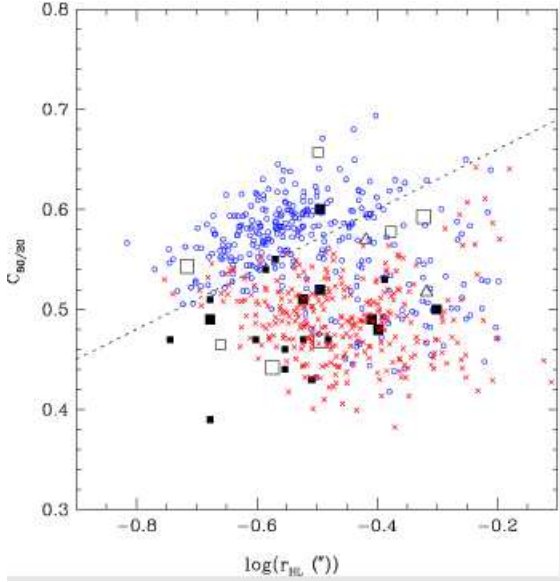}
\end{minipage}
\figcaption{
$r_{\rm HL}$ vs. $C_{80/20}$ of the $z\sim3$ LBGs (filled
squares), DRGs (open squares), and 4C28.58 (open triangle).
Large (small) symbols indicate objects brighter (fainter)
than $K=21.5$ mag.
The simulated $z\sim3$ galaxies are shown
with small symbols. Red crosses (blue open circles) for galaxies
with $n<2$ ($n>2$) in the original images.
The dotted line is drawn arbitrarily to give an eye-guide for the division
of the regions occupied by the $n<2$ and $n>2$ objects.
\label{LBG_DRG_C_HL}}
\end{figure}

\begin{figure}
\begin{minipage}{0.9\linewidth}
\plottwo{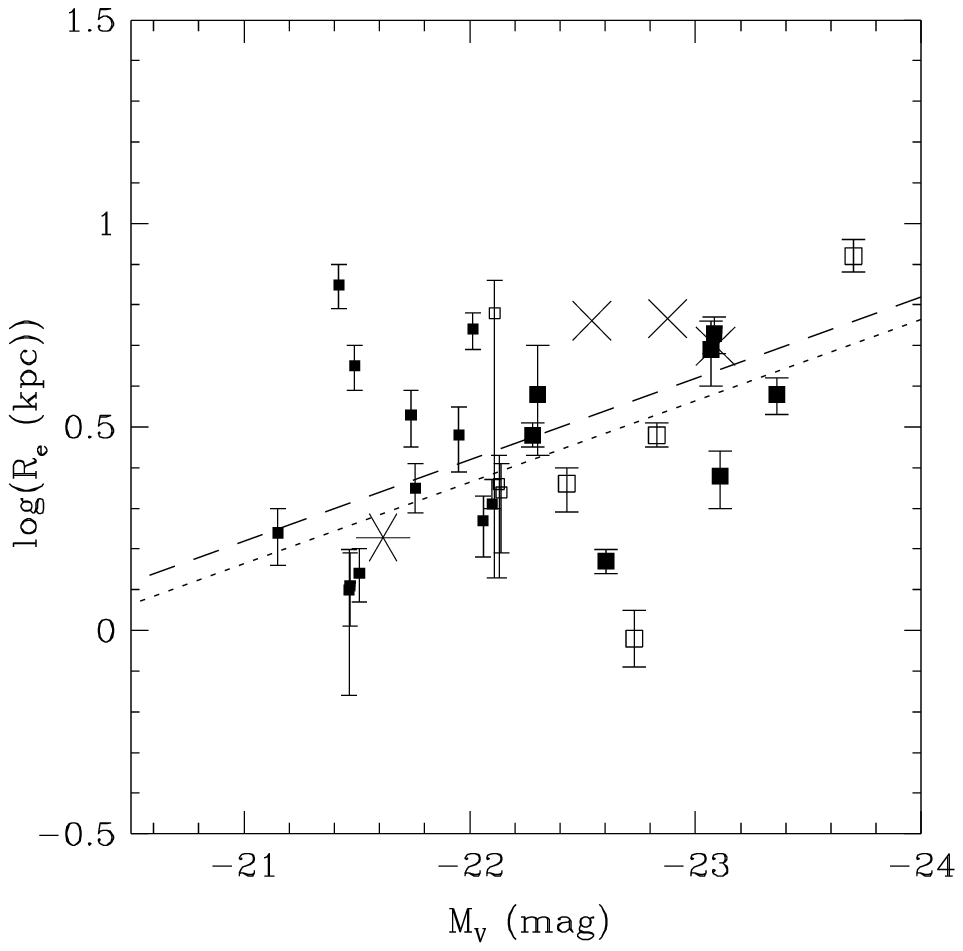}{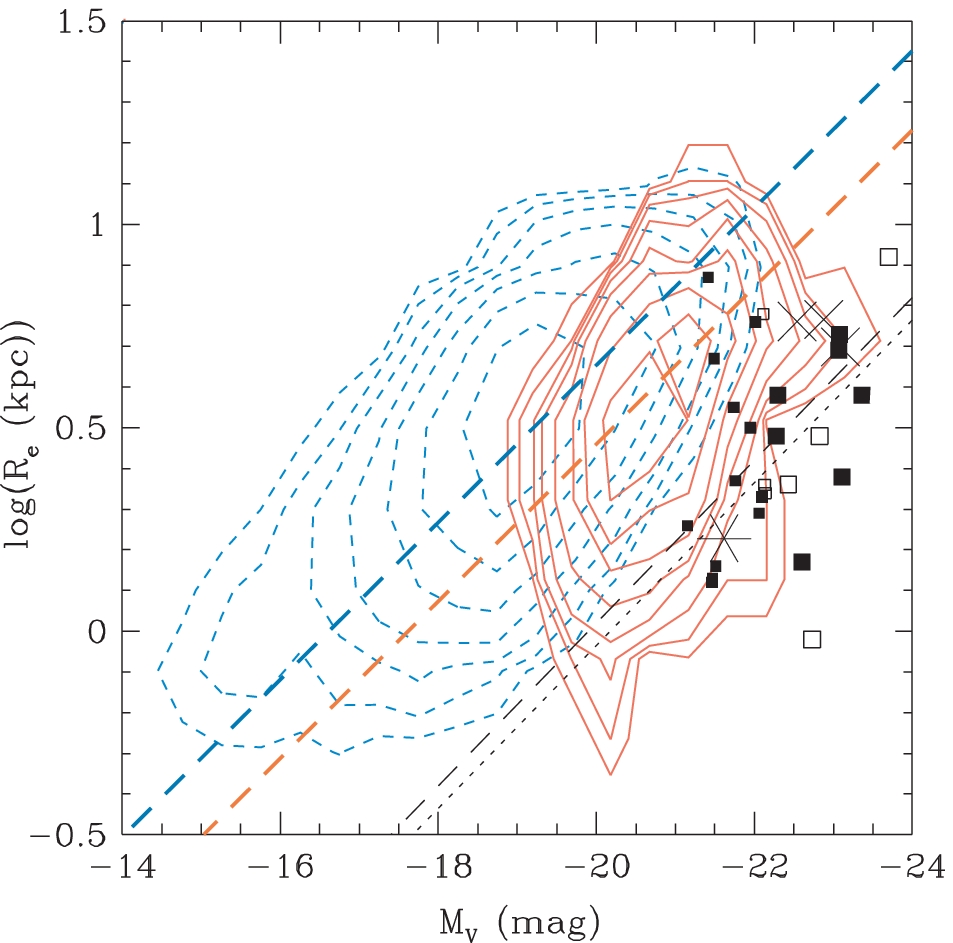}
\end{minipage}
\figcaption{
Left) $M_{V}$ vs. $R_{e}$ for the LBG sample (filled squares).
Large and small symbols indicate $R_{e}$ obtained with the 
S\'ersic profile fitting with free $n$ and fixed $n=1$, respectively.
Dotted and dashed lines indicate the average $<\mu_{V}>$ determined
from the 4 $>M_{V}^{*}$ LBGs (18.38 mag arcsec$^{-2}$)
and from all of the LBGs (18.67 mag arcsec$^{-2}$), respectively.
Open squares show the DRG sample. Small and large symbols are
same for LBGs.
For comparison, we plot LDGs 
at $2.5<z<3.0$ from Labb\'e et al. (2003) with large crosses and
a disk galaxy of old stars at $z\sim2.5$ from Stockton et al. (2004)
with a large asterisk.
Right) Same figure overlaid with the distributions of $z=0$ (dashed blue contour)
and $z=1$ (solid red contour) disk-galaxies from Barden et al. (2005).
The thick blue and red dashed lines show $<\mu_{V}>$ of disk galaxies
at $z=0$ (20.84 mag arcsec$^{-2}$) and $z=1$ (19.84 mag arcsec$^{-2}$), respectively.
\label{LBG_DRG_Mv_Re}}
\end{figure}

\begin{figure}
\begin{minipage}{0.9\linewidth}
\plottwo{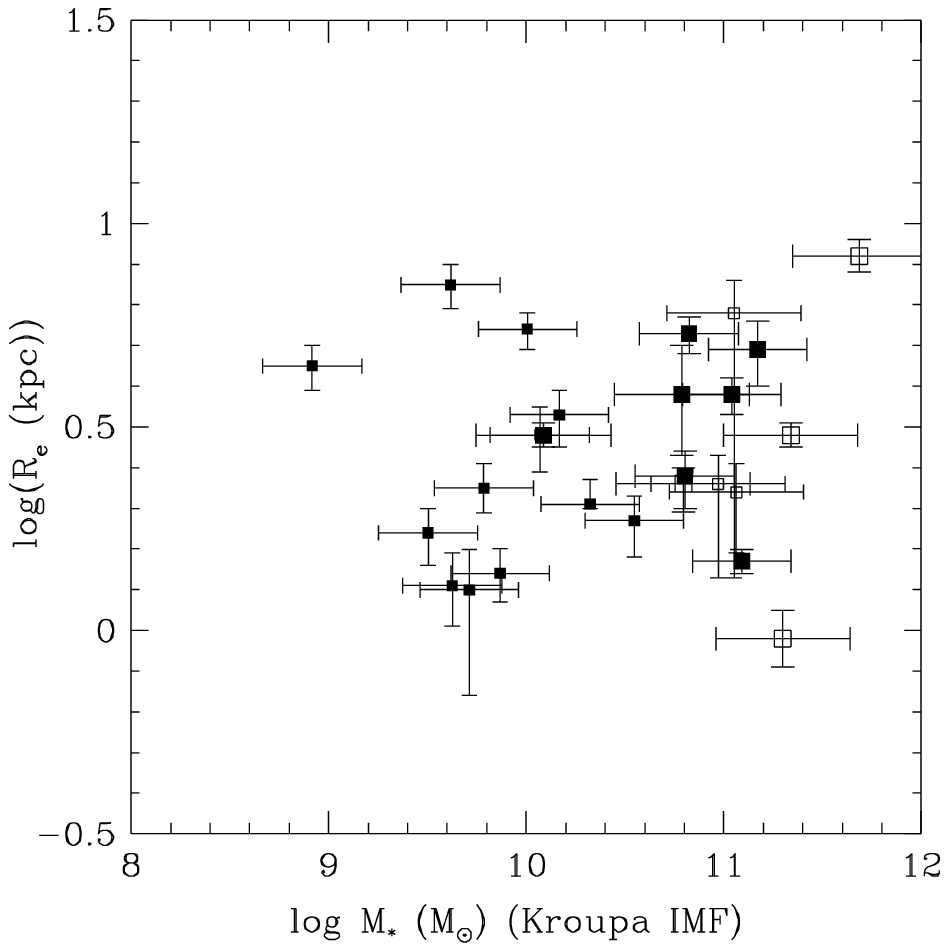}{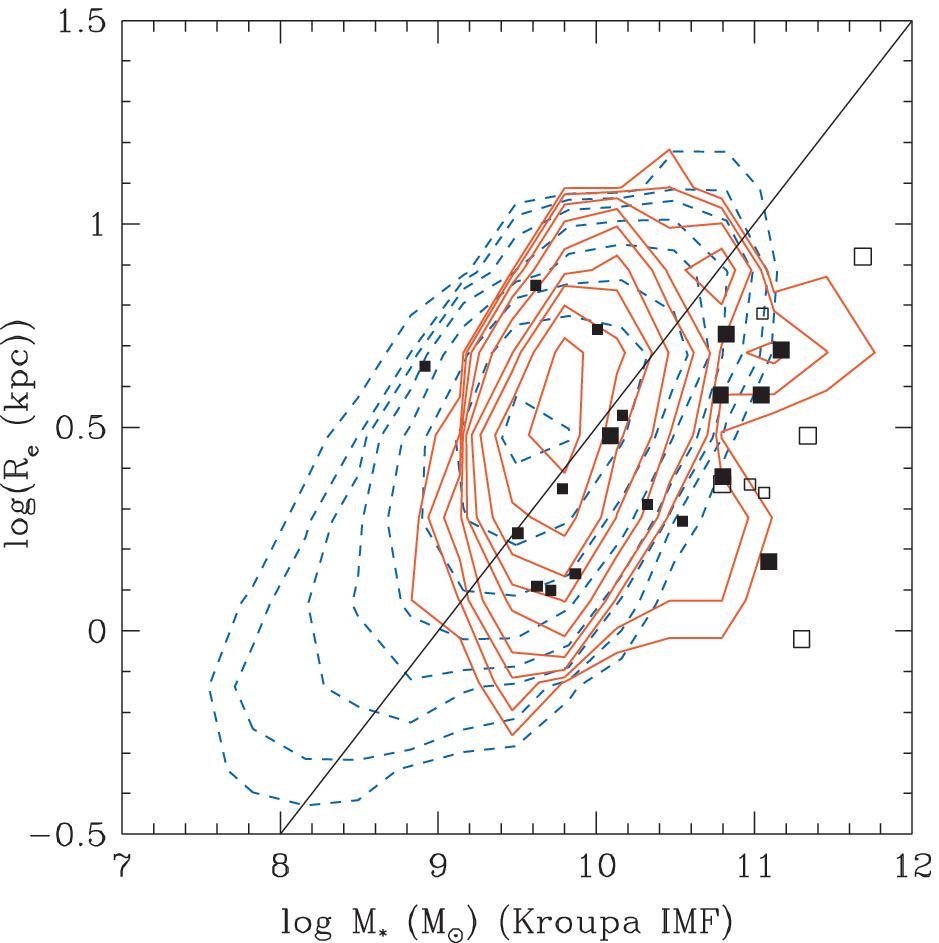}
\end{minipage}
\figcaption{
Left) $M_{*}$ vs. $R_{e}$ for the LBG sample (filled squares).
Large and small symbols indicate $R_{e}$ obtained with the 
S\'ersic profile fitting with free $n$ and fixed $n=1$, respectively.
Open squares show the DRG sample. Small and large symbols are same for LBGs.
Right) Same figure overlaid with the distributions of $z=0$ (dashed blue contour)
and $z=1$ (solid red contour) disk-galaxies from Barden et al. (2005).
The solid line represents the relation: $\log \Sigma_{M} (M_{\odot} {\rm kpc}^{-2}) = 8.50$ with $q=0.5$
derived from the disk-galaxies at $z=0-1$ (Barden et al. 2005).
\label{LBG_DRG_Ms_Re}}
\end{figure}

\begin{figure}
\begin{minipage}{0.9\linewidth}
\plotone{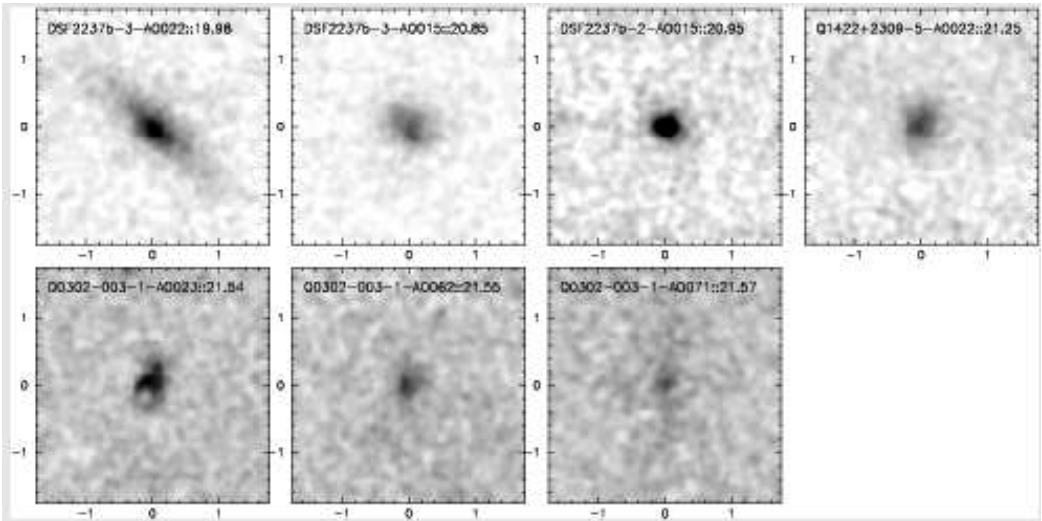}
\end{minipage}
\figcaption{
$K$-band images of the DRGs in the AO observed field.
A $3.\!^{\prime\prime}5\times3.\!^{\prime\prime}5$
FoV is shown in the order of the $K$-band magnitude. North is to the top
and east is to the left.
Gaussian convolution with $\sigma$ of 1 pixel is applied for
the images of the DRGs. 
\label{Kimage_DRG_g15}}
\end{figure}

\begin{figure}
\begin{minipage}{0.9\linewidth}
\plotone{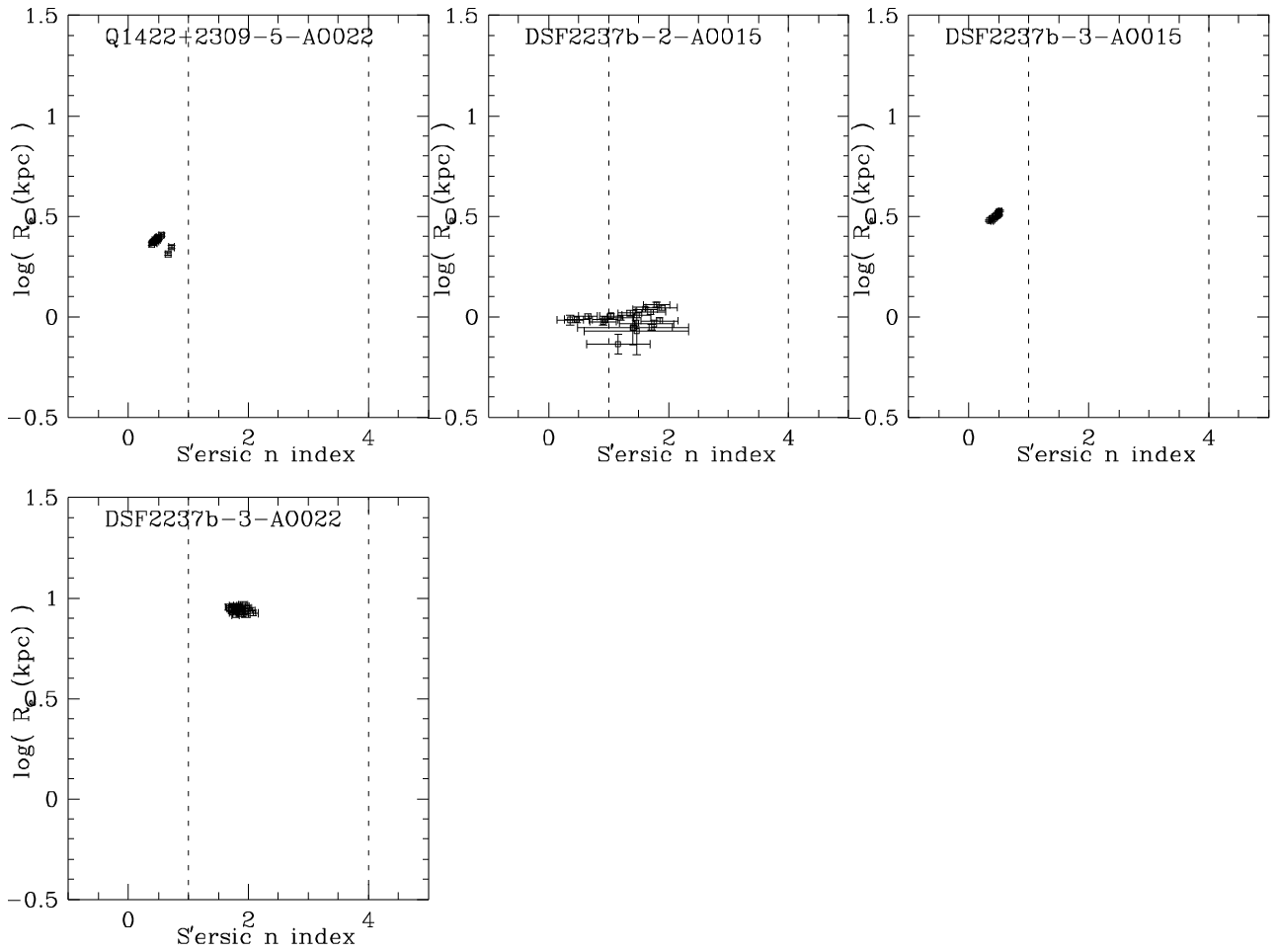}
\end{minipage}
\figcaption{
Distribution of the best fit S\'ersic $n$ index and $R_{e}$ for
DRGs. Fitting for each object is done with changing the shape of
the PSF. Each point with errorbars represents a
fitting result with one PSF model. The ranges of the shape
parameters of the PSFs covered in the fittings are same as
described in Section 5.3.
Dashed lines indicate $n=1$ (exponential-law)
and $n=4$ ($r^{1/4}$-law).
\label{DRG_FIT_ser}}
\end{figure}

\begin{figure}
\begin{minipage}{0.9\linewidth}
\plotone{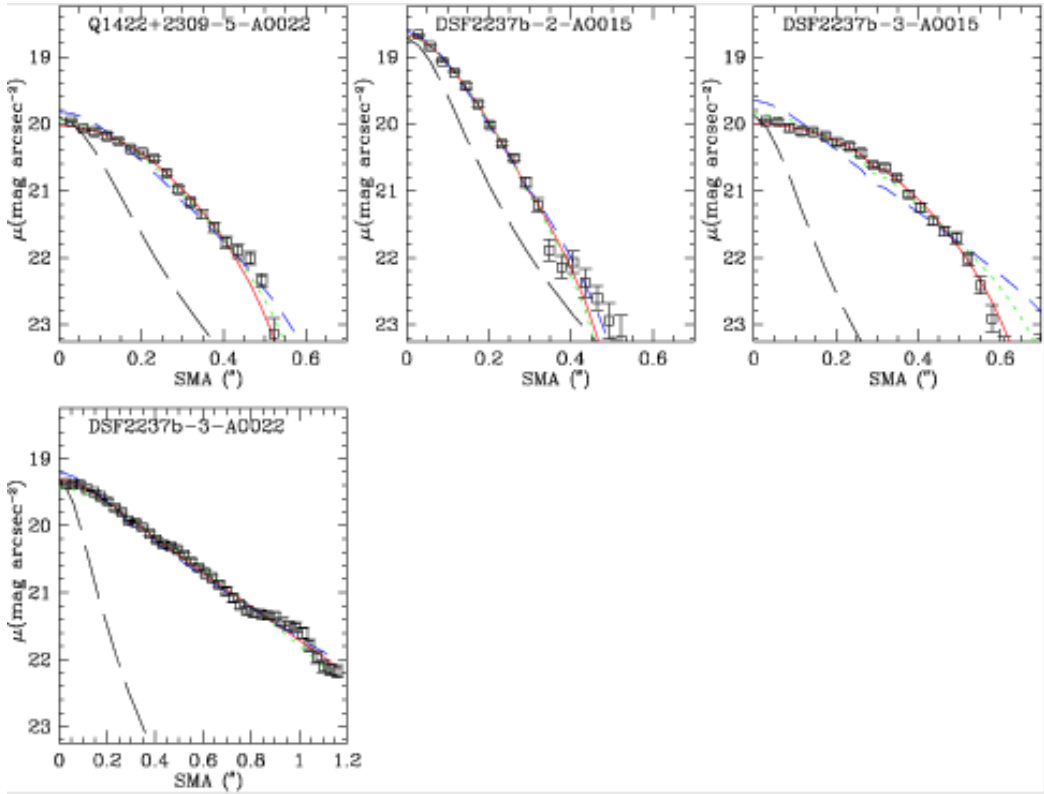}
\end{minipage}
\figcaption{
$K$-band profiles of the DRGs with $K<21.5$ mag along
the semi-major axis. The red-solid, blue-dashed, and
green-dotted lines show the best fit models with 
S\'ersic, $r^{1/4}$ (S\'ersic with $n=4$), and exponential (S\'ersic with $n=1$)
profiles, respectively.
Long dashed lines show the estimated profiles of the
PSFs at the target position.
\label{DRG_FIT_prof}}
\end{figure}

\begin{figure}
\begin{minipage}{0.9\linewidth}
\plotone{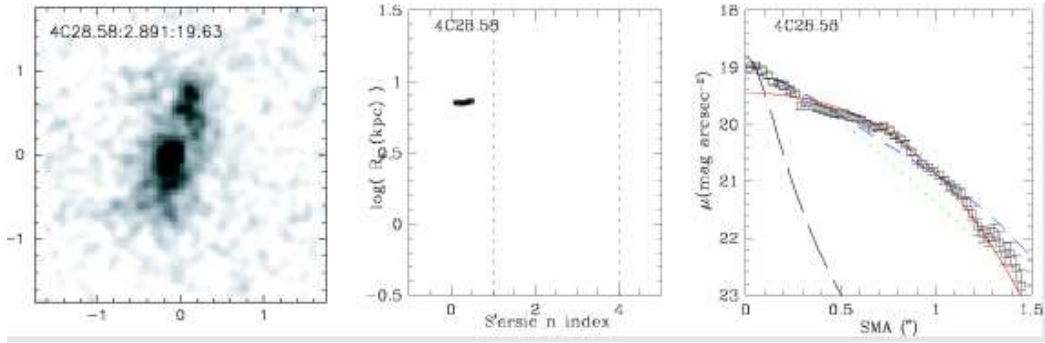}
\end{minipage}
\figcaption{
Left) $K$-band image of 4C28.58. 
A $3.\!^{\prime\prime}5\times3.\!^{\prime\prime}5$
FoV is shown.
Gaussian convolution with $\sigma$ of 1 pixel is applied for
the image. 
Middle) Distribution of the best fit S\'ersic 
$n$ index and $R_{e}$ for 4C28.58. Dashed lines
indicates $n=1$ (exponential-law) and $n=4$ ($r^{1/4}$-law)
for 4C28.58. Right) $K$-band profile of 4C28.58
along the semi-major axis. 
Red-solid, blue-dashed, and
green-dotted lines show the best fit S\'ersic model with free $n$,
$r^{1/4}$ (S\'ersic with $n=4$), and exponential (S\'ersic with $n=1$)
profiles, respectively.
Although the central part of the galaxy has a concentrated core,
the overall profile is also fitted well with the S\'ersic profile with small $n$ index of $0.3\pm0.2$.
The long dashed line show the estimated profiles of the PSF
at the position of the target.
\label{RG_FIT_ser_mod}}
\end{figure}

\end{document}